\documentclass[twocolumn, 10pt, journal]{IEEEtran}
\usepackage{color}
\usepackage{setspace}
\usepackage[cmex10]{amsmath}
\usepackage{psfrag}
\usepackage{subfigure}
\usepackage{hyperref}
\usepackage{multirow}
\usepackage{amsmath}
\usepackage{multicol}
\usepackage{setspace}
\usepackage{amsmath}
\usepackage{subfigure} 
\usepackage{multicol}
\usepackage{algorithmic}
\usepackage{algorithm}
\usepackage{epsfig}
\usepackage{cite}
\usepackage{graphicx} 

\begin{document}
\date{}
\title{%
 Geometry-Based Vehicle-to-Vehicle Channel Modeling for Large-Scale Simulation
} 
\author{
  	\IEEEauthorblockN{
  		Mate Boban\IEEEauthorrefmark{1}, %
		Jo\~{a}o Barros\IEEEauthorrefmark{2}, and %
		Ozan K. Tonguz\IEEEauthorrefmark{3}
  	}
	
	\IEEEauthorblockA{  
	 		\IEEEauthorrefmark{1}
            			NEC Laboratories Europe, NEC Europe Ltd. \\
	}
	
	\IEEEauthorblockA{ 
 			\IEEEauthorrefmark{2}
            			Instituto de Telecomunica\c{c}\~{oes}, Departamento de Engenharia Electrot\'{e}cnica e de Computadores, \\
            			Faculdade de Engenharia da Universidade do Porto\\
  	}
 	\IEEEauthorblockA{  
	 		\IEEEauthorrefmark{3}
            			Department of Electrical and Computer Engineering, 
            			Carnegie Mellon University \\
  	}

	Email: mate.boban@neclab.eu, jbarros@fe.up.pt, tonguz@ece.cmu.edu \\
	Web: \url{http://vehicle2x.net}
	\thanks{Copyright (c) 2013 IEEE. Personal use of this material is permitted. However, permission to use this material for any other purposes must be obtained from the IEEE by sending a request to pubs-permissions@ieee.org.}
	\thanks{This work was funded in part by the Portuguese Foundation for Science and Technology under grants SFRH/BD/33771/2009 and CMU-PT/NGN/0052/2008.} %

}

\maketitle
\begin{abstract}%

Due to the dynamic nature of vehicular traffic and the road surroundings, vehicle-to-vehicle (V2V) propagation characteristics vary greatly on both small- and large-scale.
Recent measurements have shown that %
both large static objects (e.g., buildings and foliage) as well as mobile objects (surrounding vehicles) have a profound impact on V2V communication. 
At the same time, system-level Vehicular Ad Hoc Network (VANET) simulators by and large 
employ simple statistical propagation models, which 
do not account for 
 surrounding objects explicitly. %
We designed GEMV$^2$ (Geometry-based Efficient propagation Model for V2V communication), which uses 
outlines of vehicles, buildings, and foliage to %
distinguish %
the following three types of links: line of sight (LOS), non-LOS due to vehicles, and non-LOS due to static objects. %
For each link, GEMV$^2$ %
calculates the large-scale signal variations deterministically, whereas the small-scale signal variations are calculated stochastically based on the number and size of surrounding objects.
We implement GEMV$^2$ in MATLAB %
and show that it scales well by using it to simulate radio propagation for city-wide networks with tens of thousands of vehicles 
on commodity hardware. %
We make the source code of GEMV$^2$ freely available. 
Finally, we validate GEMV$^2$ against extensive measurements performed in 
urban, suburban, highway, and open space environment. 

\end{abstract}%

\IEEEpeerreviewmaketitle
\section{Introduction} \label{sec:Intro}

Vehicular Ad Hoc Networks (VANET) research efforts have so far relied heavily on simulations, due to the prohibitive costs of deploying real world testbeds. %
Propagation models implemented in VANET simulators for both vehicle-to-vehicle (V2V) and vehicle-to-infrastructure (V2I) links are by and large simple statistical models (e.g., free space, log-distance path loss~\cite{rappaport96}, etc.). These models are computationally efficient and easy to implement; however, they
are applied to all links in the simulation indiscriminately, without taking into account specific link conditions (e.g., whether a link has a clear LOS path or it is obstructed by nearby objects). The reason for this simplification is that efficient modeling of the complex VANET surroundings is not trivial. 
While simple propagation models are useful for obtaining the overall statistics of VANET communication (e.g., average packet delivery rate, average communication range),
previous measurement studies have shown that these models %
 are often unable to accurately represent the link-level VANET communication, particularly in more complex environments (e.g., urban)~\cite{dhoutaut06,boban11}. Modeling the links accurately is particularly important for safety applications~\cite{bai06}, where the goal is to simulate whether the vehicle will receive a safety-critical message or not.

 On the other hand, existing geometry-based models, such as those based on ray-tracing \cite{maurer04}, yield results that are in a very good agreement with the real world. However, these models are computationally too expensive %
to be practically useful for modeling large-scale networks in VANET simulators. %
Other notable problems of ray-tracing models %
are the need for a detailed object database %
 and sensitivity to inaccuracies of the object database, which make it difficult to correctly predict the path of the reflecting and diffracting rays. %
For these reasons, ray-tracing models have not been implemented in large-scale VANET simulators.

In this study, we aim to bridge the gap between overly simplified statistical models and computationally expensive geometry-based models by performing location-specific propagation modeling with respect to large objects in the vicinity of the communicating vehicles, at the same time limiting the calculations by using only the simple representation of the objects (i.e., their outlines).
We use the real-world locations and dimensions of nearby buildings, foliage\footnote{Throughout the text, due to the lack of a more appropriate all-encompassing term, we use the term ``foliage'' for %
vegetation such as
 trees, %
bushes, shrubbery, etc.}, and vehicles to determine the line of sight conditions for each link. 
We showed in~\cite{boban11} and~\cite{meireles10} that vehicles are the most significant source of signal attenuation and variation in highway environments. In urban and suburban environments, apart from vehicles, static objects such as buildings and foliage have a significant impact on inter-vehicle communication~\cite{karedal10}. Therefore, there is a need for a model that incorporates both mobile and static objects to enable realistic modeling in different environments. %

We first %
analyze an extensive set of measurements (detailed in Section~\ref{sec:ExpSetupComplete}) to prepare the ground for designing a scalable %
V2V propagation model. We measure %
received power and packet delivery rate %
 in and around Pittsburgh, PA, USA, and Porto, Portugal, containing distinct environments where VANETs will be deployed: highway, suburban, urban, open space, and parking lot. We characterize the impact of vehicles and static objects (buildings and foliage) on the received power, packet delivery rate, and effective range.

The premise of our approach to modeling propagation for V2V communication is that line of sight (LOS) and non-LOS (NLOS) links exhibit considerably different channel characteristics. This is corroborated by numerous experimental studies %
(e.g.,~\cite{karedal10,tan08,boban11_2}), which have shown %
that the resulting channel characteristics for LOS and NLOS links are fundamentally different. %
Based on these studies and by using the findings from our previous work described in~\cite{boban11}, which identified surrounding vehicles as an important factor in V2V communication,
our approach %
is to use %
simple geographical descriptors of the simulated environment (outlines of buildings, foliage, and vehicles on the road) to %
classify V2V links into three groups:
\begin{itemize}
\item Line of sight (\textbf{LOS}) -- links that have an unobstructed optical path between the transmitting and receiving antennas;
\item Non-LOS due to vehicles (\textbf{NLOSv}) -- links whose LOS is obstructed by other vehicles;
\item Non-LOS due to buildings/foliage (\textbf{NLOSb}) -- links whose LOS is obstructed by buildings or foliage.
\end{itemize}

Based on the %
link classification, %
we design GEMV$^2$ (Geometry-based Efficient propagation Model for V2V communication), a propagation model that
separates links into LOS, NLOSv, and NLOSb link types and calculates deterministically the large-scale signal variation (i.e., path loss and shadowing) for each link type. %
Furthermore, %
GEMV$^2$ employs %
a simple geometry-based small-scale signal variation model that 
calculates stochastically the %
additional signal variation based on the information about the surrounding objects. 
GEMV$^2$ can use vehicle locations available from traffic mobility models (e.g., SUMO~\cite{krajzewicz2002sumo}) or real world traces (e.g., via GPS) and %
 the building and foliage outlines and locations that are freely available from projects such as OpenStreetMap~\cite{openstreetmap}. 
To efficiently process the object outlines, GEMV$^2$ employs computational geometry concepts suitable for representation of geographic data. Specifically, it uses R-trees~\cite{guttman84} to store information about the outlines of vehicles, buildings, and foliage. %
We provide more details on geographic data processing %
in Section~\ref{sec:SpatialTreeStructures}, whereas Section~\ref{sec:modelDescription} describes GEMV$^2$ in detail.

We validate GEMV$^2$ against extensive measurements and show that it successfully captures both small-scale and large-scale propagation effects for LOS, NLOSv, and NLOSb links in different environments (highway, urban, suburban, open space). 

We implement GEMV$^2$ in MATLAB and %
show that it scales well by simulating networks with up to tens of thousands of objects in the scene and hundreds of thousands of communicating pairs. We make the MATLAB source code of GEMV$^2$ freely available at~\url{http://vehicle2x.net}.
Since GEMV$^2$ requires minimum geographic information (the outlines of modeled objects), %
  it is well suited for implementation in VANET simulators. %
We provide the complete simulation recipe for its implementation %
  in packet-level, discrete-event VANET simulators. %

The rest of the paper is organized as follows. Details of the measurement setup are shown in Section~\ref{sec:ExpSetupComplete}. Section~\ref{sec:SpatialTreeStructures} explains the spatial tree structures we use to implement GEMV$^2$. Section~\ref{sec:modelDescription} describes GEMV$^2$ in detail, whereas %
Section~\ref{resultsComplete} presents the results validating GEMV$^2$ against measurements. We discuss the computational performance of GEMV$^2$ %
in Section~\ref{sec:Performance}. Section~\ref{sec:relWorkComplete} describes the related work, whereas Section~\ref{sec:conclusionsComplete} concludes the paper.

\section{Measurement Setup}\label{sec:ExpSetupComplete}%

\begin{figure}
  \begin{center}
  \subfigure[ Porto Downtown.]{\label{fig:DowntownPortoRoute}\includegraphics[height=0.22\textwidth]{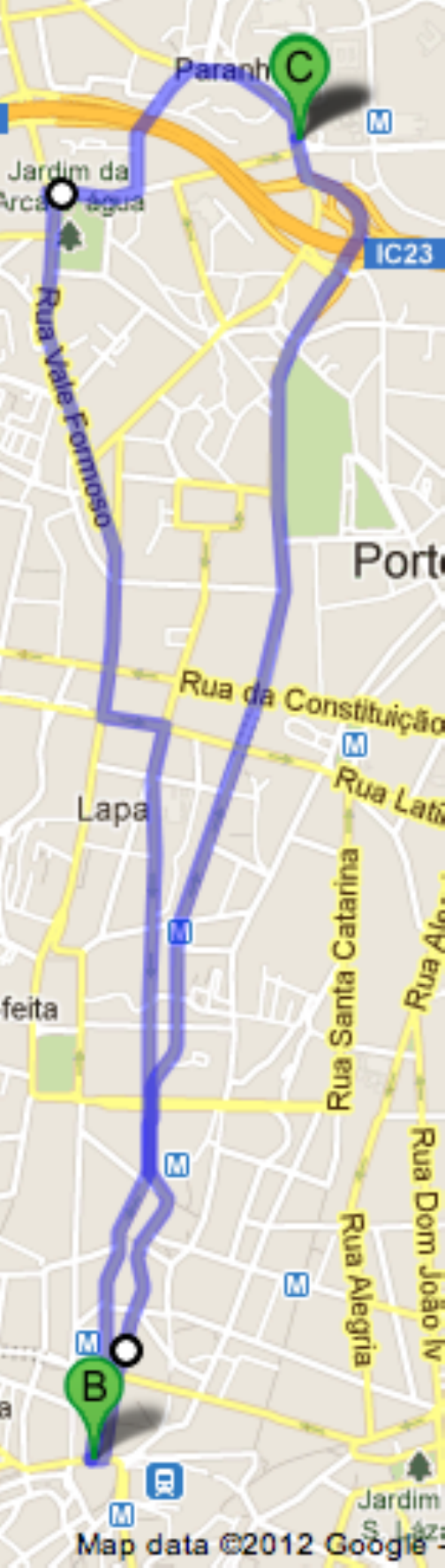}}\hspace{1mm}
  \subfigure[ Porto Open Space.]{\label{fig:LecaRoute}\includegraphics[height=0.22\textwidth]{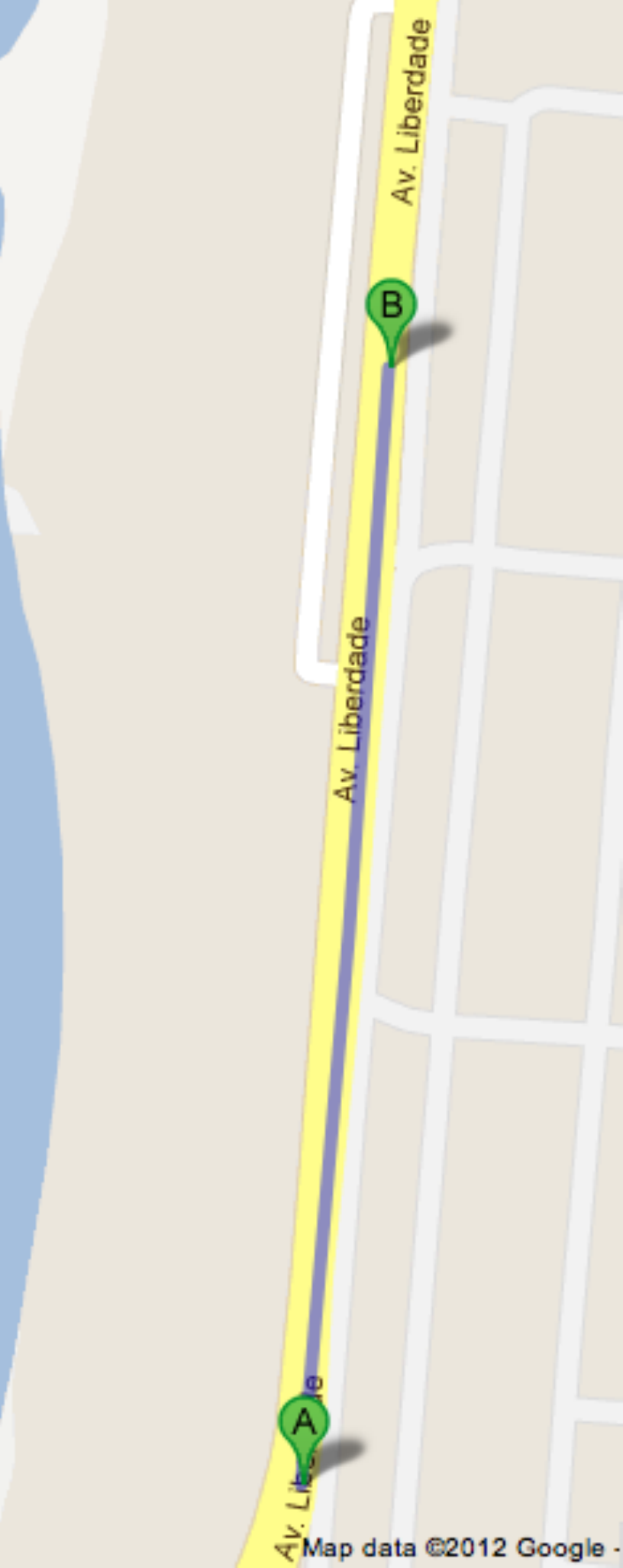}}\hspace{1mm}
\subfigure[Porto Urban Highway (VCI).]{\label{fig:VCIImgRoute}\includegraphics[height=.22\textwidth]{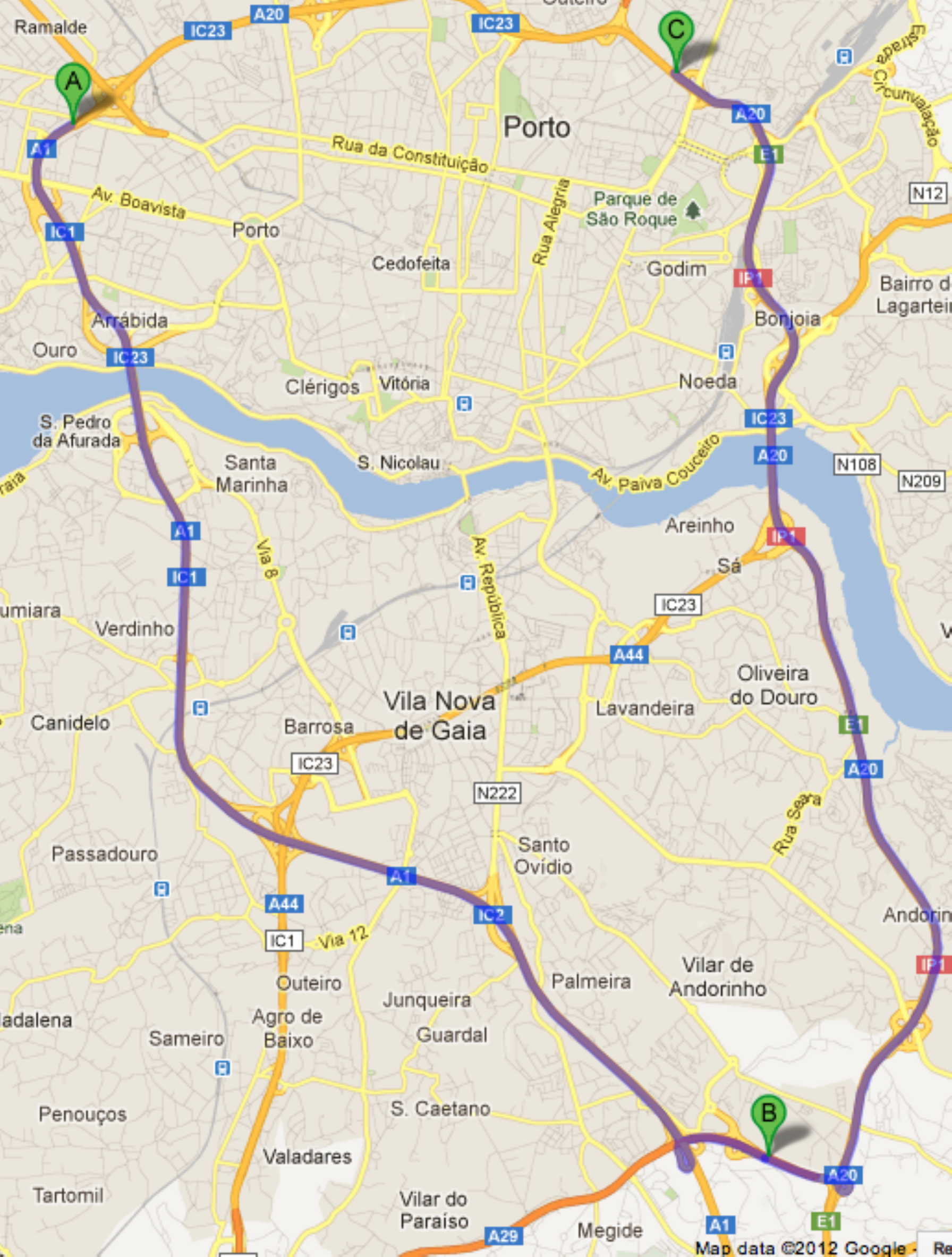}}  \hspace{1mm}
  \subfigure[Porto Highway (A28).]{\label{fig:A28ImgRoute}\includegraphics[height=.22\textwidth]{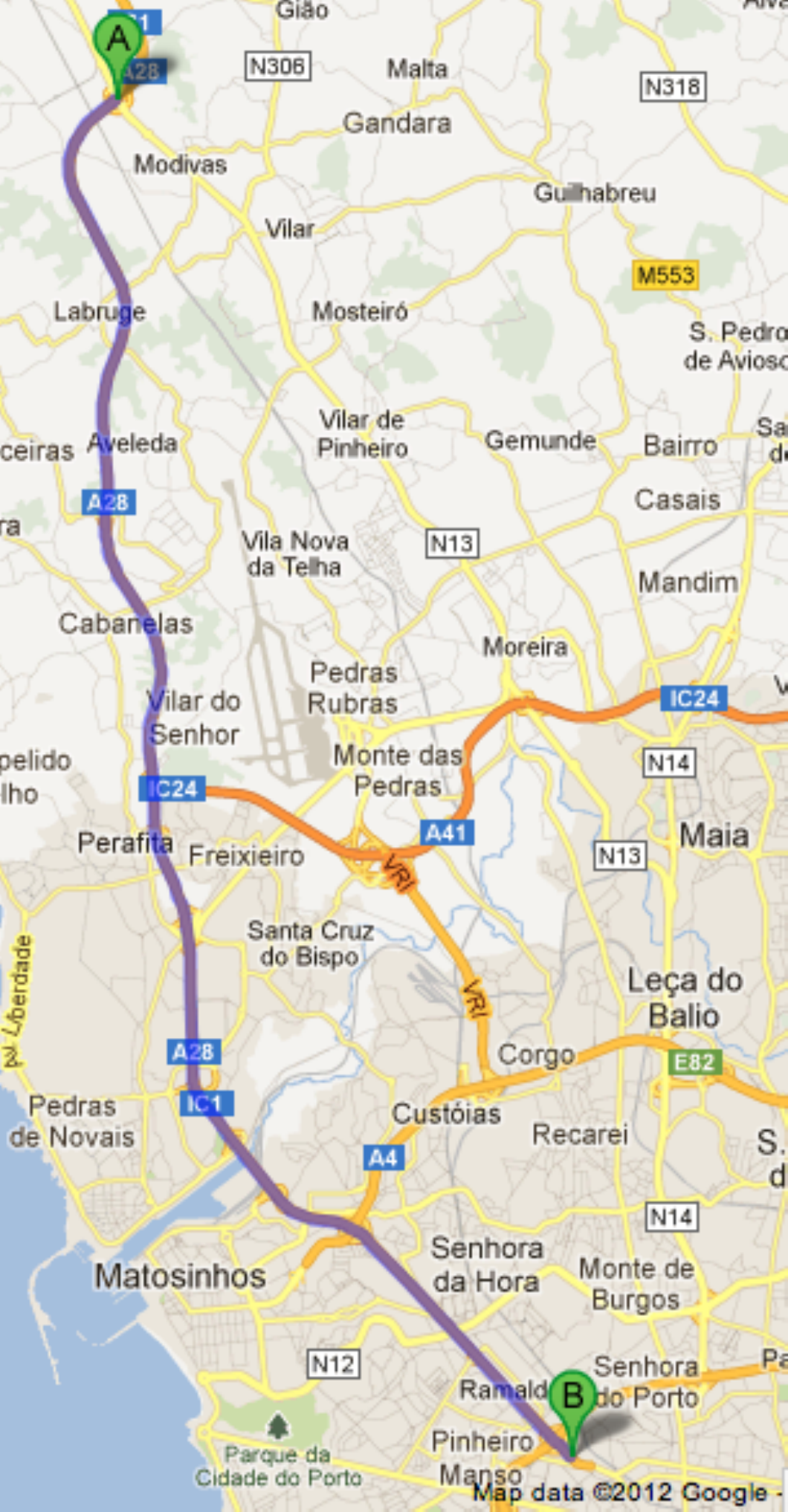}}\hspace{1mm}
  \subfigure[ Pittsburgh Suburban.]{\label{fig:5thRoute}\includegraphics[height=0.122\textwidth]{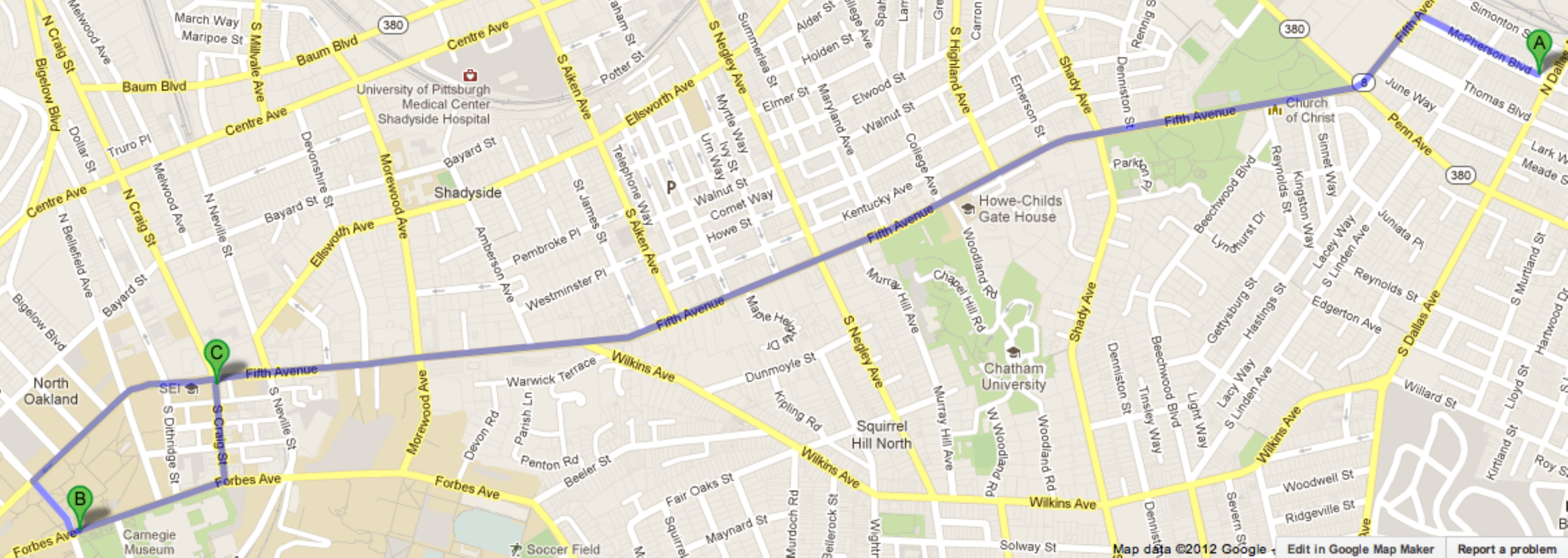}}\hspace{1mm}
  \subfigure[ Pittsburgh Open Space.]{\label{fig:HomesteadRoute}\includegraphics[height=0.122\textwidth]{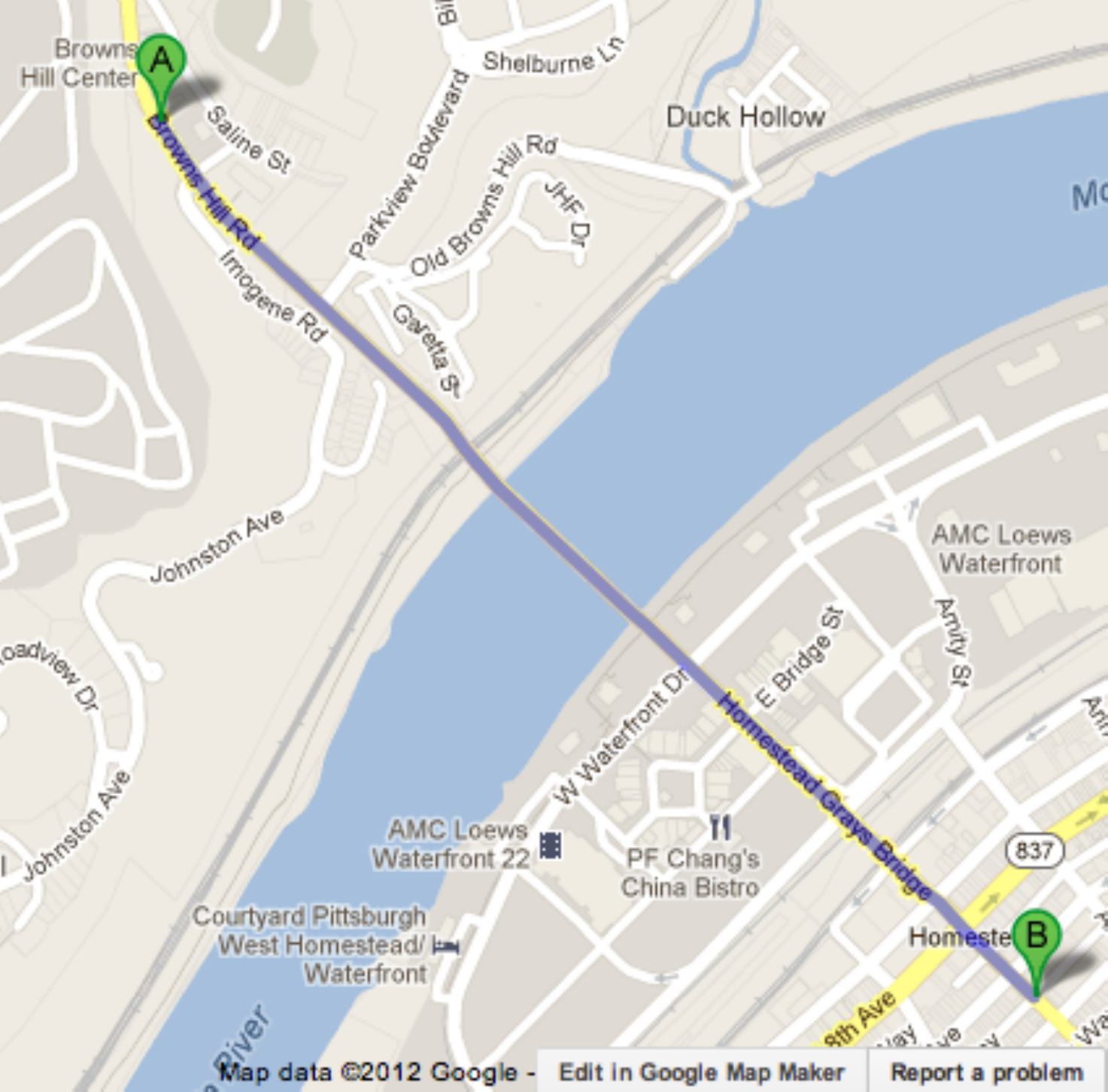}}
     \caption[Locations of the measurements]{Measurement locations with indicated routes. Figures are not in the same scale. }%
      \label{fig:ExperimentRoutes}
   \end{center}
\end{figure}

As a baseline for the validation of the model and to inform its design, %
we performed measurements in the following locations: %
\begin{itemize}
\item Porto Downtown  -- 9~km route shown in Fig.~\ref{fig:DowntownPortoRoute}, going from the Paranhos parish to the Avenida dos Aliados in downtown Porto and back. Approximate coordinates (lat,~lon): 41.153673, -8.609913; %
\item Porto Open Space -- 1~km route shown in Fig.~\ref{fig:LecaRoute}. Approximate coordinates (lat,~lon): 41.210615, -8.713418;
\item Porto Urban Highway (VCI) -- 24~km route shown in Fig.~\ref{fig:VCIImgRoute}. Approximate coordinates (lat,~lon): 41.1050224 -8.5661420; 
\item Porto Highway (A28) -- 13.5~km route shown in Fig.~\ref{fig:A28ImgRoute}. Approximate coordinates (lat,~lon): 41.22776, -8.695148; 
\item Porto Outlet -- shown in Fig.~\ref{fig:OutletOverlayReflDiffr}. Approximate coordinates (lat,~lon): 41.300137, -8.707385; 
\item Pittsburgh Suburban (5th Ave) -- 7~km route shown in Fig.~\ref{fig:5thRoute}. %
Approximate coordinates (lat,~lon):  40.4476089, -79.9398574; 
\item Pittsburgh Open Space (Homestead Grays Bridge) -- 2~km route shown in Fig.~\ref{fig:HomesteadRoute}. Approximate coordinates (lat,~lon):  40.4103279, -79.9181137). 
\end{itemize}

\begin{figure}
  \begin{center}
  \subfigure[ Porto Downtown]%
  {\label{fig:DowntownPorto}\includegraphics[height=0.112\textwidth]{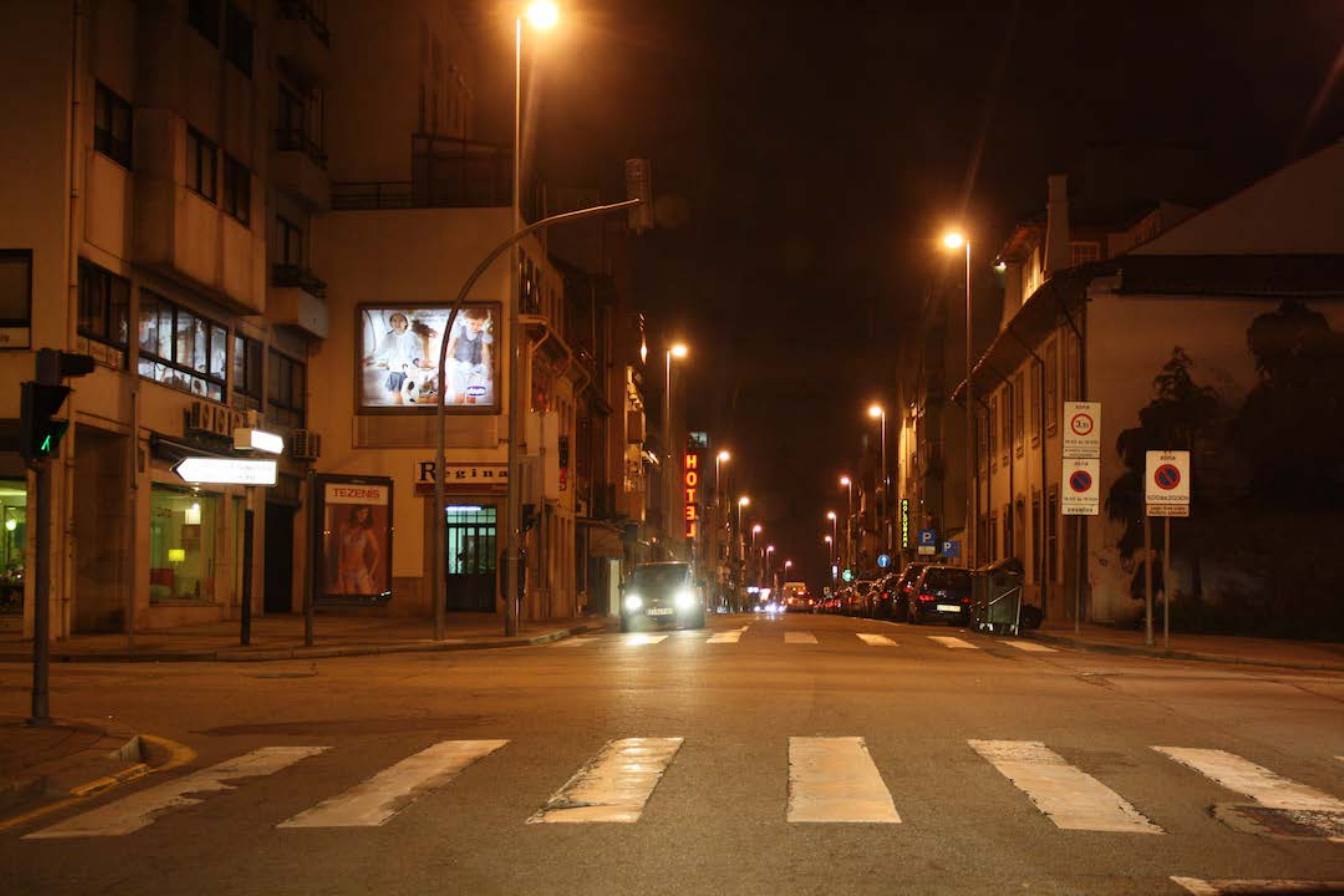}}
  \subfigure[ Porto Open Space]%
  {\label{fig:Leca}\includegraphics[height=0.112\textwidth]{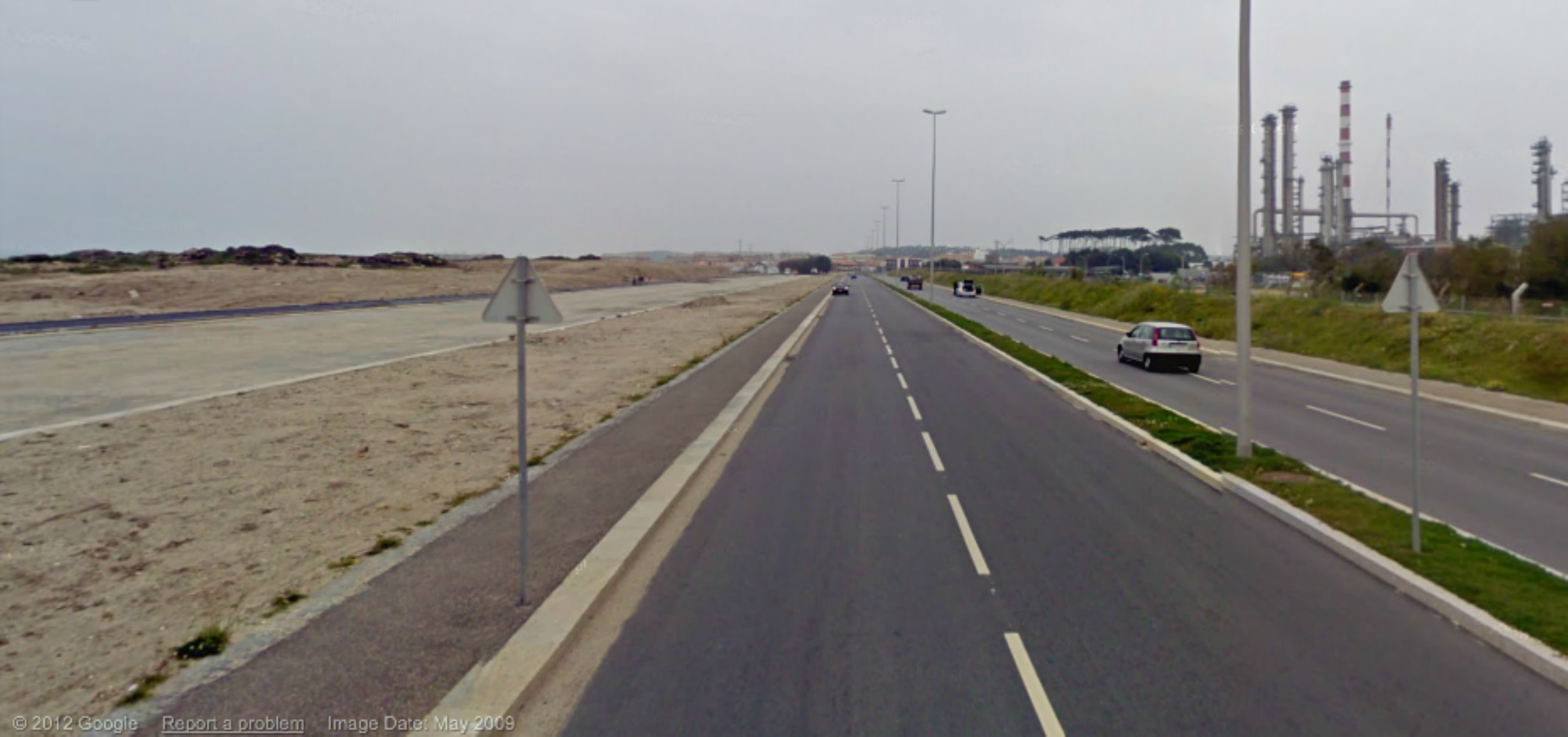}}
\subfigure[Porto Urban Highway (VCI)] %
{\label{fig:VCIImg}\includegraphics[height=0.102\textwidth]{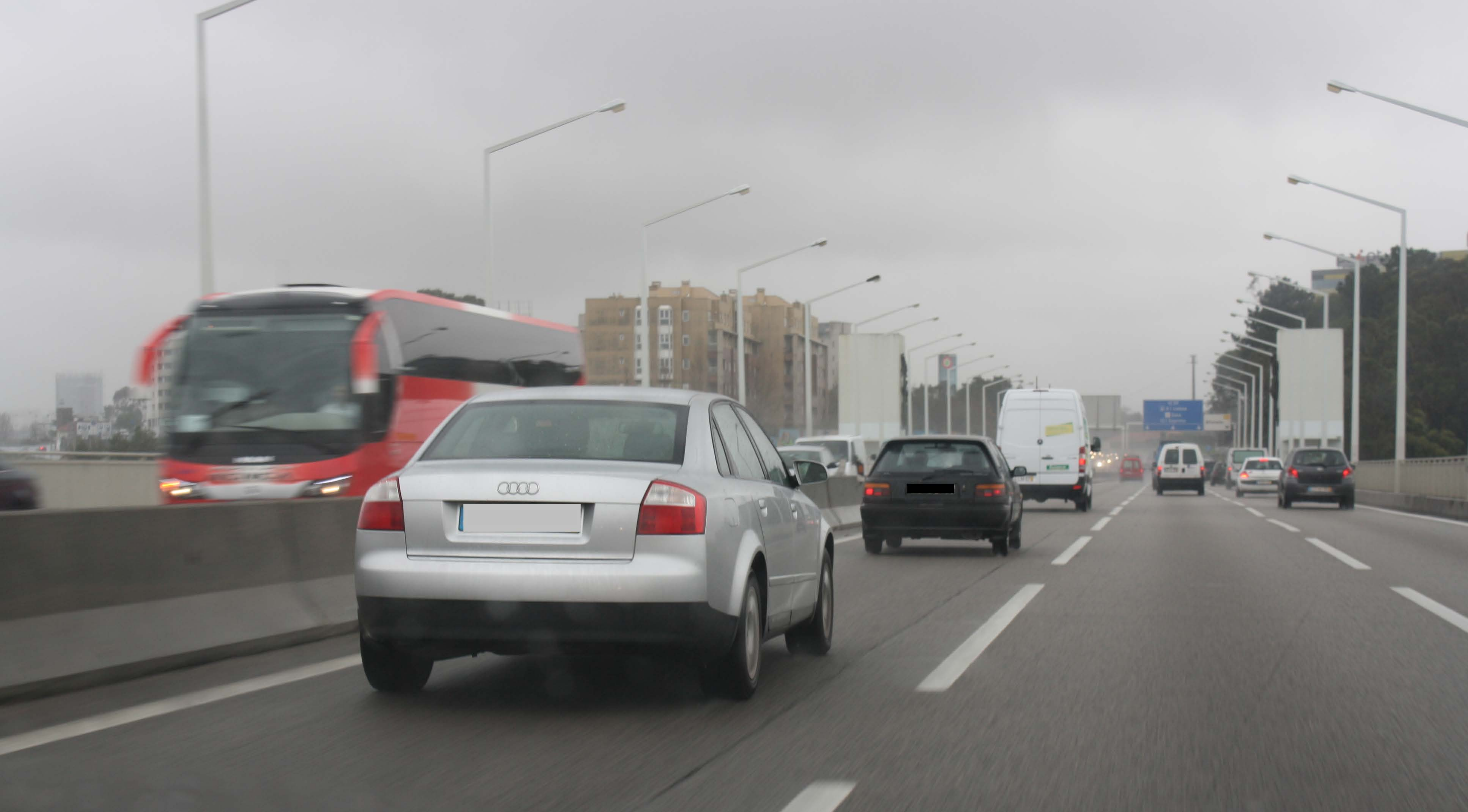}}  
  \subfigure[Porto Highway (A28)] %
  {\label{fig:A28Img}\includegraphics[height=0.102%
\textwidth]{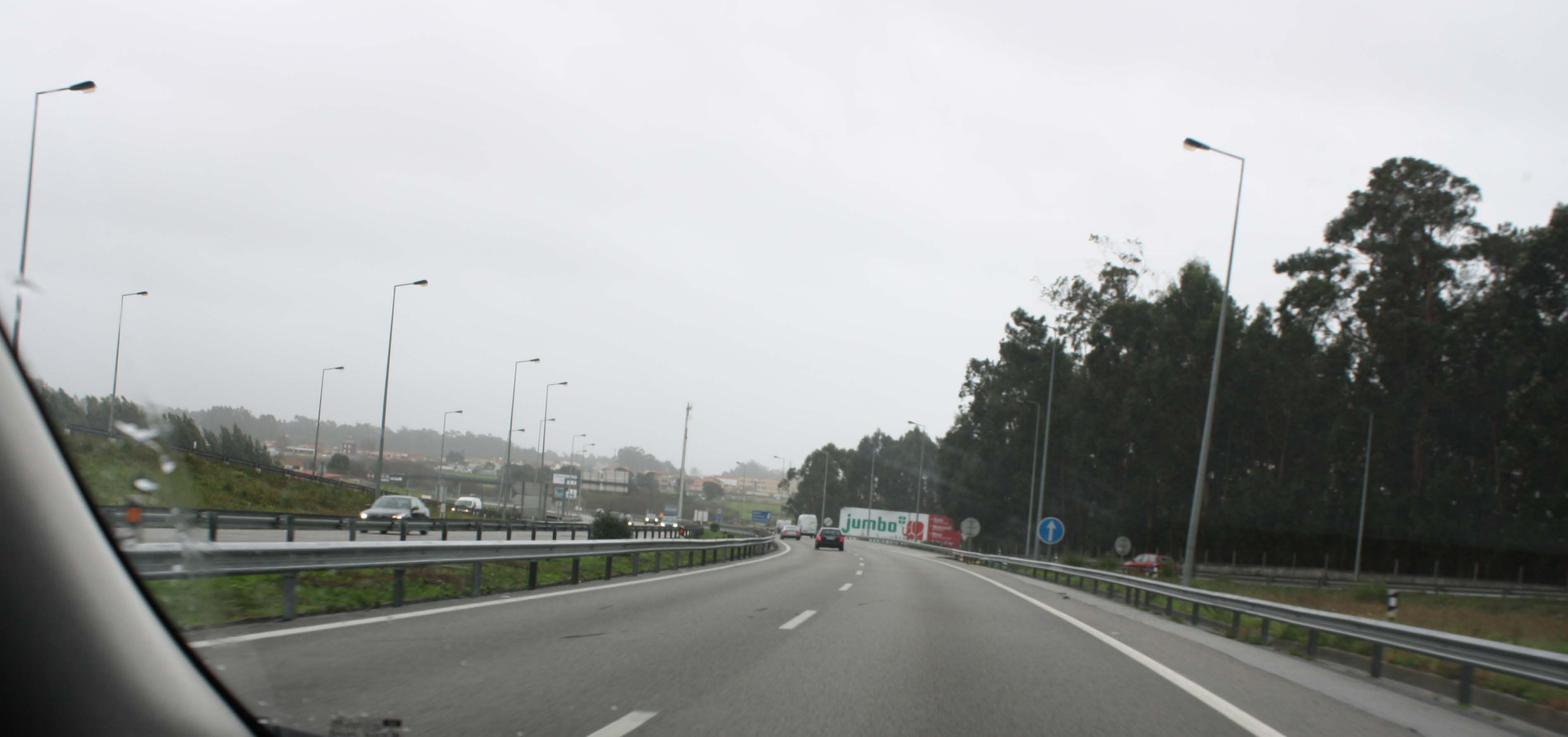}}
  \subfigure[ Pittsburgh Suburban] %
  {\label{fig:5th}\includegraphics[height=0.1\textwidth]{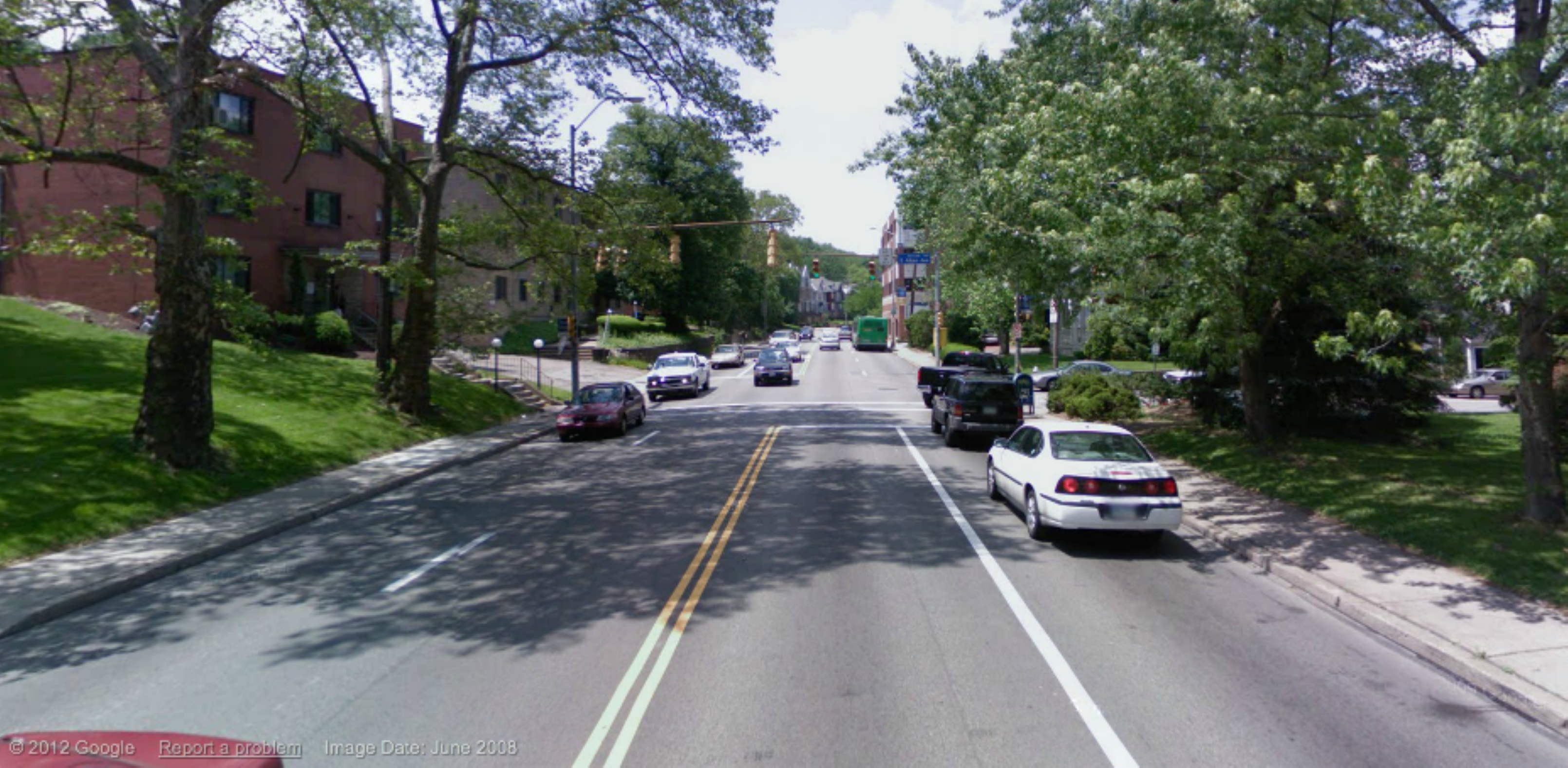}}
  \subfigure[ Pittsburgh Open Space] %
  {\label{fig:Homestead}\includegraphics[height=0.1\textwidth]{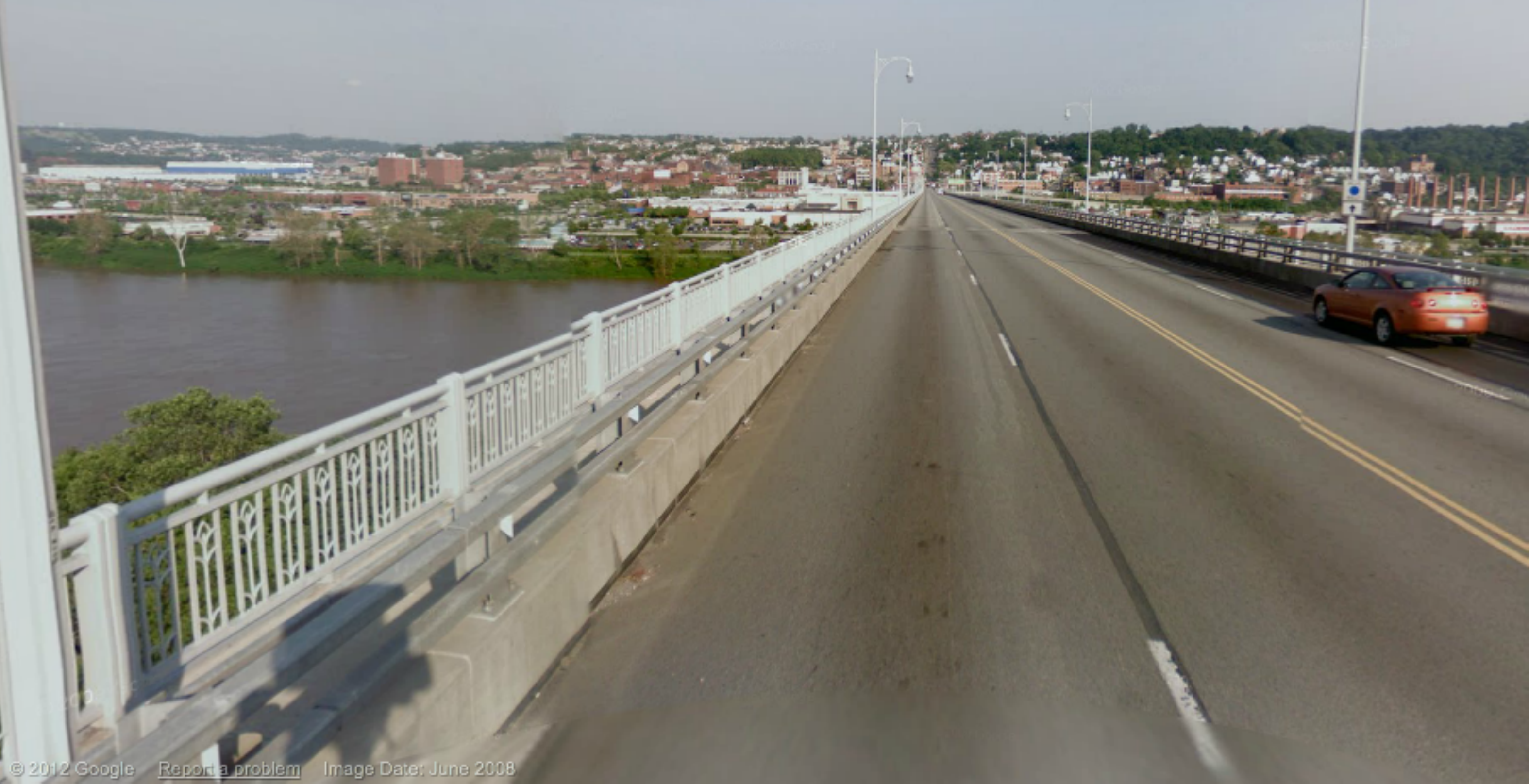}}
     \caption[Snapshots of the experiment locations]{Snapshots of the experiment locations.} %
      \label{fig:ExperimentLocations}
   \end{center}
\end{figure}

\begin{figure}
  \begin{center}
    \includegraphics[width=0.25\textwidth]{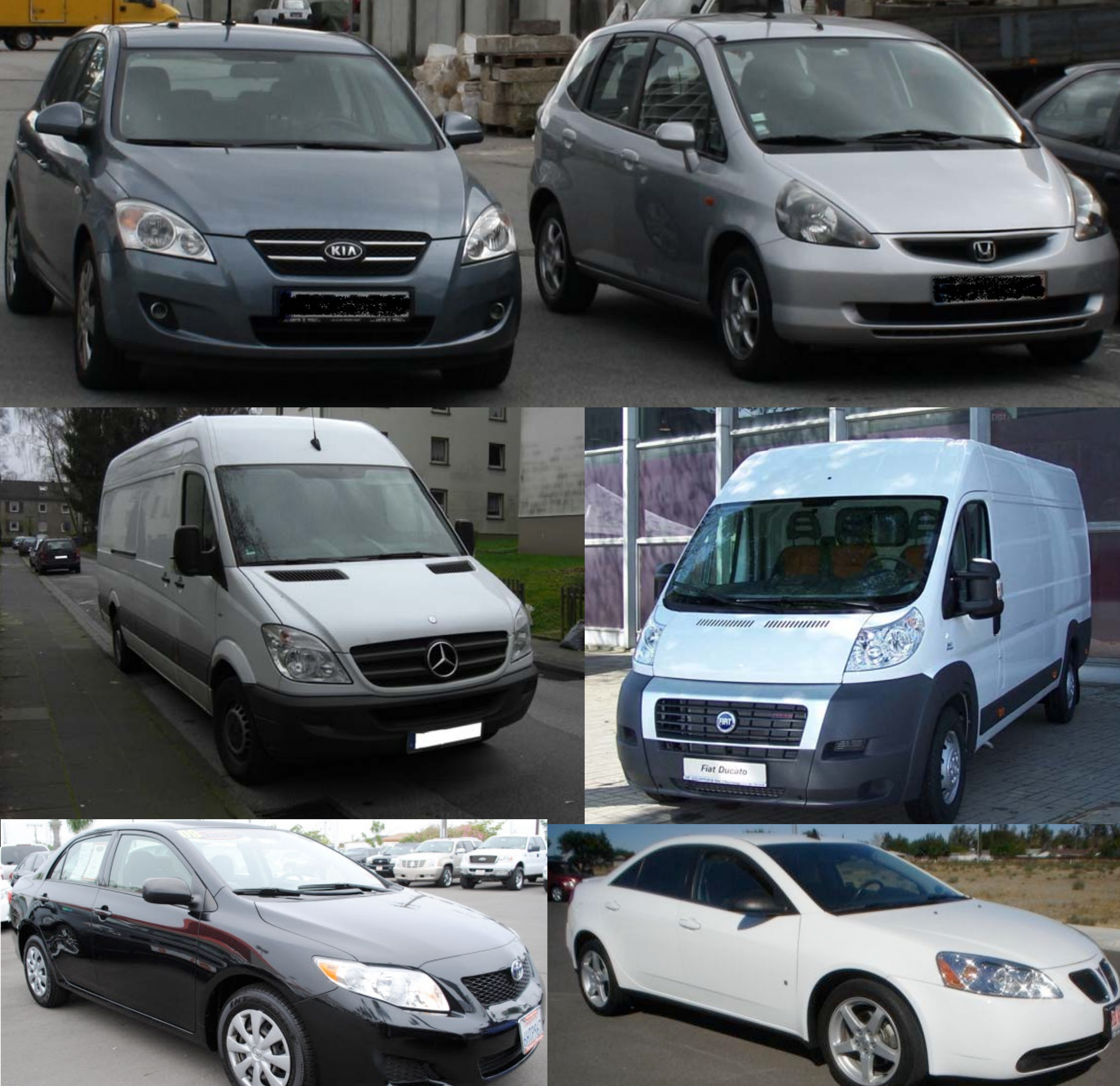}
     \caption[Vehicles used in the measurements]{Vehicles used in the measurements. First row: Kia Cee'd and Honda Jazz; second row: Mercedes Sprinter and Fiat Ducato; third row: Toyota Corolla and Pontiac G6.}
      \label{vehiclesCompleteModel}
   \end{center}
\end{figure}

Photographs of each of the measurement locations are shown in Fig.~\ref{fig:ExperimentLocations}. We performed measurements multiple times at each of these locations between May, 2010 and December, 2011. 
We used regular passenger cars and commercial vehicles depicted in Fig.~\ref{vehiclesCompleteModel}; %
their dimensions are listed in Table~\ref{tab:dimensionsCompleteModel}. %
Each vehicle was equipped with a NEC LinkBird-MX V3~\cite{festag08}, a development platform for vehicular communications that implements the IEEE 802.11p standard~\cite{ieee80211p}. IEEE 802.11p parameters are shown in Table~\ref{tab:hw-config}. Identical hardware setup and  parameters were used in all measurements.
We also performed measurements in downtown Pittsburgh. However, due to many high-rises taller than 100~meters, the GPS reception suffered from multipath that occasionally generated location errors in excess of 30 meters. %
Therefore, we did not include these results in our analysis. The buildings in downtown Porto are significantly lower, thus the GPS location information was more accurate: we observed only slight deviations of the vehicles from the roads they were traveling on. Furthermore, since the update frequency of the GPS is 1~Hz (i.e., one location update per second) and the vehicles occasionally moved at high speeds (up to 30 meters per second), to get a more accurate location of the vehicles, we performed dead reckoning on the GPS location data. For each received packet, we linearly interpolated the position between the two consecutive GPS location updates that occurred before and after the packet reception.

We separated the collected data into LOS, NLOSv, and NLOSb category using the videos we recorded from the trailing vehicle. %
Based on a technique from our previous measurement studies~\cite{meireles10,boban11_2,boban13}, we synchronized the videos with the measurements using visual recognition of known geographical locations %
and GPS coordinates from the received messages. In order to avoid misalignment of %
videos and received messages,
we periodically re-synchronized the videos\footnote{Since the separation of measured data into LOS, NLOSv, and NLOSb links was done manually (by watching videos), %
there might exist occasional classification errors %
(e.g., while transitioning from one link type into another).}.

\begin{figure}
\centering
\includegraphics[width=0.45\textwidth]{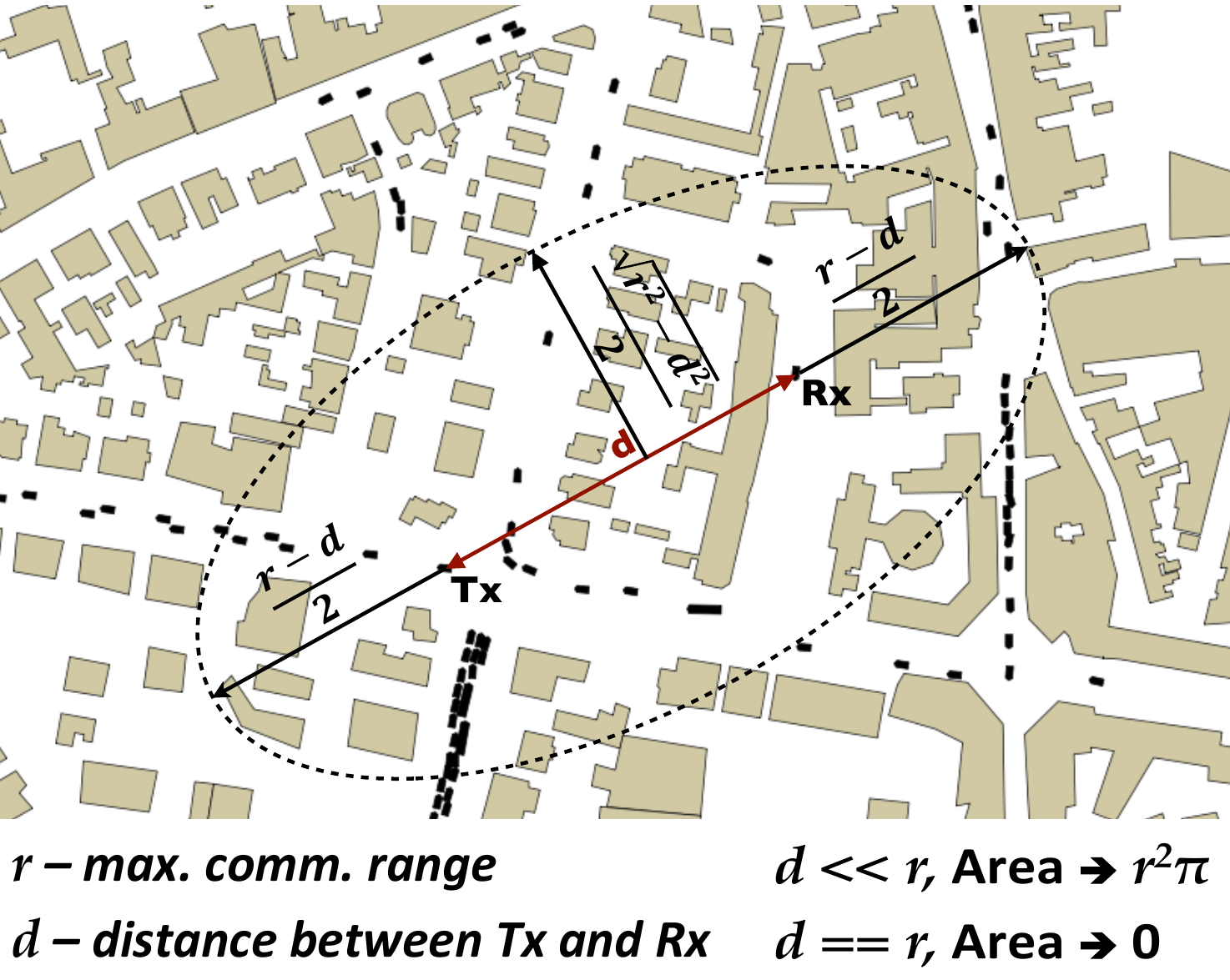} %
\caption[Explanation of the area used to determine fading, diffractions, and reflections]{Outlines of the buildings and vehicles (vehicles colored black) extracted from aerial imagery in a neighborhood of Porto, Portugal with the ellipse-bound area that the model uses to determine small- and large-scale signal variations. The search space encompasses the ellipse whose foci are the transmitting (Tx) and receiving (Rx) vehicles. This ensures that all objects whose sum of distances to Tx and to Rx (i.e., from Tx to object and from object to Rx) is less than $r$ (maximum communicating distance for a given environment) are accounted for. %
The objects in the same area are also used to calculate the small-scale signal variations. Note that the length of the major diameter of the ellipse is $r$, irrespective of the distance $d$ between Tx and Rx. The minor diameter's length is $\sqrt{r^2-d^2}$. The area of the ellipse is largest when Tx and Rx are close together, and the area goes to zero as the distance between Tx and Rx $d$ goes to $r$.
}\label{NLOSExplanation}
\end{figure}

We used %
object outlines in the city of Porto, Portugal, summarized in Table~\ref{tab:PortoDataset} and described in detail in~\cite{ferreira09}. %
A snapshot of the data is shown in Fig.~\ref{NLOSExplanation}. We also used the building and foliage outlines from OpenStreetMap~\cite{openstreetmap}. %

\begin{table}
	\begin{footnotesize}
	  \begin{center}
		\caption{Dimensions of Vehicles Used in Measurements} 
\begin{tabular}{|c c c c|} \hline
		 & \multicolumn{3}{c|}{\bf Dimensions (m)} \\ %
		 \bf Vehicle & \bf Height & \bf Width & \bf Length \\ \hline \hline
		\emph{Portugal} & & & \\ \hline
		2007 Kia Cee'd & 1.480  & 1.790 & 4.260\\ \hline
		2002 Honda Jazz & 1.525 & 1.676 & 3.845\\ \hline 
		2010 Mercedes Sprinter & 2.591 & 1.989 & 6.680\\ \hline
		2010 Fiat Ducato & 2.524 & 2.025 & 5.943\\ \hline
		\emph{USA} & & & \\ \hline
		2009 Toyota Corolla & 1.466 & 1.762 & 4.539\\ \hline
		2009 Pontiac G6 & 1.450 & 1.793 & 4.801\\ \hline 
		\end{tabular} 
\label{tab:dimensionsCompleteModel}
  \end{center}
	\end{footnotesize}
\end{table}

\begin{table}
	\begin{footnotesize}
	\begin{center}
			\caption{Hardware configuration parameters}
		\begin{tabular}{|l|c|c|}
			\hline
			\textbf{Parameter} & \textbf{802.11p}\\
			\hline
			Channel & 180\\
			\hline
			Center frequency (MHz) & 5900\\
			\hline
			Bandwidth (MHz) & 20\\
			\hline
			Data rate (Mbps) & 6\\
			\hline
			Tx power %
			& 10\\
			\hline
			Antenna gain (dBi) & 5\\
			\hline
			Beacon frequency (Hz) & 10\\
			\hline
			Beacon size (Byte) & 36\\
			\hline
		\end{tabular}

		\label{tab:hw-config}
	\end{center}
\end{footnotesize}
\end{table}

\begin{table} 
	\begin{footnotesize}
	\centering
		\caption[Porto Downtown Buildings and Vehicle Dataset]{Porto Downtown Buildings and Vehicle Dataset (more details available in~\cite{ferreira09})}
		\begin{tabular}{|c c c c c|}
 \hline \textbf{City area} & \textbf{\# buildings} & \textbf{Area} & \textbf{\# vehicles} & \textbf{\# tall vehicles}  \\
			& &  \textbf{of buildings} & & \\
			\hline
\hline 41.3~km$^2$ & 17346 & 8.6~km$^2$ & 10566 & 595 (5.6\%) \\ 
\hline
\end{tabular}
	\label{tab:PortoDataset}
		\end{footnotesize}
\end{table}

\section{Using R-trees for Efficient VANET Object Manipulation}
\label{sec:SpatialTreeStructures}

Before we discuss the structure of GEMV$^2$, we introduce the spatial tree structure we use for efficient VANET object manipulation. 
For a description of the modeled area, 
we use the outlines of vehicles, buildings, and foliage. Outlines of buildings and foliage are available through free geographic databases such as OpenStreetMap~\cite{openstreetmap}. Such sources of %
 geographical descriptors have become available recently, with a crowdsourced approach to geographic data collection and processing. %
Apart from the outlines of buildings and foliage available in such databases, we also use outlines and locations of the vehicles. Locations of the vehicles can be obtained from vehicular mobility models (e.g.,~\cite{krajzewicz2002sumo}), GPS logs, or aerial photography~\cite{ferreira09}, whereas the dimensions of vehicles can either be measured or drawn from statistical distributions~\cite{boban11}. %

\begin{figure}
\centering
\includegraphics[width=0.45\textwidth]{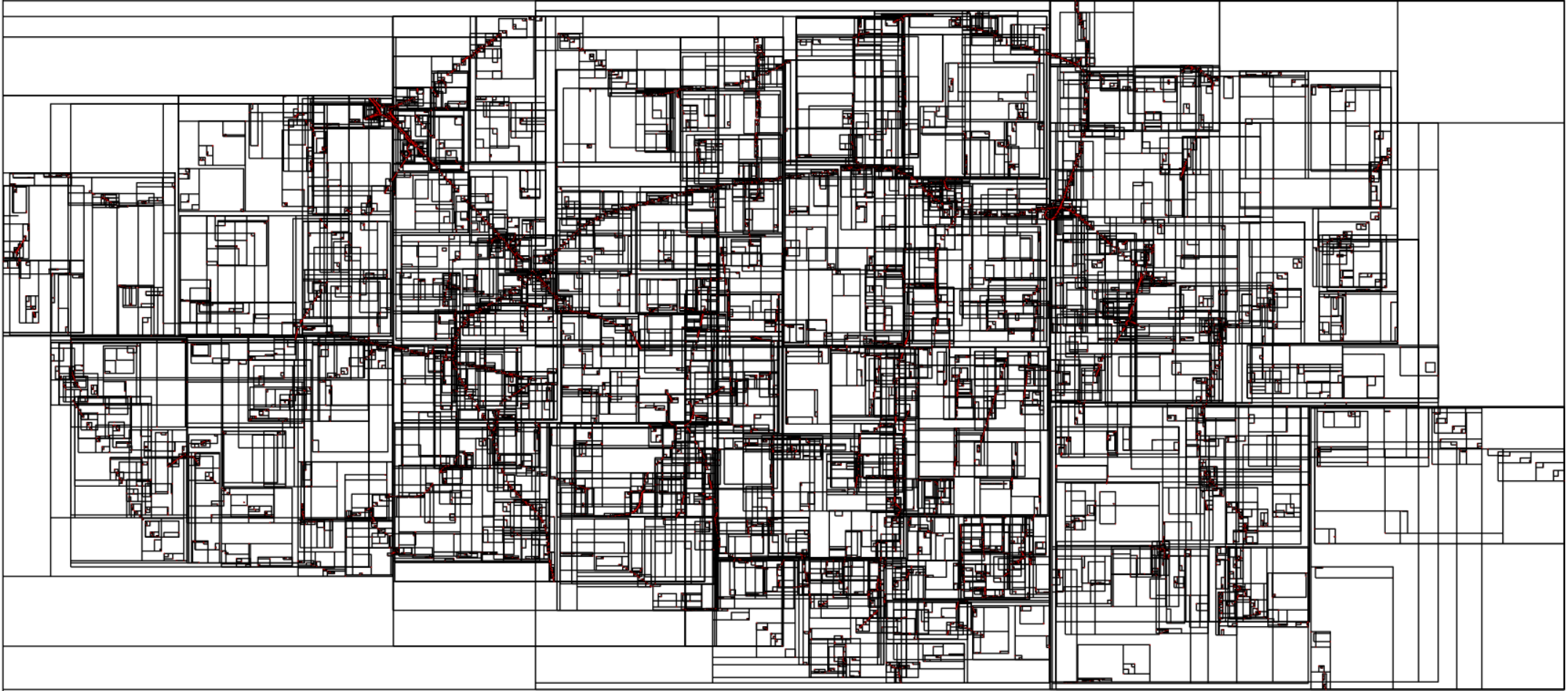} %
\caption{R-tree built atop vehicles in the city of Porto, Portugal}\label{BVHPorto}
\end{figure}

In networks with hundreds (or thousands) of vehicles, checking whether two nodes can communicate using a na\"{\i}ve approach (i.e., checking each node against each other node) is computationally too expensive. Therefore, in order to model large networks, efficient data structures are required. Based on the outlines of the objects, we form R-trees%
~\cite{guttman84}. %
R-tree is a tree data structure in which objects in the field are bound by rectangles and structured hierarchically based on their location in space. %
VANET-related geometric data lends itself to an efficient R-tree representation, due to its inherent geometrical structure (namely, relatively simple, non-overlapping object outlines). 
R-trees are often used to store spatial objects (streets, buildings, geographic regions, counties, etc.) in geographic databases. Even though they do not have good worst-case performance\footnote{When bounding rectangles of all objects overlap in a single point/area, the operation of checking the object intersection %
is quadratic in the number of objects in the R-tree (i.e., it is the same as the na\"{\i}ve approach that checks for intersection of every object with every other object). However, such extreme situations do not occur when modeling vehicular environments.}, in practice they were shown to have good tree construction and querying performance, particularly when the stored data has certain properties, such as limited object overlap~\cite{theodoridis96}. 
We store vehicle outlines in a separate R-tree. The main difference in storing the outline of vehicles when compared to buildings and foliage is that, unlike vehicles, buildings and foliage do not move, therefore their R-tree needs to be computed only once. %
On the other hand, the vehicle R-tree changes at each simulation time-step. %
Figure~\ref{BVHPorto} shows the R-tree built on top of the outlines of vehicles in the city of Porto for one time snapshot obtained through aerial photography. 

We construct each tree using a top-down approach, whereby the algorithm starts with all objects (i.e., vehicles, buildings, or foliage) and splits them into two child nodes (i.e. we use a binary R-tree). %
To keep the tree balanced, we sort the objects at each node splitting based on the currently longer axis so that each created child node contains approximately half of the objects. We note that similar tree data structures, such as \emph{k}-d tree and quadtree/octree, could be used instead of R-tree, with consideration to the specific application at hand and limitations and advantages of a specific data structure (for details, see de Berg et.~al~\cite{berg97}).

\section{Description of GEMV$^2$ }\label{sec:modelDescription}

\begin{figure}
\centering
\includegraphics[width=0.45\textwidth]{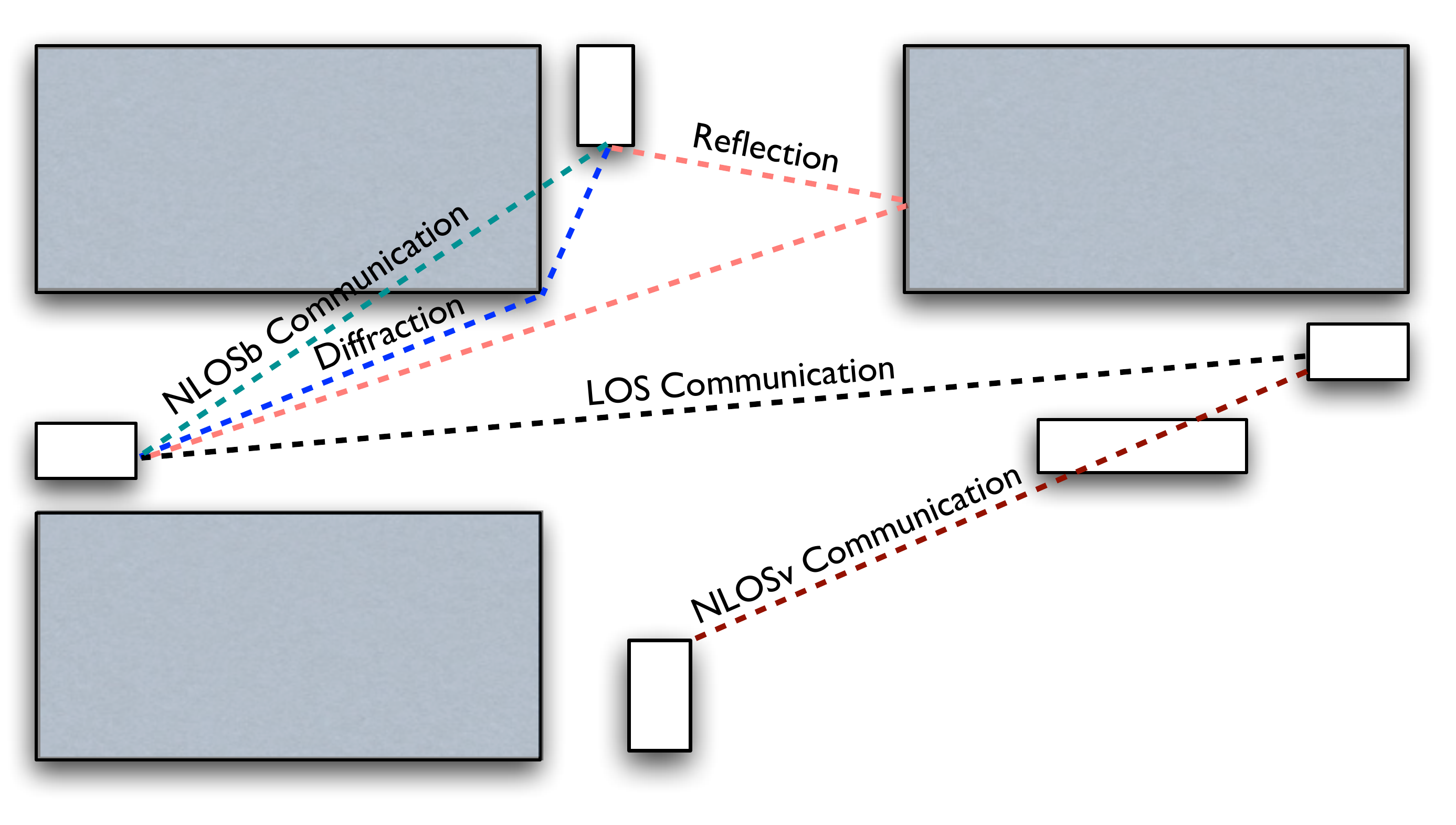} %
\caption{Link types and propagation effects captured by GEMV$^2$. White rectangles represent vehicles; gray rectangles represent buildings.}\label{modelDescription}
\end{figure}

In addition to LOS propagation, GEMV$^2$ incorporates the following propagation effects (shown in Fig.~\ref{modelDescription}): 1) transmission (propagation through material); 2) diffraction; and 3) reflection. %
We focus on modeling the impact of vehicles, buildings, and foliage (as opposed to smaller objects such as traffic signs, traffic lights, etc.) for two reasons. First, on highways, obstructing vehicles are the most important objects for modeling the V2V channel, %
as the roads are predominantly straight and the largest portion of communication happens over the face of the road~\cite{boban11}. In urban areas, obstructing vehicles have a significant impact for communicating pairs that are on the same street~\cite{meireles10}. 
However, when vehicles are on different streets, 
static obstructions such as buildings and foliage also play an important role~\cite{cardote11,mangel11}. Buildings and foliage are the main source of obstructions for communication on the intersections and across different streets~\cite{karedal10}, whereas buildings and vehicles are the main sources of reflections and diffractions~\cite{paier07_2}. Furthermore, other static objects such as lamp posts, street signs, railings, etc.,  are neither readily available in geographic databases, nor  would it be computationally feasible to model them due to their number, shape, and size. %
We validate GEMV$^2$ against extensive measurements performed in typical VANET environments. Based on the measurements, we limit the model's complexity to a point where it represents  
the real world well, but %
requires orders of magnitude less computations than more complex geometry-based models (as discussed in detail in Section~\ref{sec:Performance}). %

Propagation models for V2V communication need to incorporate different propagation mechanisms, typically divided by their scale into: %
\textit{path loss} (distance-dependent attenuation); \textit{large-scale fading} (variations including, but not limited to, shadowing by objects significantly larger than the carrier wavelength); and \textit{small-scale fading} (variations due to multipath and/or Doppler spread). 
As will become apparent in this section, parts of GEMV$^2$ do not directly translate %
into these three groups, since for the three link types, GEMV$^2$  handles these mechanisms differently. For this reason, in the rest of the paper, we use the term \textit{large-scale signal variations} for effects roughly pertaining to path loss and large-scale fading, and the term \textit{small-scale signal variations} for all effects that cause variations of the signal over distances of up to a few tenths of wavelengths. %
For example, small-scale signal variations can include a combination of the following: multipath due to single and higher order diffractions and reflections, scattering, Doppler spread, variation in the type of the obstruction object (e.g., different shape of the obstructing vehicle in case of NLOSv), etc. Accordingly, we classify the models used for these two groups of effects as those modeling large- or small-scale signal variations.

\subsection{Classification of link types}
As mentioned previously, %
we distinguish three types of links: 1) line of sight (LOS); 2) non-LOS due to vehicles (NLOSv); and 3) non-LOS due to buildings/foliage (NLOSb).
Using the insights from measurements we performed in different environments (Section~\ref{sec:ExpSetupComplete}), we apply different propagation models for each of the three link types and for both large- and small-scale signal variations.  Table~\ref{tab:linkTypes} shows the employed models. Specifically, through measurements in open space, urban, suburban, and highway environments, as well as by consulting the existing V2V measurements (e.g.,~\cite{kunisch08,boban13ACM,karedal11TVT}), we concluded that large-scale signal variations for LOS links are well approximated by a two-ray ground reflection model, whereas for NLOSv links we use an experimentally validated model %
from~\cite{boban11}, which we extend to include diffractions off the sides of vehicles, in addition to diffractions over the vehicle roofs. %
For NLOSb links, we calculate single-interaction reflections and diffractions to account for the ``around the corner'' communication, and log-distance path loss~\cite{rappaport96} for cases where single-interaction rays are either non-existent or carry low power. To increase the execution speed, GEMV$^2$ allows for utilizing only log-distance path loss for NLOSb links (i.e., without calculating reflections and diffractions). For small-scale signal variations, we design a simple stochastic model described in Section~\ref{subsec:fadingModel}, which takes into account the number and density of the objects around the communicating pair.

\begin{table} 
	\centering
	\begin{footnotesize}
\caption[Modeling different link types]{Propagation models used for different link types}
		\begin{tabular}{|c c c|}
\hline \textbf{Link Type} & \multicolumn{2}{c|}{\textbf{Propagation Model}}\\
  &  \textbf{Large-scale} & \textbf{Small-scale} \\
		 	\hline 
\hline \textbf{LOS} & Two-ray ground reflection with & Section~\ref{subsec:fadingModel}\\
 & effective reflection coefficient~\cite{kunisch08,boban13ACM} & \\
\hline \textbf{NLOSv} & Vehicles-as-obstacles \cite{boban11} & Section~\ref{subsec:fadingModel} \\
 & with side diffractions (Section~\ref{subsubsec:NLOSv}) &  \\
\hline \textbf{NLOSb} & Log-distance path loss %
& Section~\ref{subsec:fadingModel} \\
 & \multicolumn{2}{c|}{Optional: reflections \& diffractions (Section~\ref{sec:practicaConsLinks})} \\
\hline
\end{tabular}
	\end{footnotesize}
	\label{tab:linkTypes}
\end{table}

\subsection{Modeling large-scale signal variation}\label{sec:practicaConsLinks}

\subsubsection{LOS communication}

For LOS links, we implement the complete two-ray ground reflection model given by~\cite[Chap. 3]{rappaport96}:

\begin{align}
|E_{TOT}| 	&=\frac{E_0d_0}{d_{LOS}}\cos\left(\omega_c\left(t-\frac{d_{LOS}}{c}\right)\right)  \\
			&+ R_{ground}\frac{E_0d_0}{d_{ground}}\cos\left(\omega_c\left(t-\frac{d_{ground}}{c}\right)\right), \nonumber
\end{align}
where the reflection coefficient $R_{ground}$ and distance $d_{ground}$ for the ground-reflected ray are calculated according to the exact antenna heights
(i.e., we do not assume that the distance between transmitter and receiver is large compared to heights of the vehicles, as is often done in simulators~\cite{ns2},~\cite{ns3}). As will become apparent from our results (Section~\ref{resultsComplete}), using the exact height of the antennas is important, since even a centimeter-grade difference in height of either Tx or Rx results in significantly different interference relationship between the LOS and ground-reflected ray. %

In calculating $R_{ground}$, we model the relative permittivity $\epsilon_r$ to obtain the ``effective'' range of the reflection coefficient for the road.  As pointed out in~\cite{kunisch08}, the idealized two-ray model is an approximation of the actual V2V channel, since the reflection coefficient is affected by the antenna location, diffraction over the vehicle roof below antenna, and the roughness of the road. %
Therefore, we set the $\epsilon_r$ value used to generate the LOS results to 1.003, as this value %
minimized the mean square error for Porto Open Space dataset (see Section~\ref{resultsComplete}). Then, we use the same $\epsilon_r$ value for LOS links in all environments. Similar concept of effective reflection coefficient range calculation was used in~\cite{karedal11TVT} and~\cite{kunisch08}.

\subsubsection{NLOSv communication}\label{subsubsec:NLOSv}
For modeling NLOSv links, we employ the model described in~\cite{boban11}, which accounts for additional attenuation due to vehicles by taking into consideration their exact locations and dimensions and applying the multiple knife edge diffraction~\cite{itu07} over the rooftops of obstructing vehicles. %
In addition to diffraction over the vehicle roofs (\emph{vertical plane diffraction}), we model the diffraction off each side of the vehicles (\emph{horizontal plane diffraction}), also using the multiple knife edge diffraction (i.e., diffraction off sides of multiple vehicles in horizontal plane).

\subsubsection{Reflections}
With respect to the reflection coefficients off building walls, we apply similar reasoning on the ``effective'' range of reflection coefficients as with the two-ray ground reflection model. We match the reflection coefficient distribution to the values empirically derived by Landron et al.~\cite{landron96}, where the authors extract the reflection coefficients for brick building walls from controlled measurements. In the locations where we performed measurements (Fig.~\ref{fig:ExperimentRoutes}), the buildings were predominantly made of brick and concrete. 

Reflections are calculated off buildings and vehicles. Since all buildings are significantly taller than any vehicle, any building can reflect the signal for any communicating pair. On the other hand, in order to be a reflector, a vehicle needs to be taller than both communicating vehicles' antennas, %
since otherwise the reflected ray does not exist. In practice, this means that reflecting vehicles are predominantly tall ones.  Furthermore, tall vehicles are more likely to block reflections coming off the building walls or other vehicles, whereas short vehicles are less likely to do so, since they are less likely to be taller than the height of the line between the communicating antennas  discounted for the 60\% of the first Fresnel zone~\cite[Chap. 2.]{rappaport96}. With respect to reflections off vehicle roofs: since vehicle roofs are predominantly made out of metal, they are potential reflectors for any communicating pairs whose antennas are taller than the roof itself. However, in an effort to keep the computational complexity of GEMV$^2$ low, in our simulations we do not model this effect.

\subsubsection{Diffractions off buildings}

Similar to diffraction over the rooftops and off the sides of the vehicles, for diffractions off buildings we use the multiple knife-edge model~\cite{itu07}. %
In the case of buildings, however, %
 diffractions are calculated in the \emph{horizontal plane only}, since we assume that the buildings are too tall for diffraction over the rooftops. %

\subsubsection{Log-distance path loss in deep-fade areas}
Reflections and diffractions off buildings and vehicles are used for NLOSb links.
We limit the calculation of diffracted and reflected rays to single-interaction (single-bounce) rays, except for multiple diffraction due to vehicles. %
It was recently shown by Abbas et al.~\cite{abbas11} that single-interaction reflections and diffractions are most often the dominating propagation mechanisms in the absence of LOS. 
Similar findings are reported by  Paier et al.~\cite{paier09_2}. By determining the LOS conditions and modeling LOS and single-interaction rays, we aim to design a model that accounts for the most important rays, at the same time keeping the computational load manageable\footnote{GEMV$^2$ can be extended to (recursively) account for the higher order interactions, however at a prohibitively increasing computational cost. Furthermore, an increasingly precise geographical database would be required to model higher-order interaction rays correctly.}.

However, communicating pairs that are not located on the same street or adjacent orthogonal streets 
 most often do not have strong single-interaction reflected or diffracted rays, but are still often able to communicate.
For such communicating pairs, multiple interaction reflections and scattering are the dominant contributors of power at the receiver~\cite{00parsons}; calculating such rays  incurs prohibitively high computations and a geographical database with a high level of detail. Furthermore, our measurement results and those reported in similar studies (e.g.,~\cite{cardote11, meireles10,mangel11}) show that communication range in NLOSb %
 conditions using IEEE 802.11p radios operating in the 5.9~GHz frequency band is limited to approximately 200~meters, even with the maximum transmit power allowed by the standard~\cite{ieee80211p}. Thus, in order to avoid costly geometric computations which predominantly yield power levels below reception threshold, %
at the same time allowing for communication in deeply faded areas, we determine the received power as follows. %
We calculate the received power using both the single-interaction diffractions and reflections through the described model and using the log-distance path loss model~\cite{00parsons}. %
The log-distance path loss $PL$ (in dB) for distance $d$ is given by %
\begin{equation}
PL(d)\;=\;PL(d_0)\;+\;10\gamma\;\log_{10} \left( \frac{d}{d_0}\right),\
\label{eq:PathLoss}
\end{equation}
where $\gamma$ is the path loss exponent and $PL(d_0)$ is the path loss at a reference distance $d_0$.
For the log-distance path loss model, the received power $Pr_{PL}$ (in dB) at a distance $d$, assuming unit antenna gains, is given by %
\begin{equation}
Pr_{PL}(d)\;=\;Pt\;-\;PL(d),\
\label{eq:PathLossFinal}
\end{equation}
where $Pt$ is the transmitted power in dB. %

In our simulations, we used $\gamma=$~2.9, which we extracted from the Porto Downtown dataset for the NLOSb conditions where there were no significant single-interaction reflections/diffractions. Previous studies reported similar values: $\gamma=$~2.9 by Durgin et.~al.~\cite{durgin98} (NLOSb environment) %
and 2.44~$\leq \gamma \leq$~3.39 by Paschalidis et~al.~\cite{paschalidis11} (urban environment -- various (N)LOS conditions). %

In case of NLOSb links, we determine the received power as the maximum of: 1) received power calculated by using reflections and diffractions  %
and 2) log-distance path loss (eq.~\ref{eq:PathLoss}).
The maximum is taken so that the log-distance path loss model (with a comparatively high path loss exponent) is used in case no strong one-interaction reflections/diffractions are present.

\subsubsection{Transmission through foliage}

For transmission through foliage, we use the attenuation-through-transmission model based on the measurements described in~\cite{goldhirsh98, benzair91,durgin98}. Specifically, we use the empirically-derived formulation from~\cite{goldhirsh98}, where attenuation for deciduous trees is calculated per meter of transmission as %

\begin{equation}
MEL = 0.79 f^{0.61},
\end{equation}
where $MEL$ is mean excess loss per meter of transmission through trees and $f$ is frequency in GHz~\cite{benzair91}. For IEEE 802.11p frequency centered at 5.9~GHz, this results in attenuation of 2.3~dB per meter of transmission through trees, which is in line with the measurement results in the 5.85~GHz band reported in~\cite{durgin98}. Similar calculations can be performed for coniferous trees as well as for seasonal changes when trees are not in full foliage~\cite{goldhirsh98}. Decision on which kind of trees to model (deciduous or coniferous) and the level of foliage (e.g., due to the time of the year) can be determined for the location where the simulations are carried out. 
Finally, we do not explicitly model reflections, diffractions, or scattering off foliage (i.e., only transmission is accounted for); rather, we encompass these effects with the small-scale signal variation model (Section~\ref{subsec:fadingModel}).

\subsection{Combining multiple rays: E-field and received power calculations}\label{subsec:efield}

Once all contributing rays (LOS, reflected, and diffracted) %
have been calculated, we determine their contributions in terms of the E-field and the received power for each link. %
We obtain the resultant E-field envelope as follows~\cite[Chap. 3.]{rappaport96}:
\begin{align} 
|E_{TOT}| &= |E_{LOS} + \sum_jE_{Refl_j} + \sum_kE_{diffr_k}|,
\label{eq:Etot}
\end{align}
where $E_{LOS}$, $E_{Refl}$, and $E_{diffr}$ are E-fieds of line or sight, reflected, and diffracted rays, respectively. Expanding eq.~\ref{eq:Etot}, we get

\begin{align}
|E_{TOT}| 	& = \frac{E_0d_0}{d_{LOS}}\cos\left(\omega_c\left(t-\frac{d_{LOS}}{c}\right)\right) \label{eq:EtotExpanded} \\
			& + \sum_{j}R_j\frac{E_0d_0}{d_{j}}\cos\left(\omega_c\left(t-\frac{d_{j}}{c}\right)\right) \nonumber \\
			& + \sum_{k}D_k\frac{E_0d_0}{d_{k}}\cos\left(\omega_c\left(t-\frac{d_{k}}{c}\right)\right), \nonumber 
\end{align}
where $\frac{E_0d_0}{d_{LOS}}$ is the envelope E-field at a reference distance $d_0$, $\omega_c$ is the angular frequency ($\omega_c = 2\pi f$), $t$ is the time at which the E-field is evaluated, $d_x$ represents distance traversed by ray $x$,  $R_j$ is the reflection coefficient of reflected ray $j$, and $D_k$ is the diffraction coefficient of diffracted ray $k$. When the originating medium is free space, the reflected coefficient $R$ is calculated as follows for vertical and horizontal polarization, respectively~\cite[Chap. 3.]{rappaport96}:

\begin{equation}
R_{||} = \frac{-\epsilon_r \sin\theta_i + \sqrt{\epsilon_r - \cos^2\theta_i}}{\epsilon_r \sin\theta_i + \sqrt{\epsilon_r - \cos^2\theta_i}}
\end{equation}
and
\begin{equation}
R_{\perp} = \frac{\sin\theta_i - \sqrt{\epsilon_r - \cos^2\theta_i}}{\sin\theta_i + \sqrt{\epsilon_r - \cos^2\theta_i}},
\end{equation}
where $\theta_i$ is the incident angle and $\epsilon_r$ is the relative permittivity of the material. 

Regarding diffractions, we do not calculate the diffraction coefficient directly; we approximate the E-field for diffracted rays using the knife-edge model~\cite{itu07}. 

The ensuing received power $Pr$ (in watts), assuming unit antenna gains, is calculated as follows:
\begin{equation}
Pr = \frac{|E_{TOT}|^2 \lambda^2}{480\pi^2},%
\label{eq:totPower}
\end{equation}
where $\lambda$ is the wavelength. Note that $Pr$ accounts for the 
large-scale signal variation 
of the LOS links, whereas for NLOSv and NLOSb links there can also exist contributions in terms of multipath generated by multiple diffractions around vehicles for NLOSv links (horizontal and vertical multiple knife-edge diffractions) and single-interaction reflections and diffractions for NLOSb links.

\subsection{Modeling small-scale signal variations}\label{subsec:fadingModel}

To account for small-scale signal variation inherent in V2V communication (e.g., due to multipath, scattering, first or higher order diffractions and reflections, etc.), %
using the insights obtained through measurements, we designed a small-scale signal variation
 model that aims to capture the richness of the propagation environment surrounding the communicating pair. %
We first characterize the per-bin signal variations in the collected measurements; %
next, we design a simple model to include the per-bin signal variation that complements the previously described components of GEMV$^2$ that deal with large-scale signal variations.

\subsubsection{Small-scale signal variations in measurement datasets}
\begin{figure}
  \begin{center}
  
  \subfigure[ LOS data from Porto Open Space.] %
  {\label{fig:LOSFading}\includegraphics[trim=0cm 7cm 0cm 7cm,clip=true,width=0.23\textwidth]{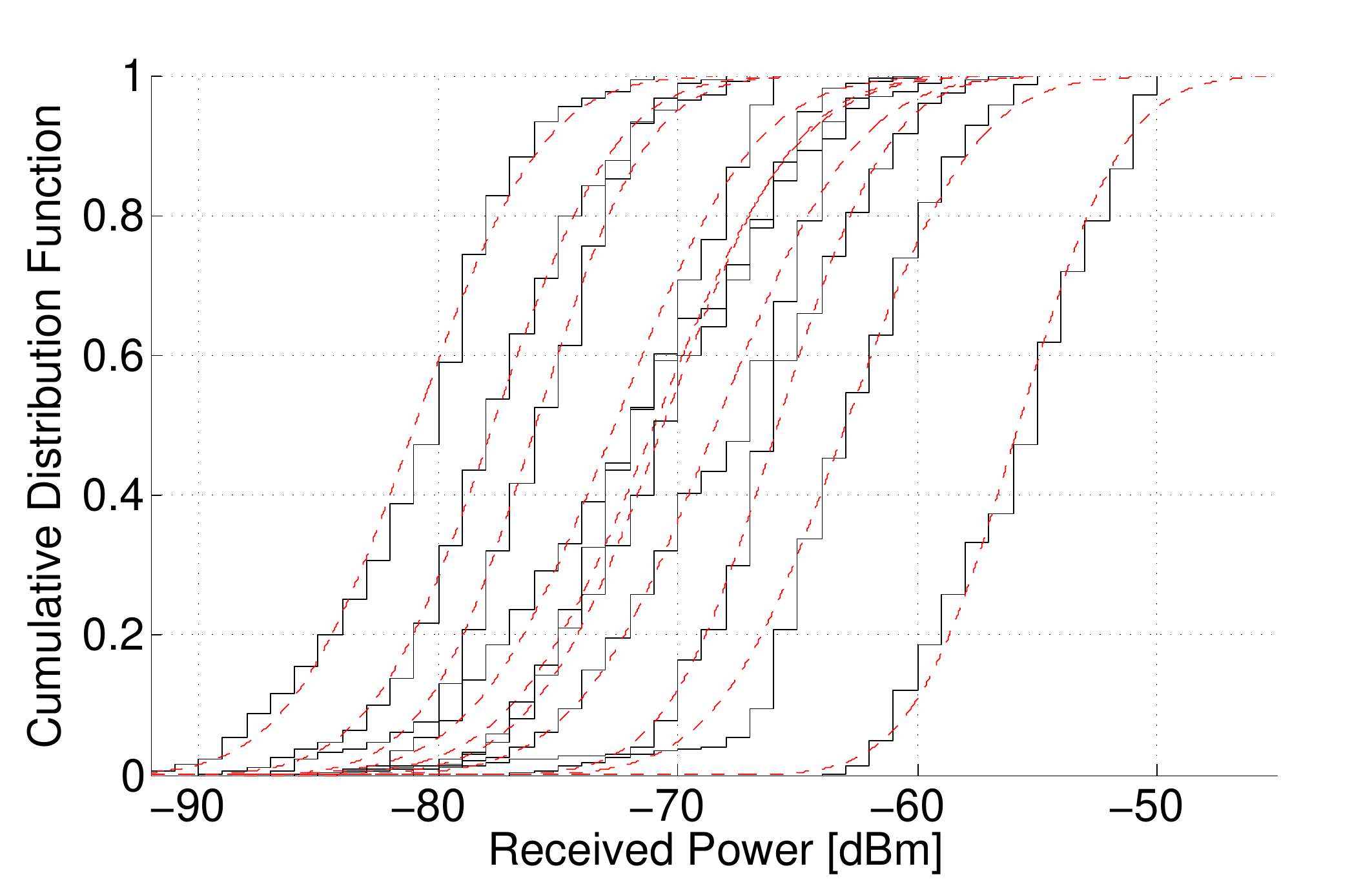}}
  \hspace{1mm}
    \subfigure[ LOS data from Porto Downtown.] %
     {\label{fig:LOSFadingUrb}\includegraphics[trim=0cm 7cm 0cm 7cm,clip=true,width=0.23\textwidth]{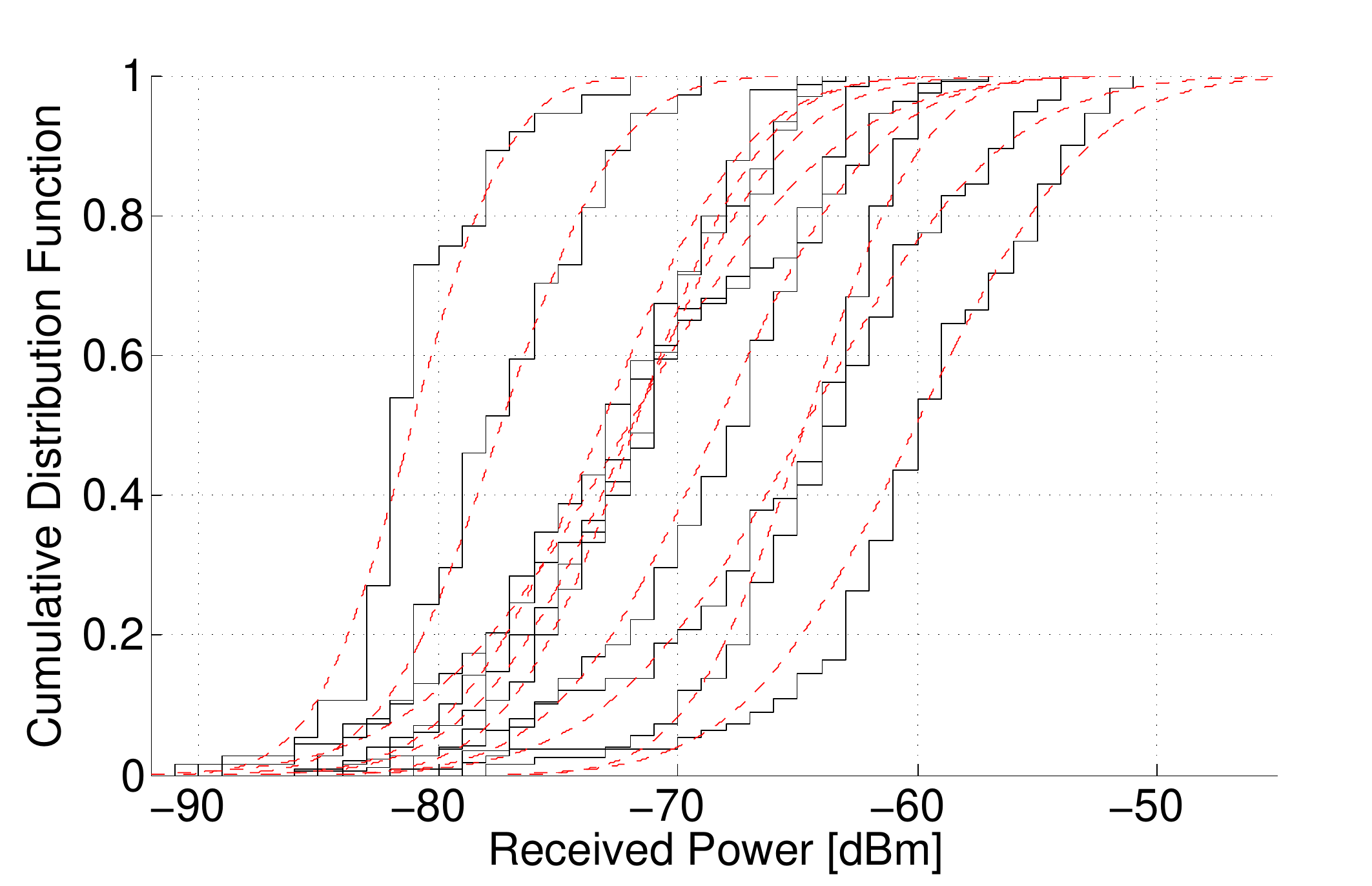}}\\
\subfigure[ NLOSv data from Porto Downtown.] %
{\label{fig:NLOSvFading}\includegraphics[trim=0cm 7cm 0cm 7cm,clip=true,width=0.23\textwidth]{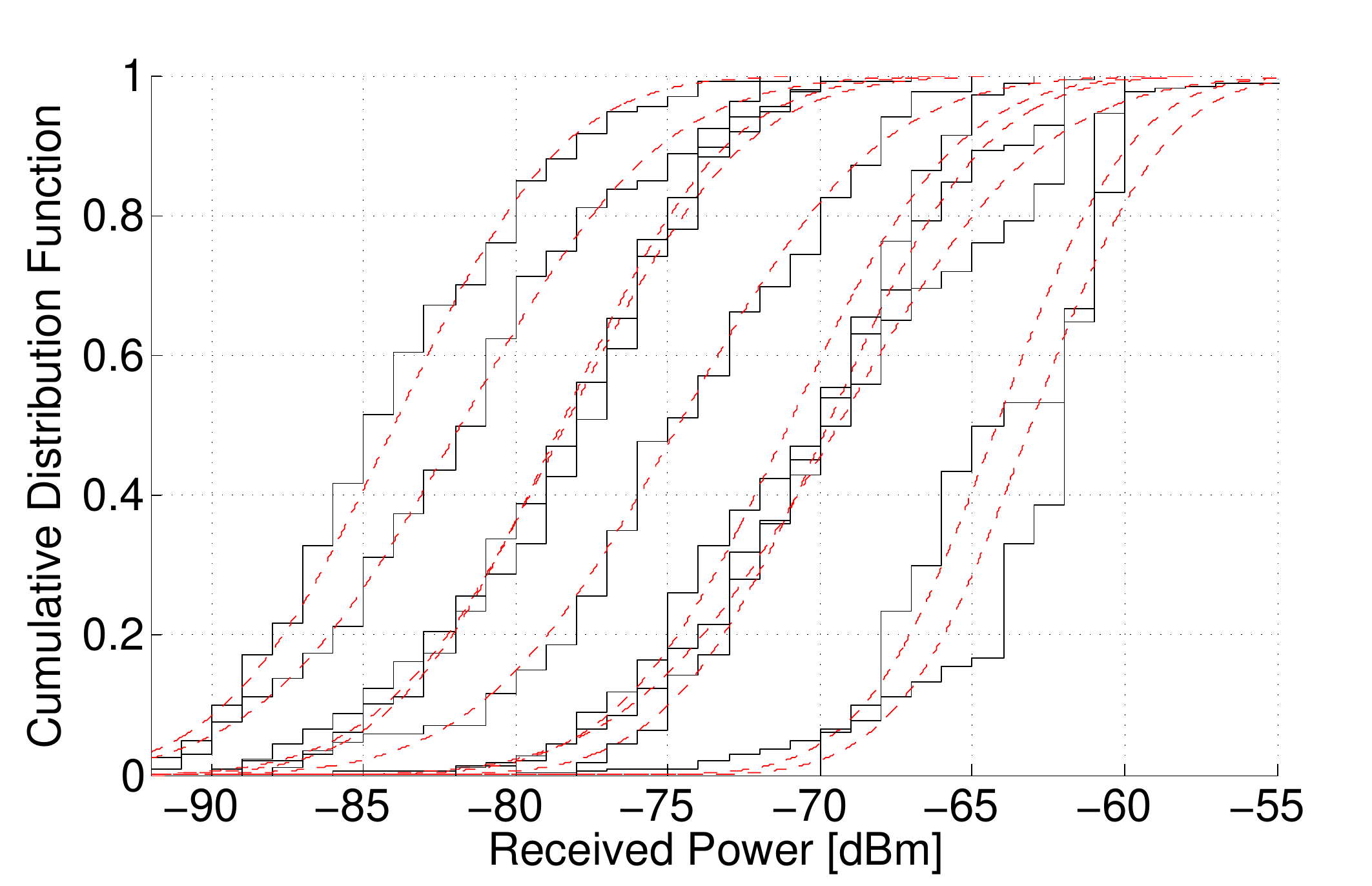}}
\hspace{1mm}
\subfigure[ NLOSb data from Porto Downtown.] %
{\label{fig:NLOSbFading}\includegraphics[trim=0cm 7cm 0cm 7cm,clip=true,width=0.23\textwidth]{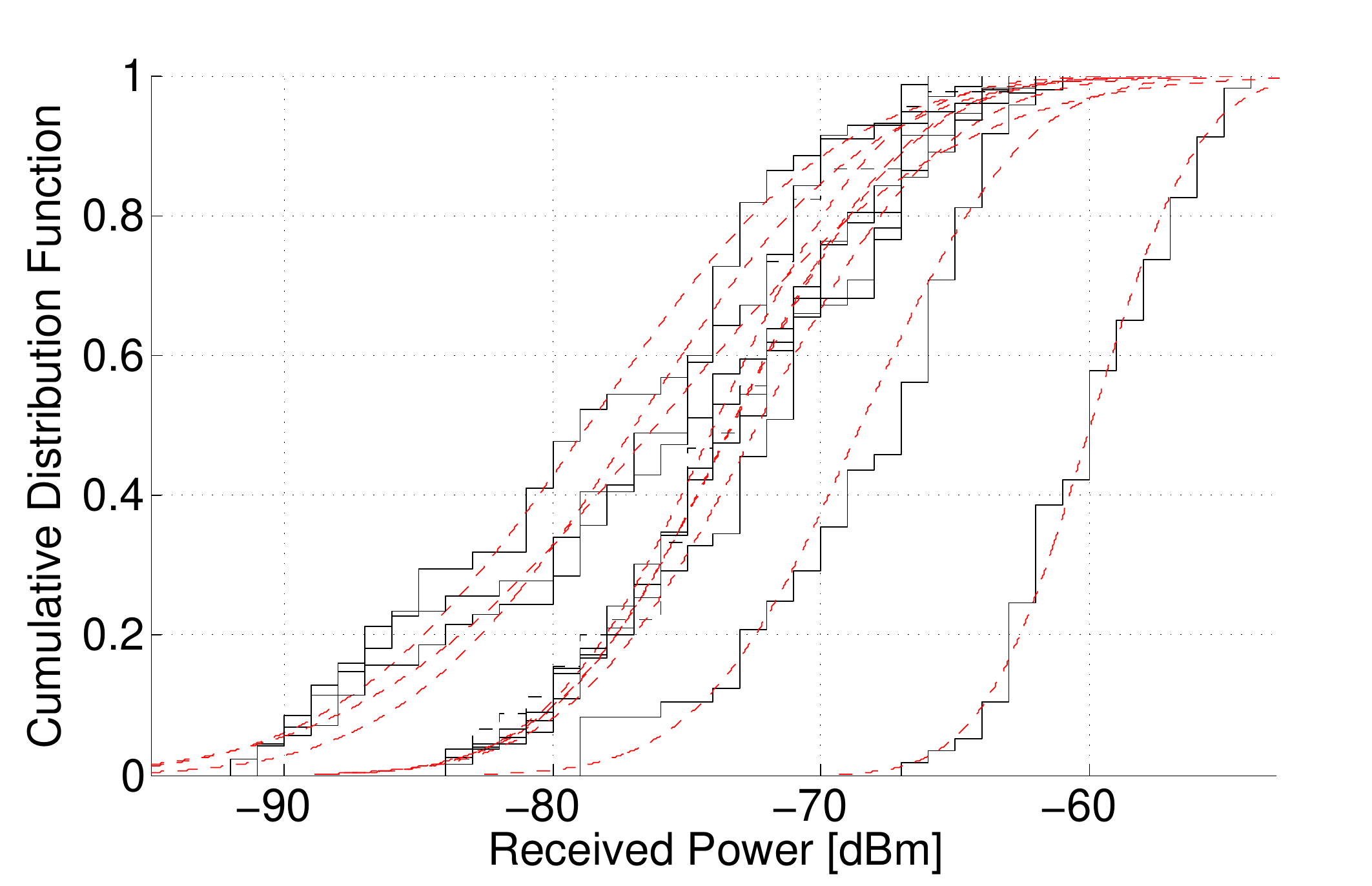}}
     \caption[CDF of the received power for two-meter distance bins]{Cumulative distribution functions (CDF) of the received power for selected two-meter distance bins with best-fit normal distributions (dashed red lines). 
     All plotted bins contain at least 40 data points. 
     For LOS (Figs.~\ref{fig:LOSFading} and~\ref{fig:LOSFadingUrb}) and NLOSv data (Fig.~\ref{fig:NLOSvFading}), the bins are centered at decades from 10 to 100 meters  (i.e., the curves represent the following bins, left to right: [99-101], [89-91], ..., %
[9-11] meters). For NLOSb data (Fig.~\ref{fig:NLOSbFading}), due to lack of data points at lower distances, the two-meter bins centered at the following distances are shown (left to right): 90, 85, 70, 65, 60, 55, 50, 45, and 15 meters. %
All plots are for measurements involving passenger (short) vehicles.}
      \label{fig:FadingTen}
   \end{center}
\end{figure}

We used the collected measurements to characterize small-scale signal variation for different LOS conditions, environments, and with different levels of vehicular traffic (i.e., temporal variation). %
For each collected measurement, we separate the links into LOS, NLOSv, and NLOSb, using the videos recorded during the measurements. Then, we divide the collected data into two-meter distance bins. %
We selected two meter bins because %
they are small enough not to incur significant distance-related path loss dependence, at the same time containing enough data points to allow for a meaningful statistical characterization. %
Figure~\ref{fig:FadingTen} shows the distribution of received power for two-meter distance bins. For the LOS links, the normal distribution seems to fit the data reasonably well, with a better fit for the open space environment LOS links (Fig.~\ref{fig:LOSFading}) than the urban LOS links (Fig.~\ref{fig:LOSFadingUrb}), due to the richer reflection environment in the case of the latter.
Normal fit for the NLOSv and NLOSb links is less accurate due to the variety of conditions that are encompassed (e.g., different number of obstructing vehicles in case of NLOSv, deep or slight building obstruction in case of NLOSb).
Based on the measured data, we choose to use %
zero-mean %
normal distribution $N(0,\sigma)$ %
 to describe the small-scale signal variation for all three link types. %

\subsubsection{Accounting for small-scale signal variations} %

Apart from establishing the distribution of signal variation, we also need to determine its parameter -- i.e., the standard deviation ($\sigma$) of the normal distribution -- since different environments and link types experience different levels of signal variation, as shown in Fig.~\ref{fig:FadingTen}. Therefore, we implement a simple model that accounts for the additional small-scale signal variation due to the objects in the area around the communicating pair as follows. Using the communication ellipse for each pair as explained in Fig.~\ref{NLOSExplanation}, we count the \emph{number of vehicles} and sum the \emph{area of static objects}  in the ellipse. We chose the area of the static objects rather than their number because, unlike the size of vehicles, their area varies greatly (see Fig.~\ref{NLOSExplanation}). Since a large-area building/foliage is more likely to impact the communication than a smaller one, %
we use their area instead of their number in the calculations. In terms of different link types, the objects in the ellipse have the following effects: 1) for the communicating pairs located on the same street (i.e., LOS or NLOSv links), the objects inside the ellipse will include the vehicles along that street and buildings and foliage lining the street -- arguably, these are the most important sources of %
multipath 
for such links; %
2) similarly, for NLOSb links (i.e., links between vehicles on different streets and with buildings/foliage blocking the LOS), the ellipse will include buildings, foliage, and vehicles that generate significant reflections, diffractions, and scattering (see Fig.~\ref{NLOSExplanation}).

Next, we set the minimum and maximum $\sigma$ for a given LOS condition based on the collected measurements. We do not extract the minimum and maximum $\sigma$ for each experiment location, since we aim to determine a single pair of values for each of the three link types (LOS, NLOSv, and NLOSb), which could then be used across a number of different locations. Therefore, we utilize minimum and maximum $\sigma$ as calculated from the measurements and shown in Table~\ref{tab:fadingValues}, which we obtained by averaging $\sigma$ for all two-meter bins with more than 40 samples in that dataset. %
For simplicity, we use a single pair of minimum/maximum values for both short and tall vehicles. 
For LOS and NLOSv links, we extract the minimum $\sigma$ from the least variable environments in terms of small-scale signal variation (Porto Open Space and Porto Highway, respectively). 
On the other hand, when reflections and diffractions are calculated, the minimum $\sigma$ for NLOSb links is set to zero, since the most significant reflected and diffracted rays for these links are already accounted for. Alternatively, %
when only log-distance path loss is used for NLOSb links, we set the minimum $\sigma$ based on~\cite{mangel11}.
The maximum values for all three link types have been taken as the environment with the most variable small-scale signal variation from the collected datasets. %
Note that minimum and maximum $\sigma$ values can be different for other environments; if this is the case, the values of $\sigma$ different than those in Table~\ref{tab:fadingValues} can be used. %

\begin{table} 
	\begin{footnotesize}
	\centering
\caption[Minimum and fading $\sigma$ extracted from experimental data]{Minimum and maximum values of the small-scale signal deviation $\sigma$ extracted from experimental data }
\footnotesize{
		\begin{tabular}{|c c c|}
\hline \textbf{Link Type} & $\sigma_{min}$ (source) & $\sigma_{max}$ (source) \\
		 	\hline 
\hline \textbf{LOS} & 3.3~dB (Porto Open space) & 5.2~dB (Porto Downtown)\\
\hline \textbf{NLOSv} 	%
						& 3.8~dB (Porto Highway) & 5.3~dB (Porto Downtown) \\
\hline \textbf{NLOSb} & 0~dB (when using refl./diffr.)  & 6.8~dB (Porto Downtown)\\ %
 &  4.1~dB (based on~\cite{mangel11}) & \\
\hline
\end{tabular}}
	\label{tab:fadingValues}
		\end{footnotesize}
\end{table}

\begin{table} 
	\centering
		\begin{footnotesize}
\caption{Max. communication ranges used for different link types}
		\begin{tabular}{|c c|}
\hline \textbf{Link Type} &  \textbf{Max. comm. range} \\
		 	\hline 
\hline \textbf{$r_{LOS}$} - urban & 500\\
\hline \textbf{$r_{LOS}$} - outside urban & 1000\\
\hline \textbf{$r_{NLOSv}$} & 400 \\
\hline \textbf{$r_{NLOSb}$} & 300 \\
\hline
\end{tabular}
	\label{tab:rLOS}
		\end{footnotesize}
\end{table}

\begin{figure*}
  \begin{center}
    \includegraphics[height=0.95\textheight]{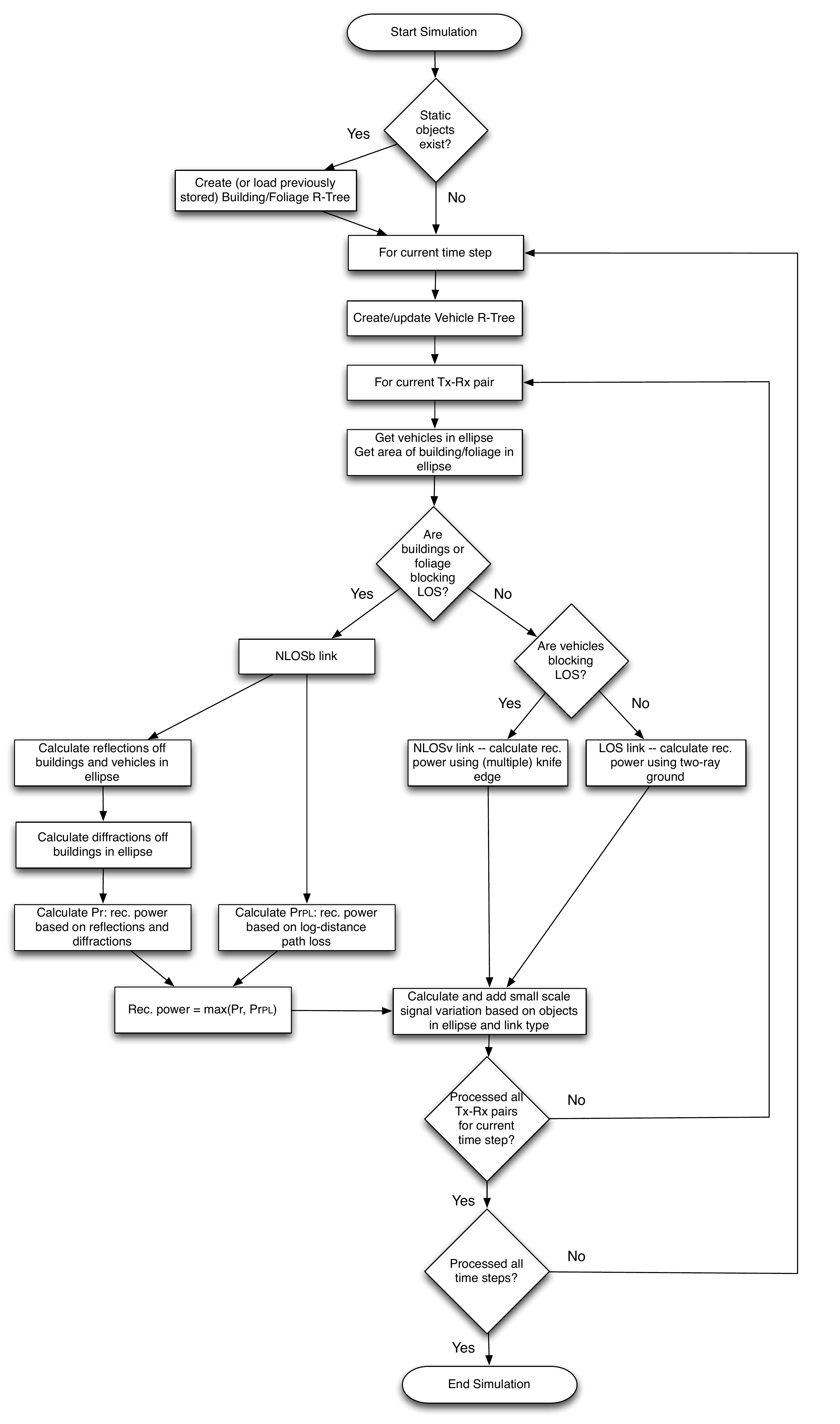}
     \caption{GEMV$^2$: simulation flow.}
      \label{fig:simulationFlowchart}
   \end{center}
\end{figure*}

We calculate the small-scale signal deviation $\sigma$ (in dB) for the communication pair $i$,  $\sigma_i$, as %
\begin{equation}
\sigma_i = \sigma_{min} + \frac{\sigma_{max}-\sigma_{min}}{2} \cdot \left( \sqrt{\frac{NV_i}{NV_{max}}} + \sqrt{\frac{AS_i}{AS_{max}}} \right),
\label{eq:FadingCalc}
\end{equation}
where $\sigma_{min}$ is the minimum small-scale signal deviation (in dB) for $i$'s LOS type (LOS, NLOSv, or NLOSb), $\sigma_{max}$ is the maximum deviation value for $i$'s LOS type, $NV_i$ is the number of vehicles per unit area in $i$'s ellipse, $NV_{max}$ is the maximum number of vehicles per unit area, $AS_i$ is the area of static objects per unit area in $i$'s ellipse, and $AS_{max}$ is the maximum area of static objects per unit area. The value of $NV_{max}$ can be calculated a priori from historical data (e.g., maximum number of vehicles per area in a given city or a highway), whereas $AS_{max}$ can be calculated from geographical databases, such as~\cite{openstreetmap}.
In our calculations, we used $NV_{max}$ and $AS_{max}$ derived from the Porto dataset, with references defined on a square kilometer (i.e., maximum number of vehicles and maximum area of static objects in a square kilometer). For each of its constituents (vehicle-induced and static objects-induced signal variation), eq.~\ref{eq:FadingCalc} is essentially a square interpolation between minimum small-scale signal variation in an environment (e.g., open space without any objects other than communicating vehicles) and maximum variation (e.g., the downtown of a city during rush hour with a high density of vehicles and buildings/foliage). The square root in eq.~\ref{eq:FadingCalc} is taken to give comparatively more significance to lower number of vehicles than the higher number, because previous studies have shown that the difference between having no vehicles near the communicating pair and having a few vehicles nearby is more pronounced than the difference between a few and many vehicles~\cite{meireles10,abbas11}. Similar reasoning is applied to buildings and foliage. Furthermore, as shown in eq.~\ref{eq:FadingCalc}, due to the lack of a better classification, 
we give equal weights to the number of vehicles and the area of static objects when calculating the small-scale signal variation.

Once %
$\sigma_i$ is calculated for pair $i$, we combine the results of the small-and large-scale signal variation model by %
adding a normally distributed random variable $N(0,\sigma_i)$ to the previously calculated received power (eq.~\ref{eq:totPower}):
\begin{equation}
Pr_{TOT_i} = 10\log_{10}(Pr_i) + N(0,\sigma_i).
\end{equation}

\subsection{Implementation details and simulation structure of GEMV$^2$}\label{subsec:Rules}

In order to improve the performance of GEMV$^2$ and make it suitable for implementation in VANET simulators, we exploit additional information available in VANETs, along with known geometric properties of the environment. Specifically, we implement the following rules.

\begin{enumerate}
\item For each link, we first check the blockage of LOS by buildings and foliage. If there is LOS blockage, we do \emph{not} check the vehicle R-tree for LOS blockage, since obstructing buildings and foliage reduce the power at the receiver considerably more than
 obstructing vehicles (see, e.g.,~\cite{meireles10, karedal10}). For links whose LOS is not blocked by buildings or foliage, we check the R-tree containing vehicles.
\item R-trees enable efficient intersection testing and neighbor querying~\cite{guttman84}. Apart from using R-trees for link type classification, we use them %
 to efficiently search for objects around the communicating pair and to implement a variation of the method of images~\cite[Chap. 7]{00parsons} -- a technique used to geometrically determine the reflected and diffracted rays.

\item For LOS, NLOSv, and NLOSb links, we define the maximum communication range $r$ as shown in Fig.~\ref{NLOSExplanation}, which determines the threshold distance above which the received power is assumed to be insufficient to correctly decode the message at the receiver, irrespective of the channel conditions. Specifically, we define $r_{LOS}$, $r_{NLOSv}$, and $r_{NLOSb}$ for LOS, NLOSv, and NLOSb links, respectively. %
These radii are functions of transmit power, receiver sensitivity, antenna gains, and the surrounding environment. For a given set of radio parameters (reception threshold, transmit power, etc.), the ranges can be obtained either through field measurements or analytically. Radii used for the purpose of this study are shown in Table~\ref{tab:rLOS}.
\item We used the insights from the measurements to refine %
GEMV$^2$. Specifically, we tested the benefits we obtain when considering reflections and diffractions for each of link types (LOS, NLOSv, and NLOSb). As we will show in Section~\ref{resultsComplete}, the comparison between the model employing path loss propagation mechanisms shown in Table~\ref{tab:linkTypes} and the measurement results for LOS and NLOSv links showed a good match. Furthermore, adding reflections and diffractions resulted in minimal benefits in terms of accuracy, while incurring a high computational overhead. Therefore, we do not calculate reflections and diffractions for LOS and NLOSv links.
On the other hand, NLOSb links can benefit from single-interaction reflected and diffracted rays %
(see, e.g.,~\cite{abbas11}). However, when computational speed is of essence, not calculating reflections and diffractions results in a considerable speedup of the execution time (as discussed later in Section~\ref{sec:Performance}).
Therefore, for NLOSb links, we provision for calculating the received power both with explicit calculation of the single-interaction reflections and diffractions and without (i.e., relying on log-distance path loss). %
\end{enumerate}

Figure~\ref{fig:simulationFlowchart} shows the simulation execution flowchart of GEMV$^2$. The flowchart synthesizes the large-scale and small-scale propagation models, describes how the rules for reducing the complexity of GEMV$^2$ are implemented, and %
contains the basic information required to implement it in discrete-event VANET simulators. %
  
\subsection{Assumptions}
To keep the computational complexity of GEMV$^2$ low, we made the following assumptions and simplifications. %
\begin{enumerate}	
\item We assume that buildings are too tall for any meaningful amount of power to be received over them. Since even the shortest buildings are at least 5~meters taller than the vehicles, simple calculations using knife-edge diffraction~\cite{itu07} show that the losses due to diffraction over the rooftops is in excess of 30~dB (with 40+~dB loss for buildings 15 or more meters taller than vehicles), thus making the power contribution over the rooftops negligible.  %
\item Currently, vehicles, buildings, and foliage are accounted for in our model. In environments where other objects have a significant impact (e.g., lamp posts, signs, railing, etc.), the model would need geographical information about these objects as well. However, such objects are currently not readily available in geographic databases. Furthermore, the additional gains in realism would need to be compared with the increase in the computational complexity due to the additional objects, particularly if the number of such objects is large.

\item Due to the limited precision of the databases and the focus on simulating large vehicular networks, we do not model scattering (i.e., dispersion of radio signal by objects significantly smaller than the wavelength of the carrier wave). Because of their potentially large number, scattering objects could significantly increase the complexity of the calculations that need to be performed. %

\item Currently, we assume that the terrain is flat. For locations with significant elevation changes, GEMV$^2$ would need to be adapted so that the elevation is included, %
provided such data is sufficiently accurate. %

\end{enumerate}

\section{Results}\label{resultsComplete}
In this section we 
validate GEMV$^2$ against the measurements we performed in 
locations shown in Figs.~\ref{fig:ExperimentRoutes} and~\ref{fig:ExperimentLocations}. 
For the purpose of comparison, we use the GPS locations of the vehicles recorded during measurements to perform simulations in the same locations where the communication occurred. %
We use the actual dimensions of the vehicles used for measurements (Table~\ref{tab:dimensionsCompleteModel}) and the corresponding buildings and foliage extracted from geographical databases. Furthermore, we use the communication ranges specified in Table~\ref{tab:rLOS}. These values are based on our own measurements, as well as results previously obtained in \cite{cardote11,karedal10,mangel11}.
Note that $r_{LOS}$ was set to 1000~meters outside urban areas and 500~meters in urban areas, whereas we use the same values of $r_{NLOSv}$ and $r_{NLOSb}$ in all environments.

\begin{figure}
  \begin{center}
    \includegraphics[width=0.45\textwidth]{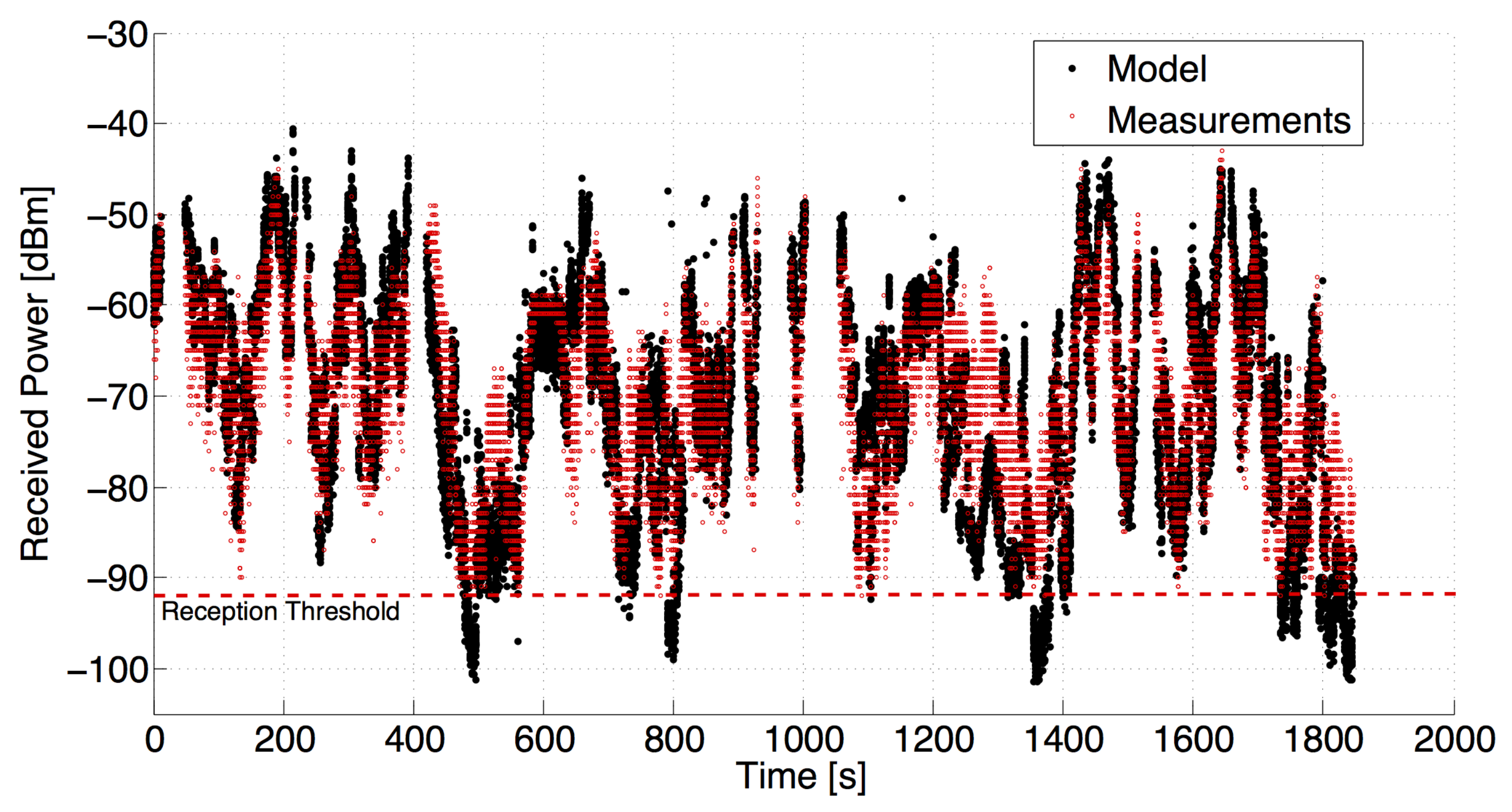}

     \caption[Received power for a 30-minute experiment in downtown Porto]{Received power for a 30-minute experiment in downtown Porto, Portugal, along with the received power predicted by GEMV$^2$. Number of data points: 16500.}
      \label{fig:PortoComplete}
   \end{center}
\end{figure}

\begin{figure*}
\centering
\subfigure[ Raw data from the Porto Open Space dataset collected through measurements and generated by GEMV$^2$.]{\label{fig:LOSFitsLecaRaw}\includegraphics[trim=0cm 6.5cm 0cm 5.5cm,clip=true,width=0.31\textwidth]{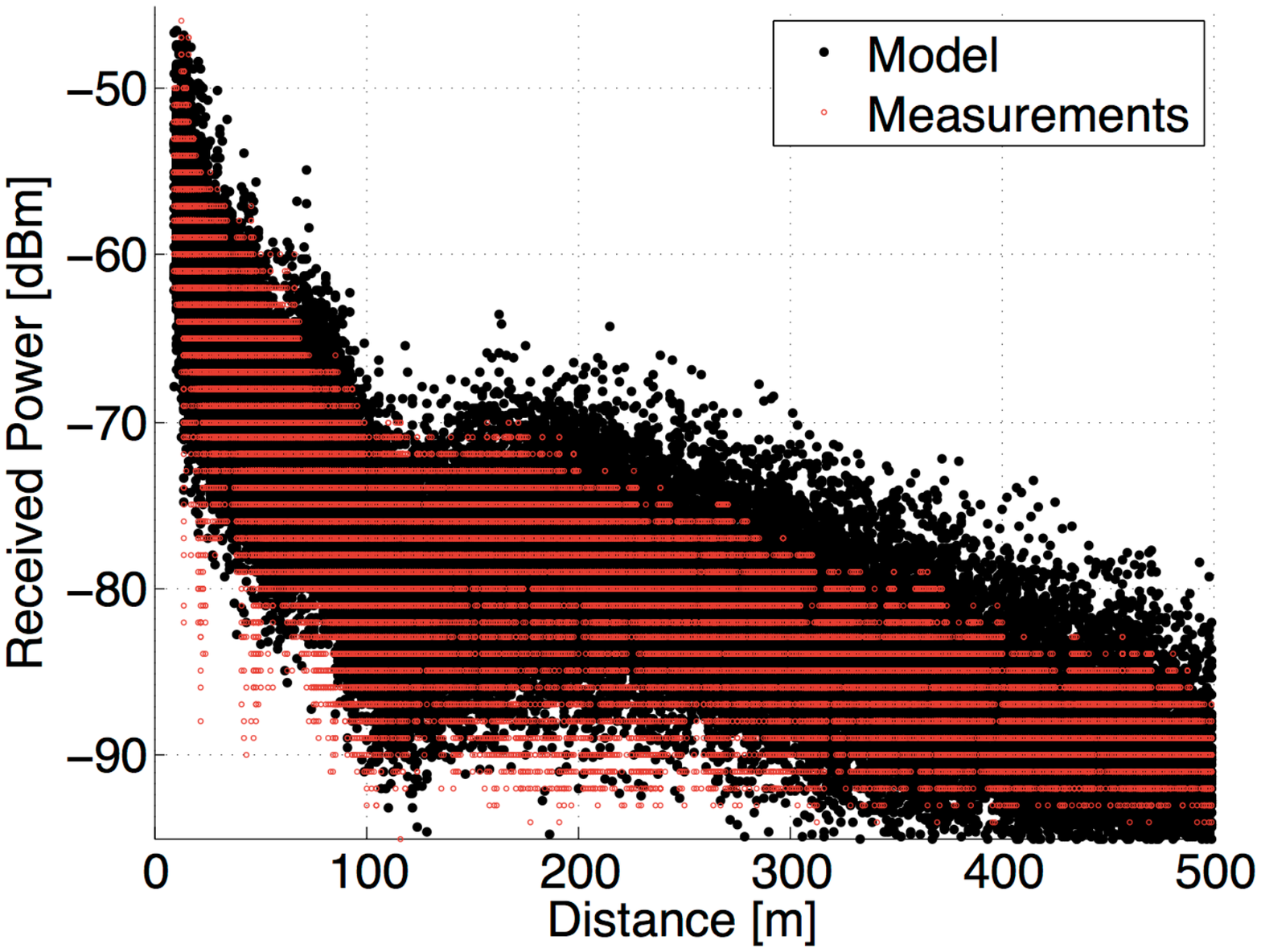}} \label{fig:LOSFitsRaw}
  \hspace{1mm}
\subfigure[ Porto Open Space -- Coordinates: 41.210615, -8.713418.  Number of data points: 61000. \textbf{Measured averaged $\sigma$: 3.3~dB}]{\label{fig:LOSFitsLeca}\includegraphics[trim=0cm 6.5cm 0cm 5.5cm,clip=true,width=0.31\textwidth]{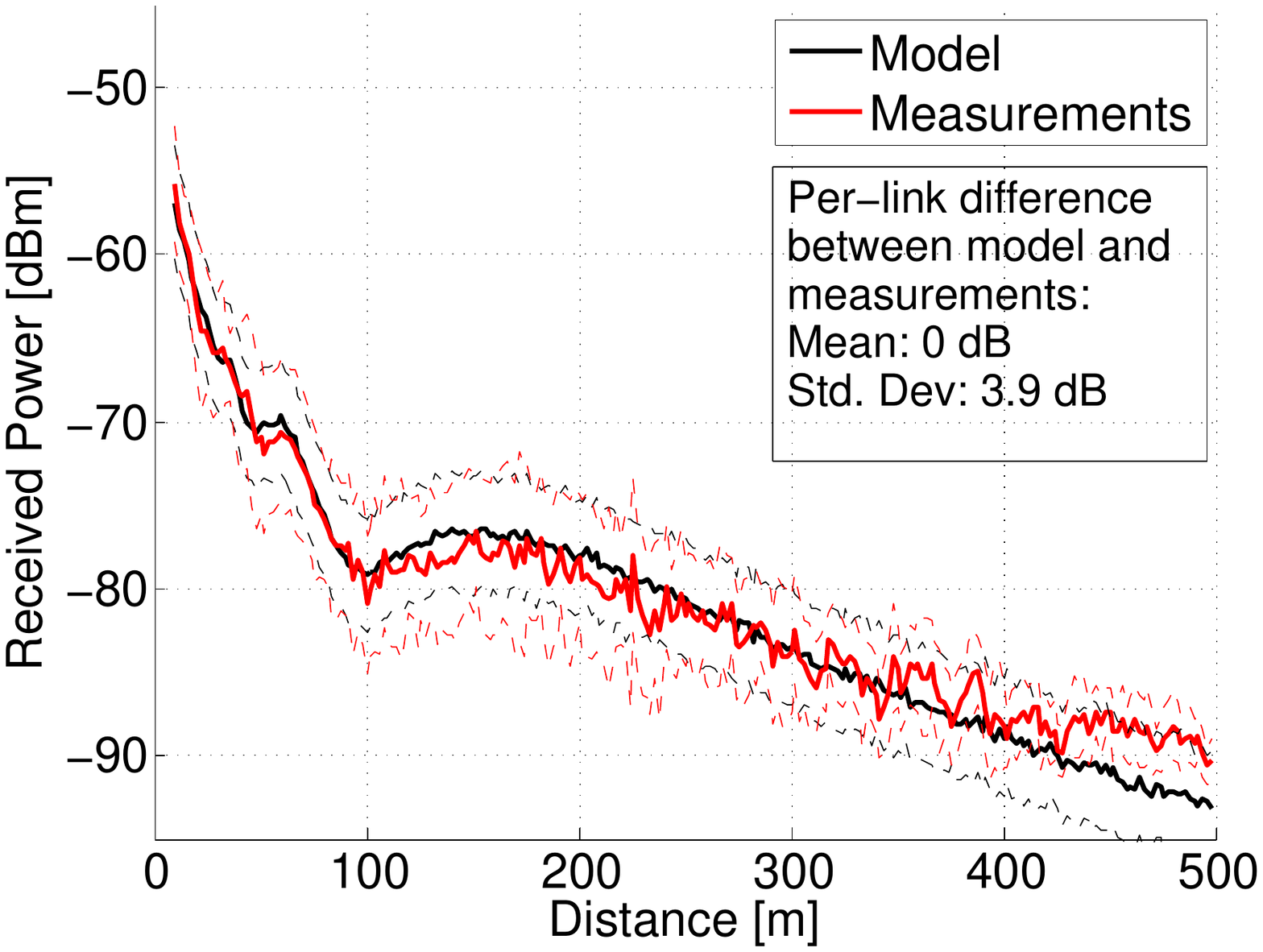}} \label{fig:LOSFitsPorOS}
  \hspace{1mm}
\subfigure[ Pittsburgh Open Space -- Coordinates:  40.4103279, -79.9181137. Number of data points: 10700. \textbf{Measured averaged $\sigma$: 4.6~dB}]{\label{fig:LOSFitsHomestead}\includegraphics[trim=0cm 6.5cm 0cm 5.5cm,clip=true,width=0.31\textwidth]{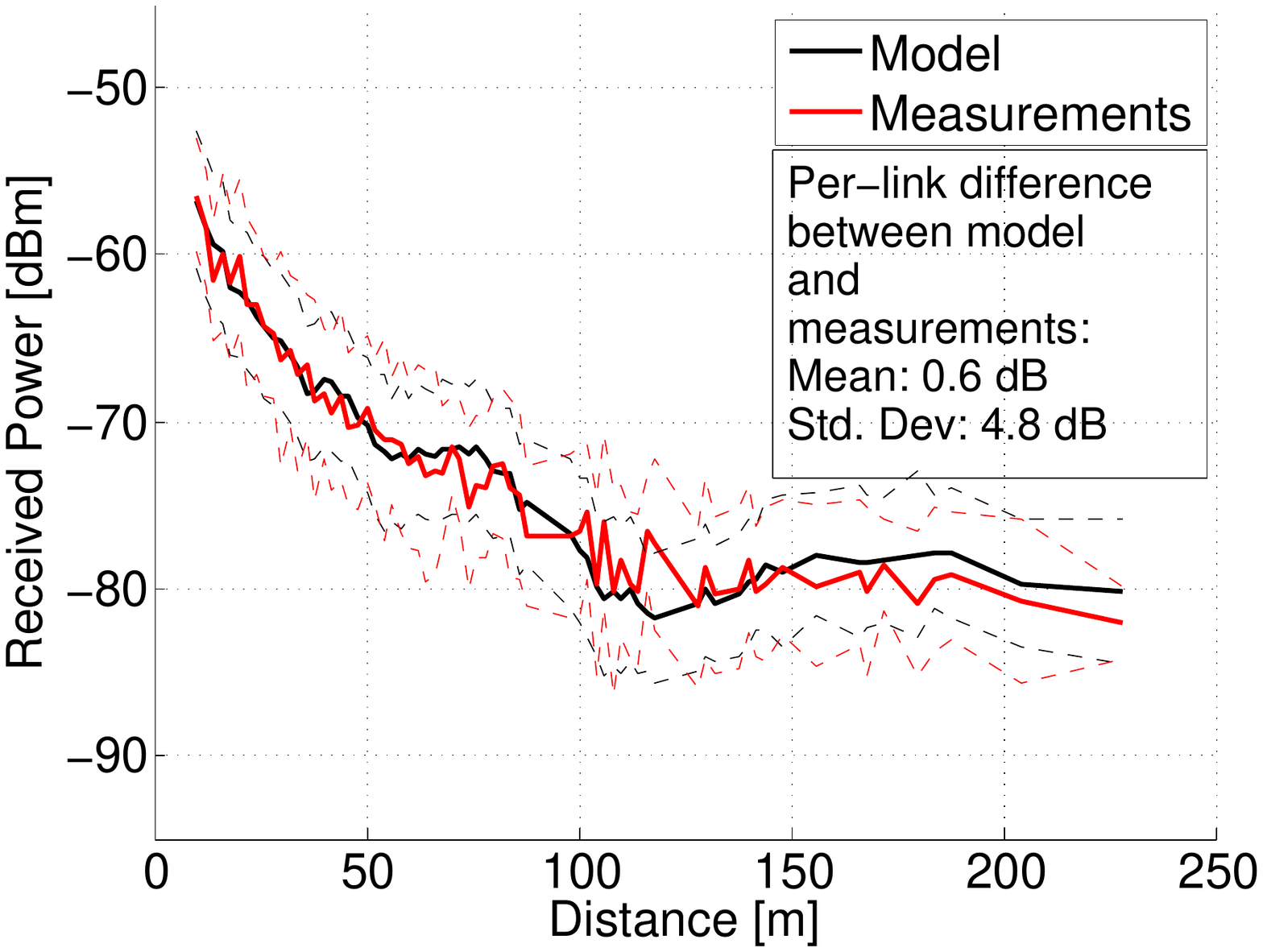}} \label{fig:LOSFitsRawPitOS}
  \hspace{1mm}
\subfigure[ Pittsburgh Suburban Nighttime measurements  -- Coordinates:  40.4476089, -79.9398574. Number of data points: 11900.  \textbf{Measured averaged $\sigma$: 4.1~dB}]{\label{fig:LOSFits5thNight}\includegraphics[trim=0cm 6.5cm 0cm 5.5cm,clip=true,width=0.31\textwidth]{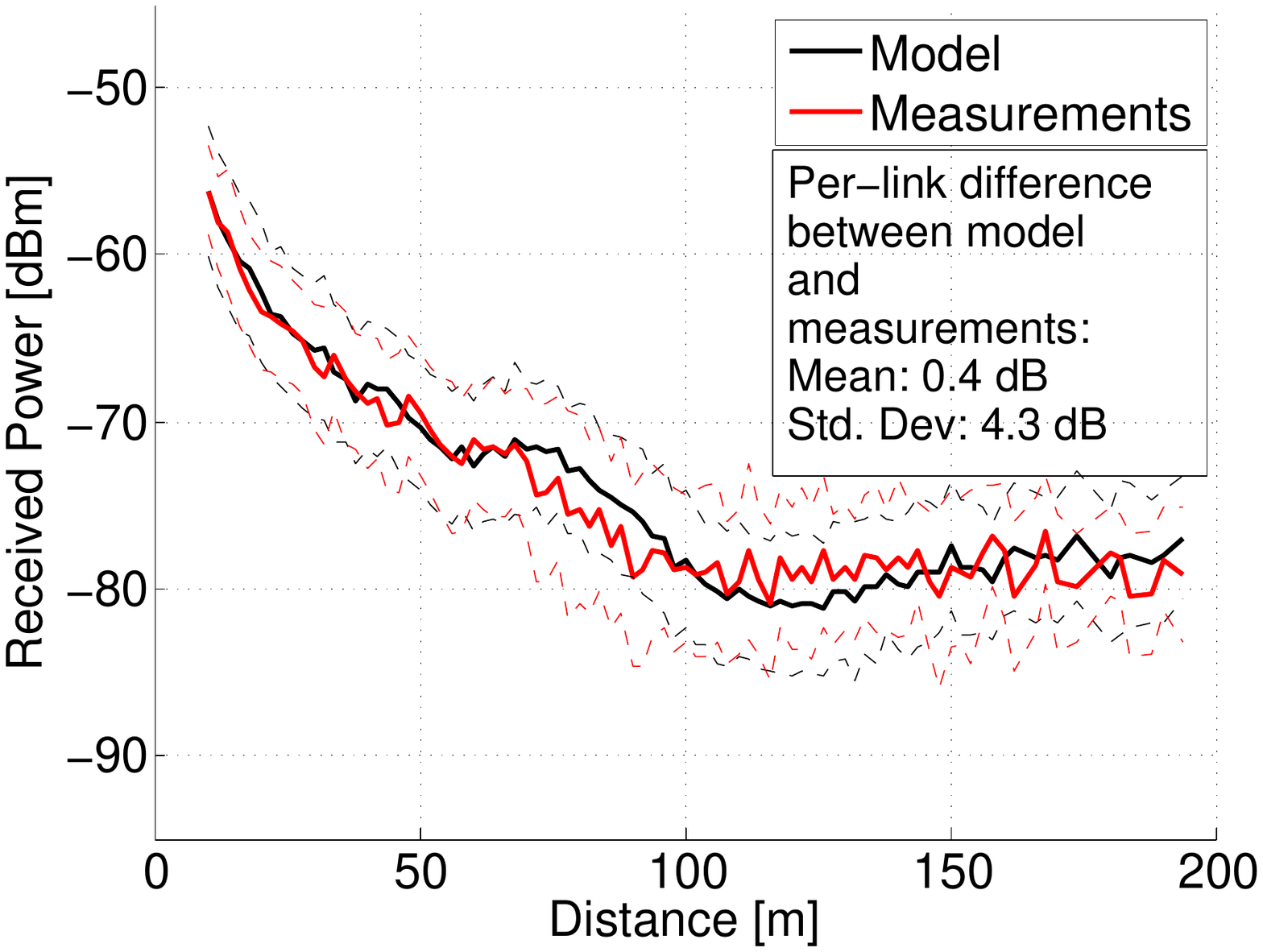}} \label{fig:LOSFitsRawPitSubN}
  \hspace{1mm}
\subfigure[ Pittsburgh Suburban Daytime measurements  -- Coordinates:  40.4476089, -79.9398574.  Number of data points: 13000. \textbf{Measured averaged $\sigma$: 4.8~dB}]{\label{fig:LOSFitsFifthDay}\includegraphics[trim=0cm 6.5cm 0cm 5.5cm,clip=true,width=0.31\textwidth]{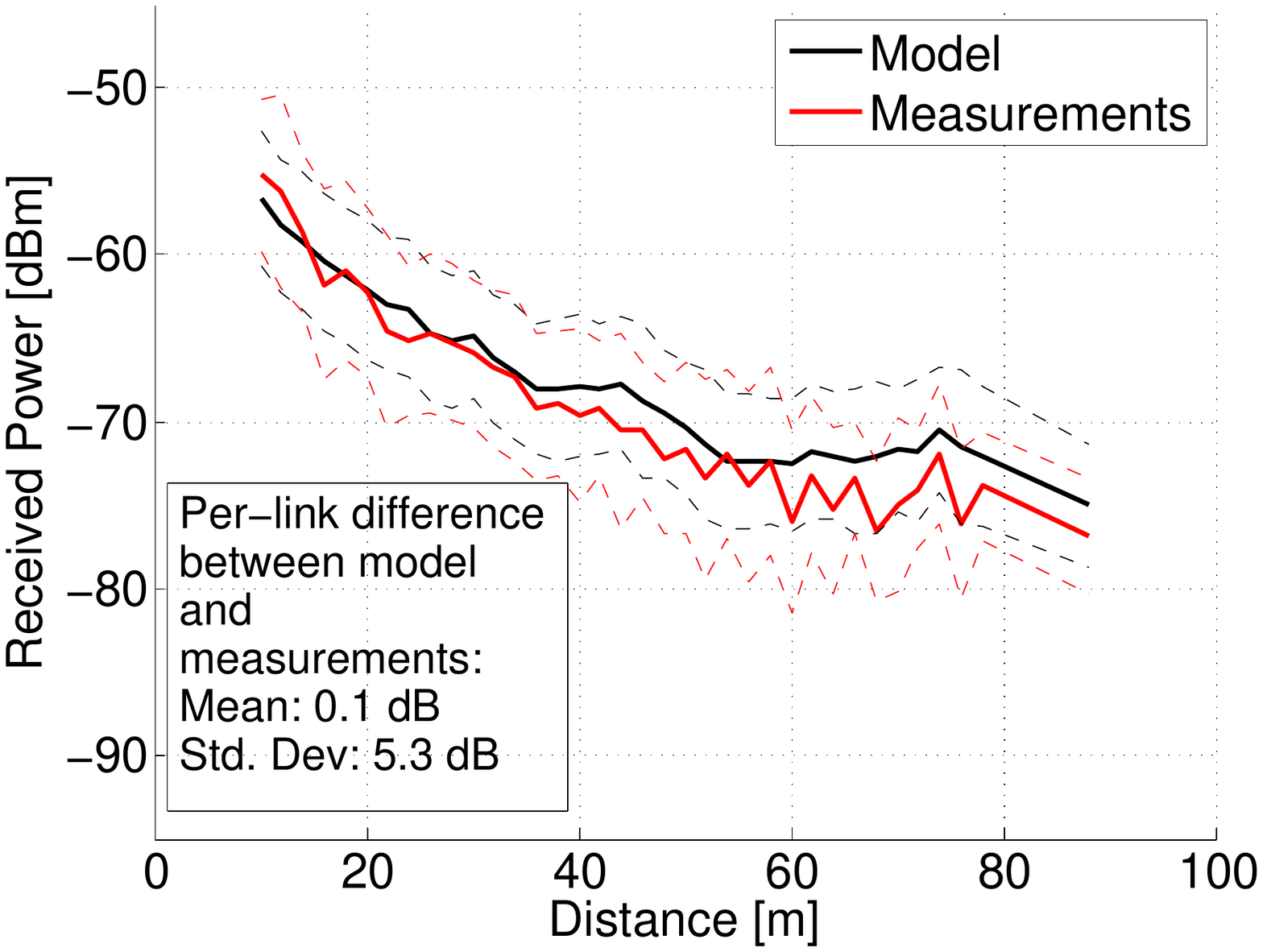}} \label{fig:LOSFitsRawPitSubD} \hspace{1mm}
\subfigure[ Porto Downtown -- Coordinates: 41.153673, -8.609913. Number of data points: 4400. \textbf{Measured averaged $\sigma$: 5.2~dB}]{\label{fig:LOSFitsUrban}\includegraphics[trim=0cm 6.5cm 0cm 5.5cm,clip=true,width=0.31\textwidth]{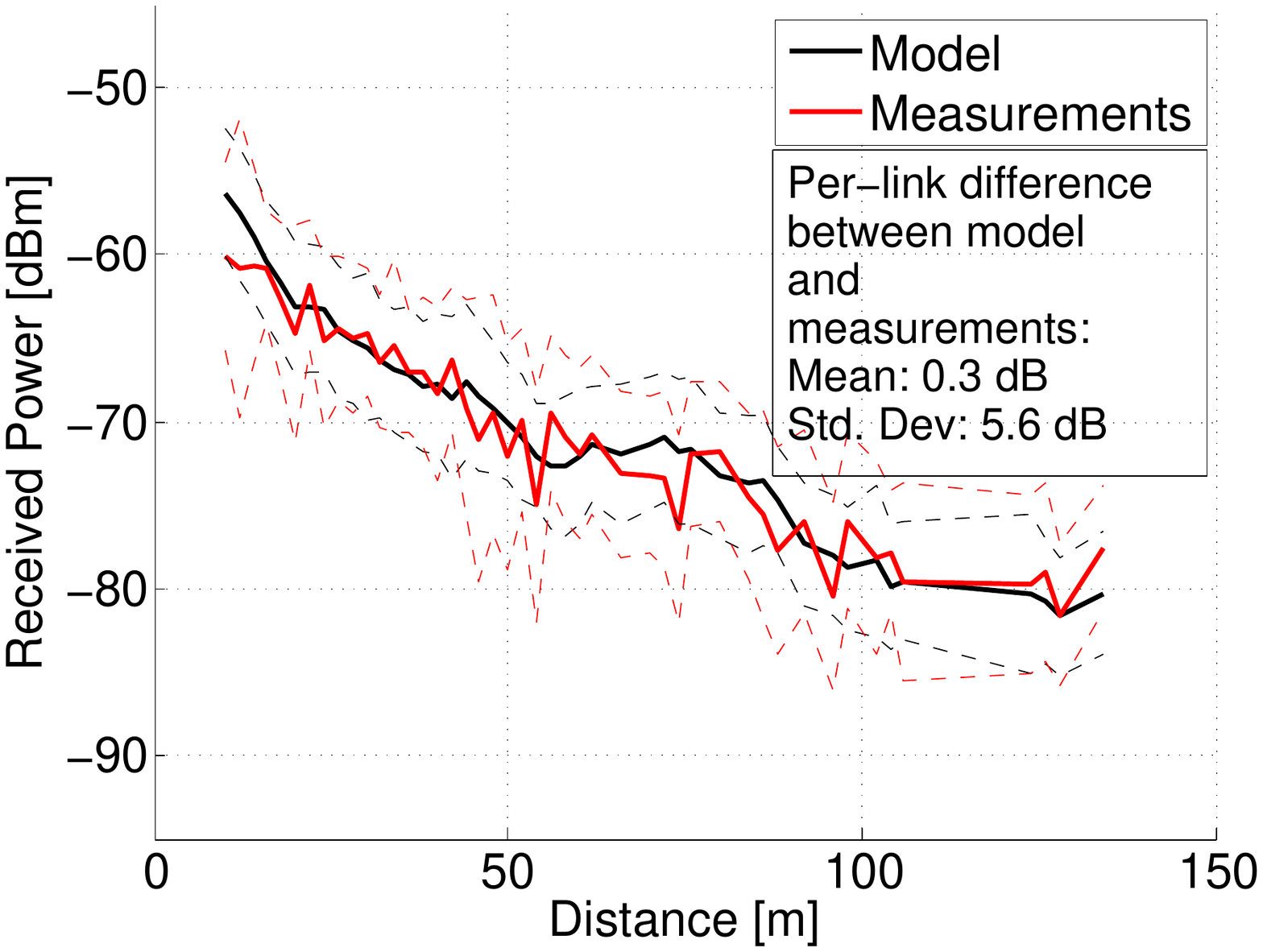}} \label{fig:LOSFitsRawPorU}
\caption[LOS data - comparison of model and measurements]{LOS data -- comparison of the received power generated by GEMV$^2$ and collected during the measurements. %
Figures~(b) through~(f) show the mean (full lines) and the standard deviation (dashed lines) around the mean received power for two-meter distance bins. %
Results are plotted for bins with at least 40 data points. %
All results show the data collected with passenger (short) vehicles. The results with tall vehicles (vans) exhibited similar behavior.} %
\label{fig:LOSFits}
\end{figure*}

Figure~\ref{fig:PortoComplete} shows the received power for a 30-minute experiment conducted on a 10~km route in downtown Porto, Portugal (Fig.~\ref{fig:DowntownPorto}), along with the results generated by GEMV$^2$. Using the videos recorded during the measurements, we separated the data into three link types: LOS, NLOSv, and NLOSb. 
This allowed us to evaluate the ability of GEMV$^2$ to simulate each link type. 

Since we extracted the small-scale signal variation from the measured datasets, %
we consider that the best performance GEMV$^2$ can have %
 is bounded by the empirically measured small-scale signal variation for the given link type. In other words, the performance of the employed large- and small-scale model is upper-bounded by the measured small-scale signal variation. %
 Therefore, for all results henceforth, apart from the mean difference in terms of received power between measured and modeled results, we also report the standard deviation around the mean %
  (i.e., the standard error of the model).  %
It is important to note that %
for each collected measurement datapoint, we calculate the mean and standard deviation using the 
per-packet received power difference between GEMV$^2$ and measurements. Furthermore, we use the same value of relative permittivity ($\epsilon_r$=1.003) in all environments to calculate the reflection coefficient for the ground reflection (i.e., we do not fit the value to a given dataset).

\begin{figure*}
\centering

\subfigure[ Porto Highway (A28) Passenger Vehicles -- Coordinates: 41.22776, -8.695148. Number of data points: 14200. \textbf{Measured averaged $\sigma$: 4.5~dB}]{\label{fig:NLOSvHighA28CC}\includegraphics[trim=0cm 6.5cm 0cm 5.5cm,clip=true,width=0.31\textwidth]{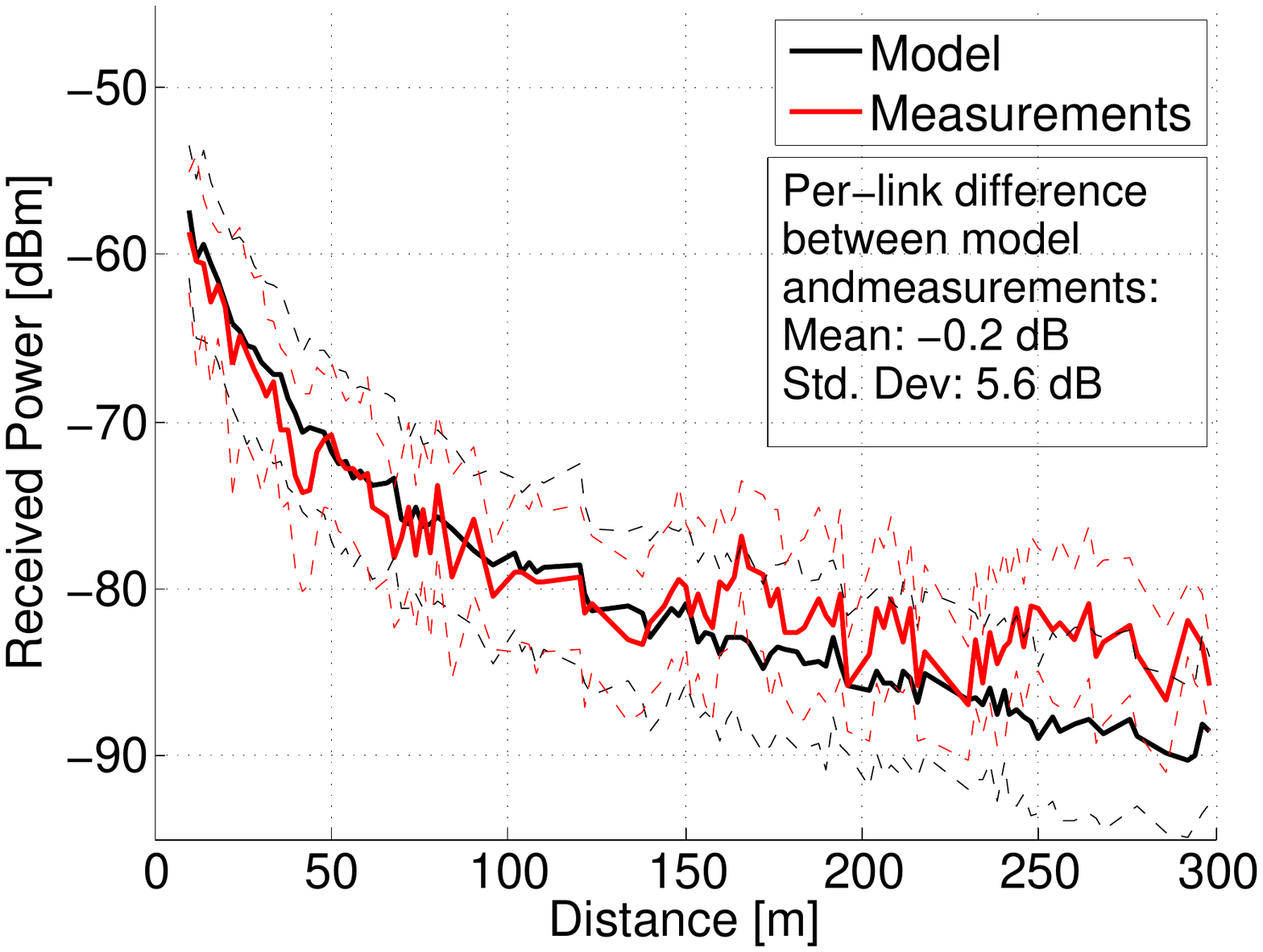}}
  \hspace{1mm}
\subfigure[ Porto Highway (A28) Tall Vehicles -- Coordinates: 41.22776, -8.695148. Number of data points: 14700. \textbf{Measured averaged $\sigma$: 3.8~dB}]{\label{fig:NLOSvHighA28VV}\includegraphics[trim=0cm 6.5cm 0cm 5.5cm,clip=true,width=0.31\textwidth]{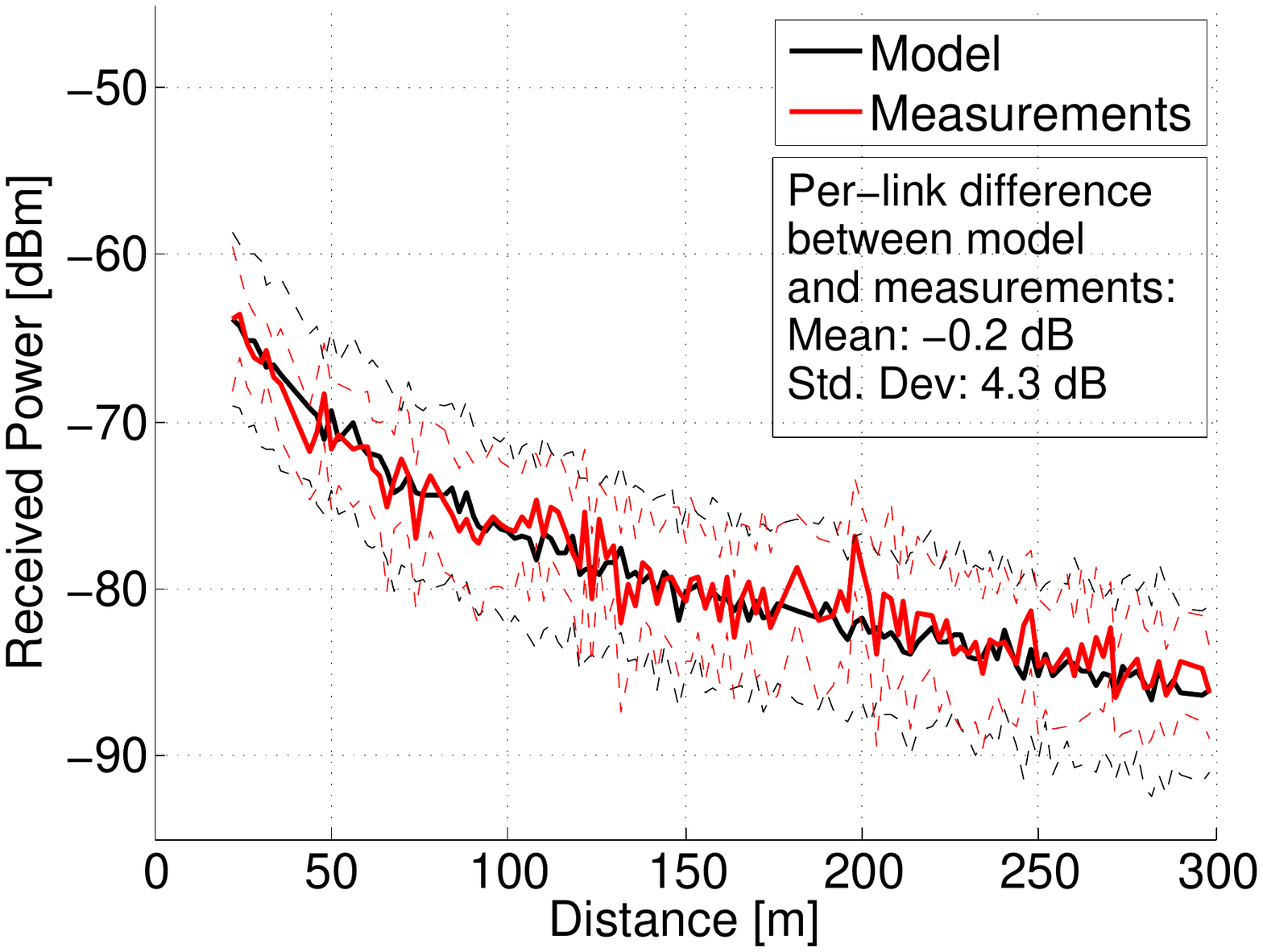}}
\subfigure[ Pittsburgh Suburban Passenger Vehicles %
-- Coordinates:  40.4476089, -79.9398574. Number of data points: 9500. \textbf{Measured averaged $\sigma$: 4.5~dB}]{\label{fig:NLOSvFits5thDay }\includegraphics[trim=0cm 6.5cm 0cm 5.5cm,clip=true,width=0.31\textwidth]{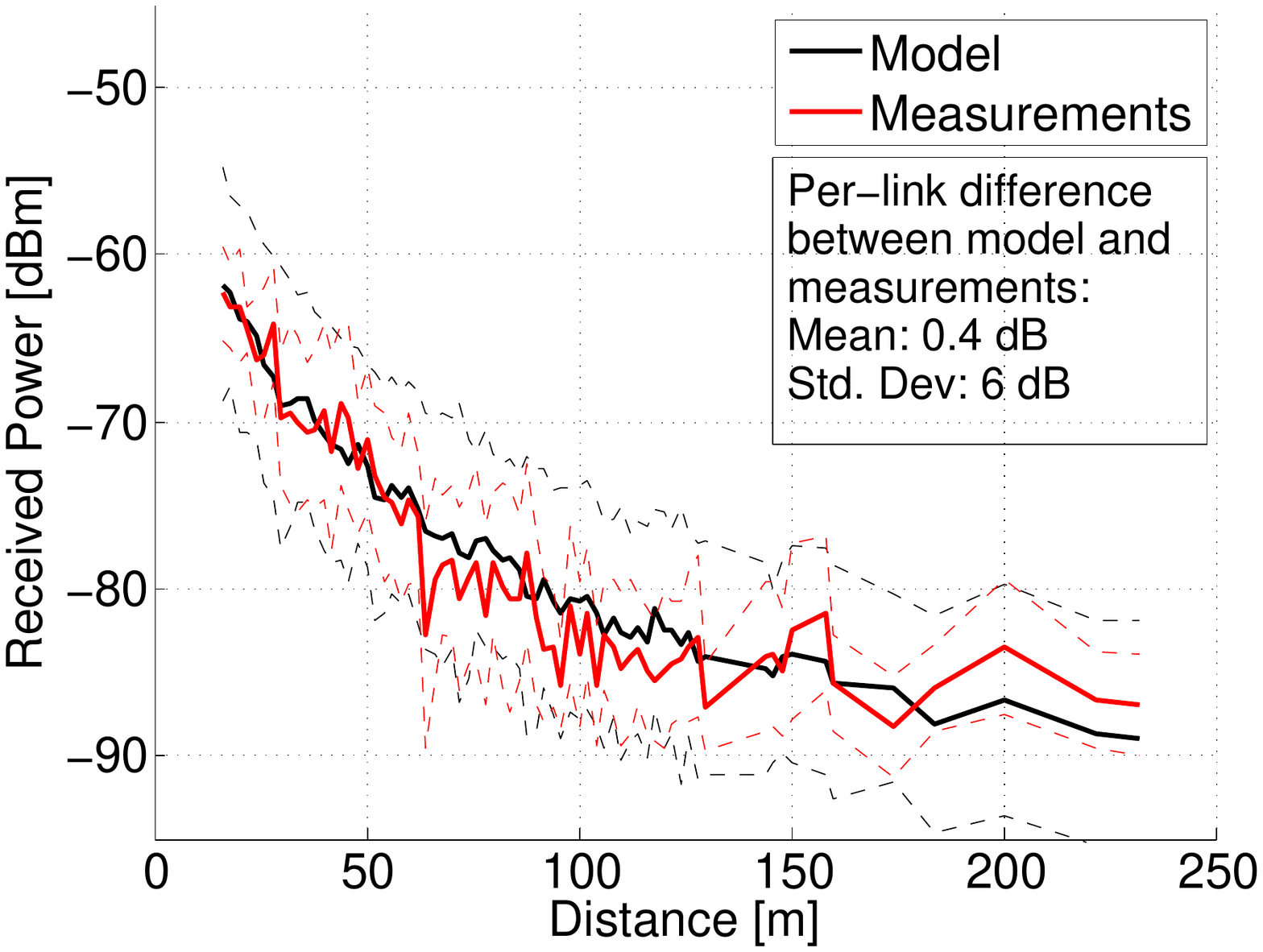}}

\subfigure[ Porto Downtown Passenger Vehicles -- Coordinates: 41.153673, -8.609913. Number of data points: 6300. \textbf{Measured averaged $\sigma$: 5.3~dB}]{\label{fig:NLOSvFitsUrbanCC}\includegraphics[trim=0cm 6.5cm 0cm 5.5cm,clip=true,width=0.31\textwidth]{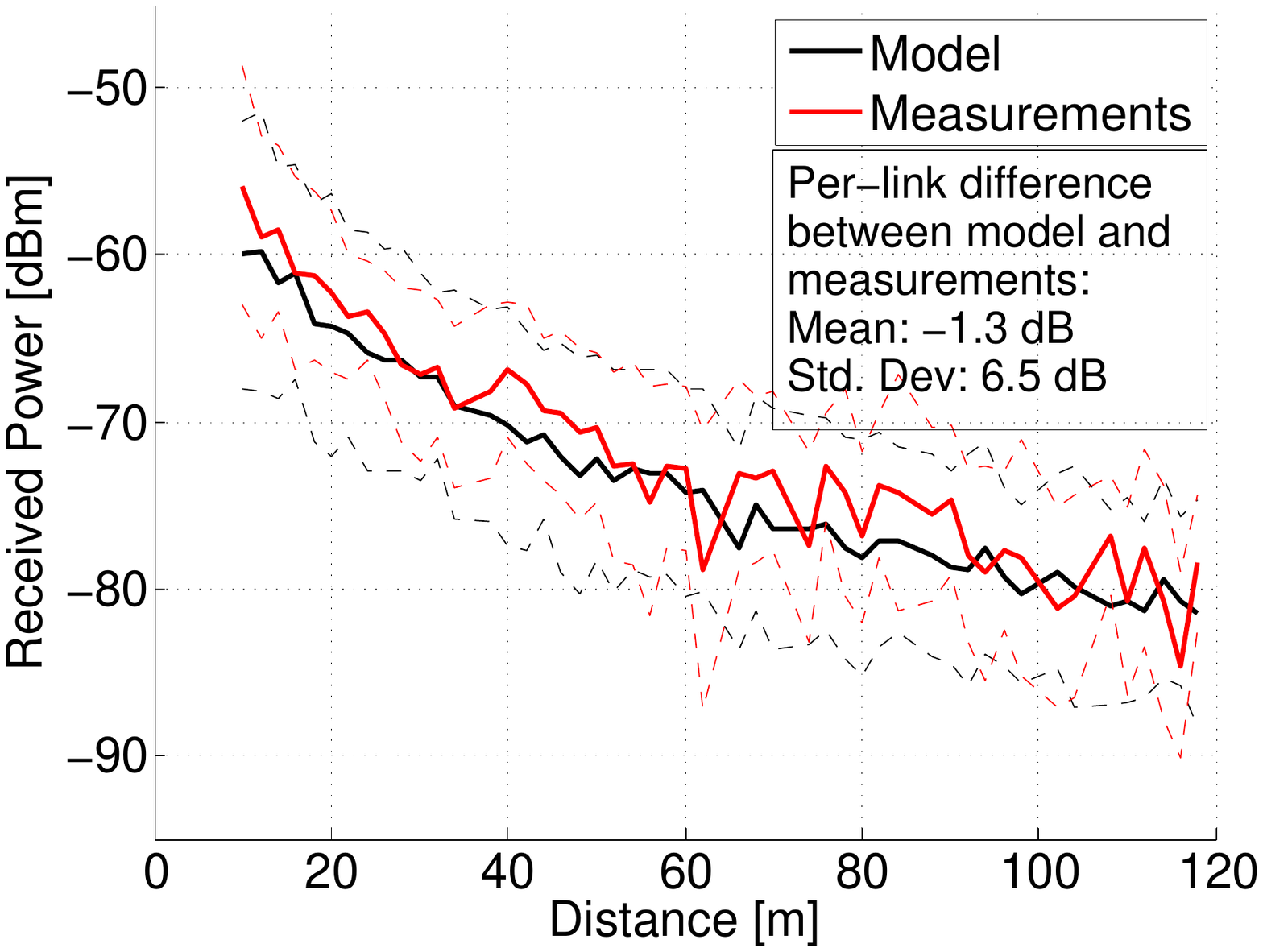}}
  \hspace{1mm}
\subfigure[ Porto Downtown Tall Vehicles -- Coordinates: 41.153673, -8.609913. Number of data points: 10500. \textbf{Measured averaged $\sigma$: 4.7~dB}]{\label{fig:NLOSvFitsUrbanVV}\includegraphics[trim=0cm 6.5cm 0cm 5.5cm,clip=true,width=0.31\textwidth]{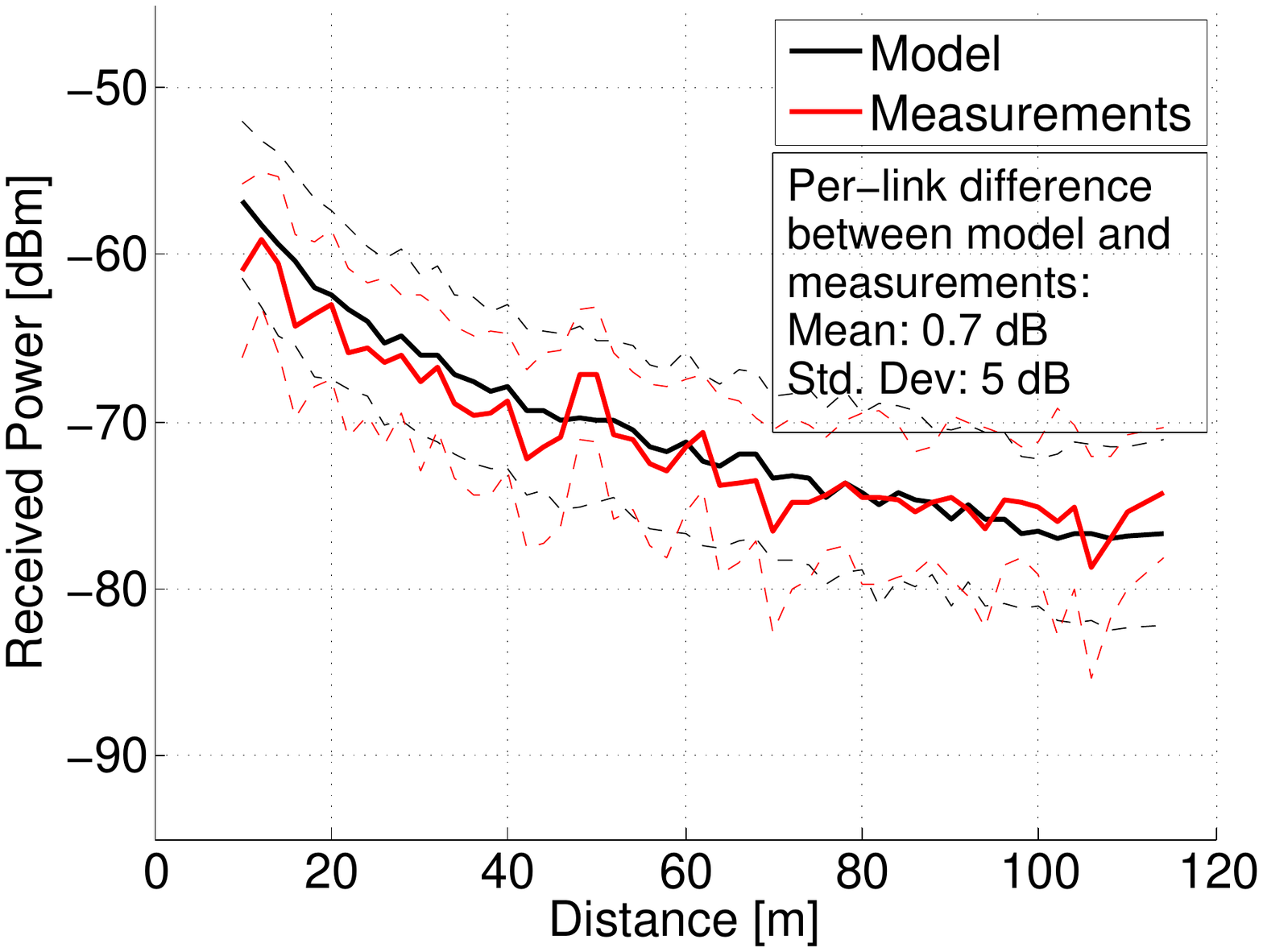}}
\hspace{1mm}
\caption[NLOSv data -- comparison of model and measurements]{NLOSv data -- comparison of the received power generated by GEMV$^2$ and collected during the measurements. Figures show the mean (full lines) and the standard deviation (dashed lines) around the mean received power for two-meter distance bins. %
Results are plotted for bins with at least 40 data points. %
 }

\label{fig:NLOSvFits}
\end{figure*}

\subsection{LOS links} \label{sec:LOSLinks}

Figure~\ref{fig:LOSFits} shows the results for the LOS links in different environments. The results generated by GEMV$^2$ fit the experimental data quite well in all environments, with the mean difference between model and measurements within 0.6~dB. Similarly, the standard error for all LOS datasets (shown in a text box in each of the subfigures of Fig.~\ref{fig:LOSFits}) is within 0.6~dB of the measured small-scale signal variation for each dataset (noted in the caption of each subfigure).
 Regarding the open space LOS results, we attribute the higher variability of the Pittsburgh Open Space dataset (Fig.~\ref{fig:LOSFitsHomestead}) compared to the Porto Open Space dataset (Fig.~\ref{fig:LOSFitsLeca}) to the guard rails and metal fence (visible in Fig.~\ref{fig:Homestead}), which did not exist in the Porto Open Space location (Fig.~\ref{fig:Leca}). 
 The daytime Pittsburgh Suburban (Fig.~\ref{fig:LOSFitsFifthDay}) and Porto Downtown scenarios (Fig.~\ref{fig:LOSFitsUrban}) have a significantly richer propagation environments due to the nearby vehicles in case of the former and both vehicles and buildings in case of the latter. This results in the increase of both the small-scale signal variation %
 and the standard error. %
More details on the suitability of the two-ray ground reflection model for LOS links in different environments, as well as the impact on the application-level performance metrics is available in~\cite{boban13ACM}.
\subsection{NLOSv links}

Figure~\ref{fig:NLOSvFits} shows the results for the NLOSv links in different environments with both passenger (short) and tall vehicles. The results generated by GEMV$^2$ fit the experimental data well in all environments, with the mean difference between model and measurements within 1.3~dB in each of the environments. Again, the standard error for NLOSv links (shown in a text box in each of the subfigures of Fig.~\ref{fig:NLOSvFits}) is within 0.9~dB of the
measured 
 small-scale signal variation of each dataset (noted in the caption of each subfigure). It is interesting to see that NLOSv results for tall vehicles (vans) experience both lower small-scale signal variation and lower standard error. %
This is due to the taller position of the antennas, %
which experience fewer significant reflected, diffracted, and scattered rays than the antennas on the shorter vehicles, thus resulting in a more stable channel.

\begin{figure}
  \begin{center}
    \includegraphics[trim=0cm 7.5cm 0cm 8cm,clip=true,width=0.45\textwidth]{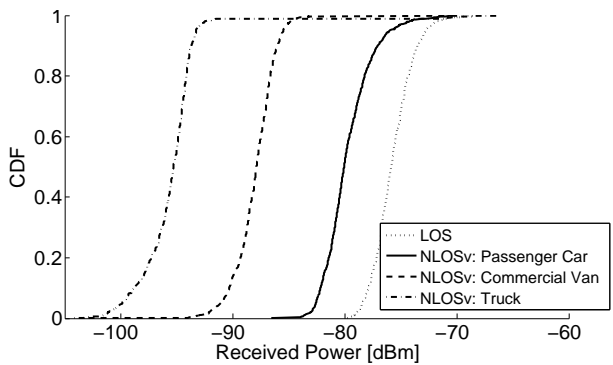}
     \caption{Received power distribution as generated by GEMV$^2$ for LOS links and NLOSv links due to three vehicle types: passenger car (mean height: 1.5 meters), commercial van (mean height: 2 meters), and truck (mean height: 3 meters); standard deviation for the height of each vehicle type was set to 0.15 meters. For each link, a single vehicle of a given type is placed between transmitter and receiver, both of which are passenger cars with the height of 1.5 meters. %
     Distance between transmitter and receiver is uniformly distributed between 75 and 125 meters. Transmit power is set to 10~dBm and gains at both transmit and receive antenna are 1~dBi.}
      \label{fig:NLOSvCarVanTruck}
   \end{center}
\end{figure}

To illustrate the impact of different types of vehicular obstructions, %
Fig.~\ref{fig:NLOSvCarVanTruck} shows the received power when the LOS is blocked by a car, a van, and a truck. For a mean distance of 100 meters between transmitter and receiver, different obstructing vehicle types attenuate the power in a distinct fashion, with the mean attenuation compared to LOS of approximately 5, 13, and 20~dB for car, van, and truck, respectively. The attenuation varies depending on the position of the obstructing vehicle: it is lowest when the obstructing vehicle is near the middle of the transmit-receive distance, while it increases as the vehicle gets closer to either the transmitter or the receiver. %
These results, which are in line with previous measurements reported in~\cite{abbas12,boban11,meireles10}, demonstrate the ability of GEMV$^2$ to incorporate the impact of different types of vehicular obstructions.  %

\subsection{NLOSb links}

\begin{figure}
\centering
\subfigure[ Reflections (green lines) and diffractions (magenta) generated by the model and overlaid on the image of the Porto Outlet location. The vehicles started close to each other with clear LOS and slowly moved along paths indicated by the arrows, thus going from LOS to NLOSb conditions. During the measurements, the two large buildings that create reflections and diffractions were the only large protruding objects in the scene, with clearance in excess of 100~meters to the nearest objects (i.e., there were no parked vehicles). Coordinates of the location: 41.300137, -8.707385]{\label{fig:OutletOverlayReflDiffr}\includegraphics[width=0.4\textwidth]{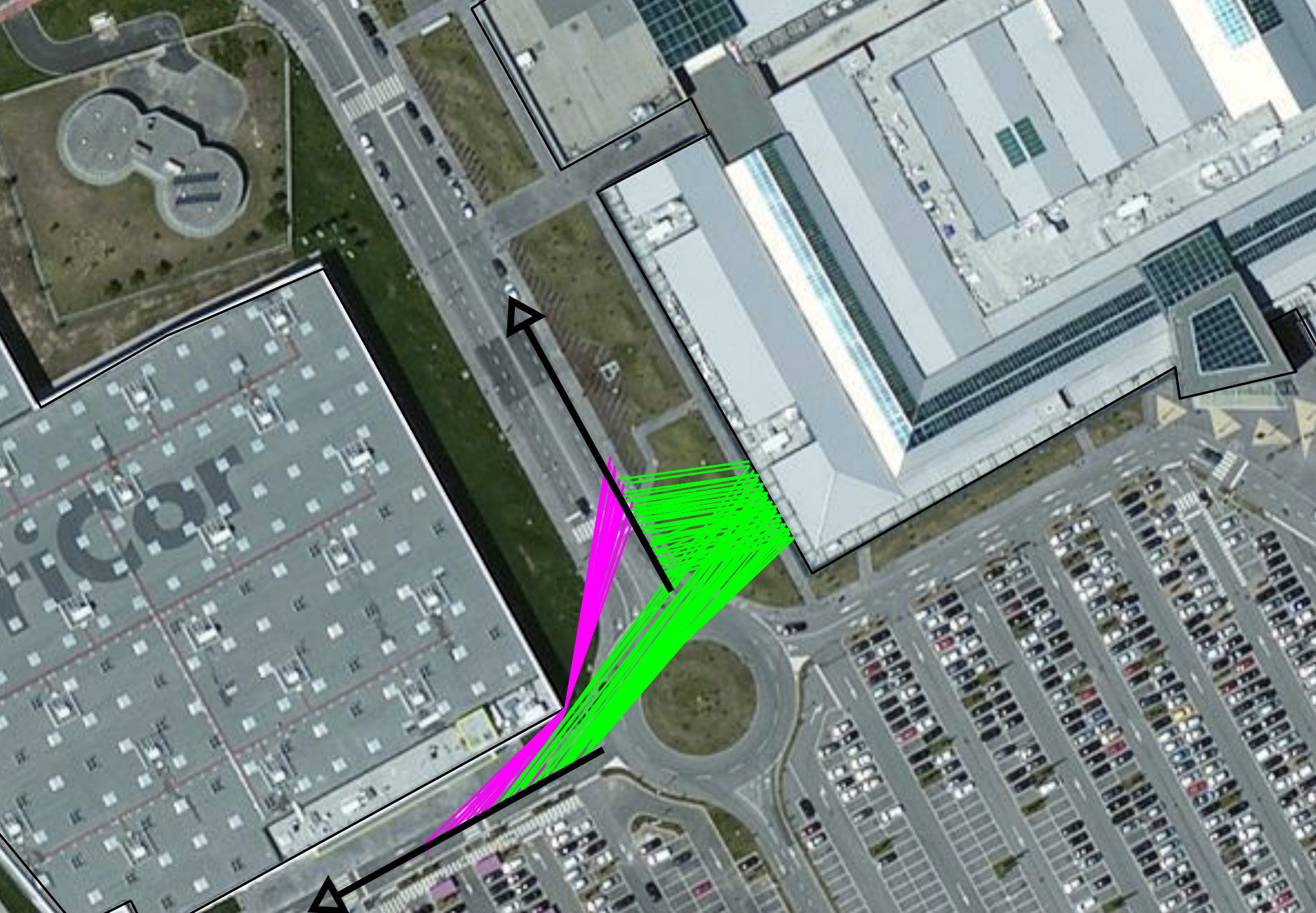}}
\subfigure[ Transition of the model between LOS and NLOSb conditions at the Porto Outlet location for a single 30-second run. The annotations refer to the LOS conditions as identified by the model.
The outliers in the top right corner are due to GPS inaccuracy while vehicles were stationary. For comparison, we plot the log-distance path loss for the same location. The log-distance path loss model was calculated by taking the reference distance of one meter and fitting the path loss exponent to the measurement data and minimizing the square residuals with respect to the received power in dBm.]{\label{fig:outletAnnotated}\includegraphics[trim=3cm 13cm 4cm 13cm,clip=true,
width=0.4\textwidth]{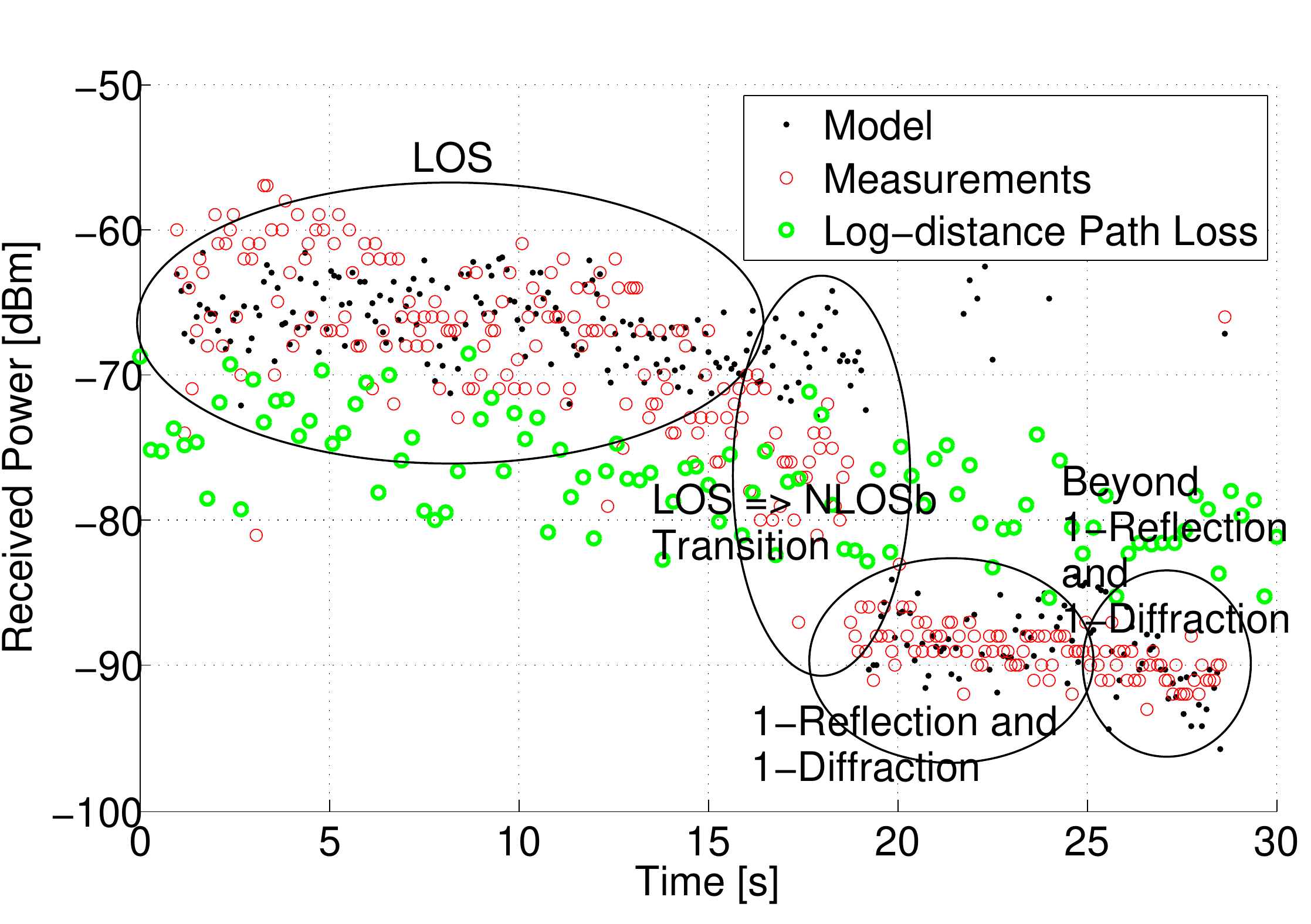}}
\caption{Porto Outlet experiment.}
\label{fig:Outlet}
\end{figure}

\begin{figure}
  \begin{center}
    \includegraphics[trim=0cm 6cm 0cm 5.5cm,clip=true,width=0.41\textwidth]{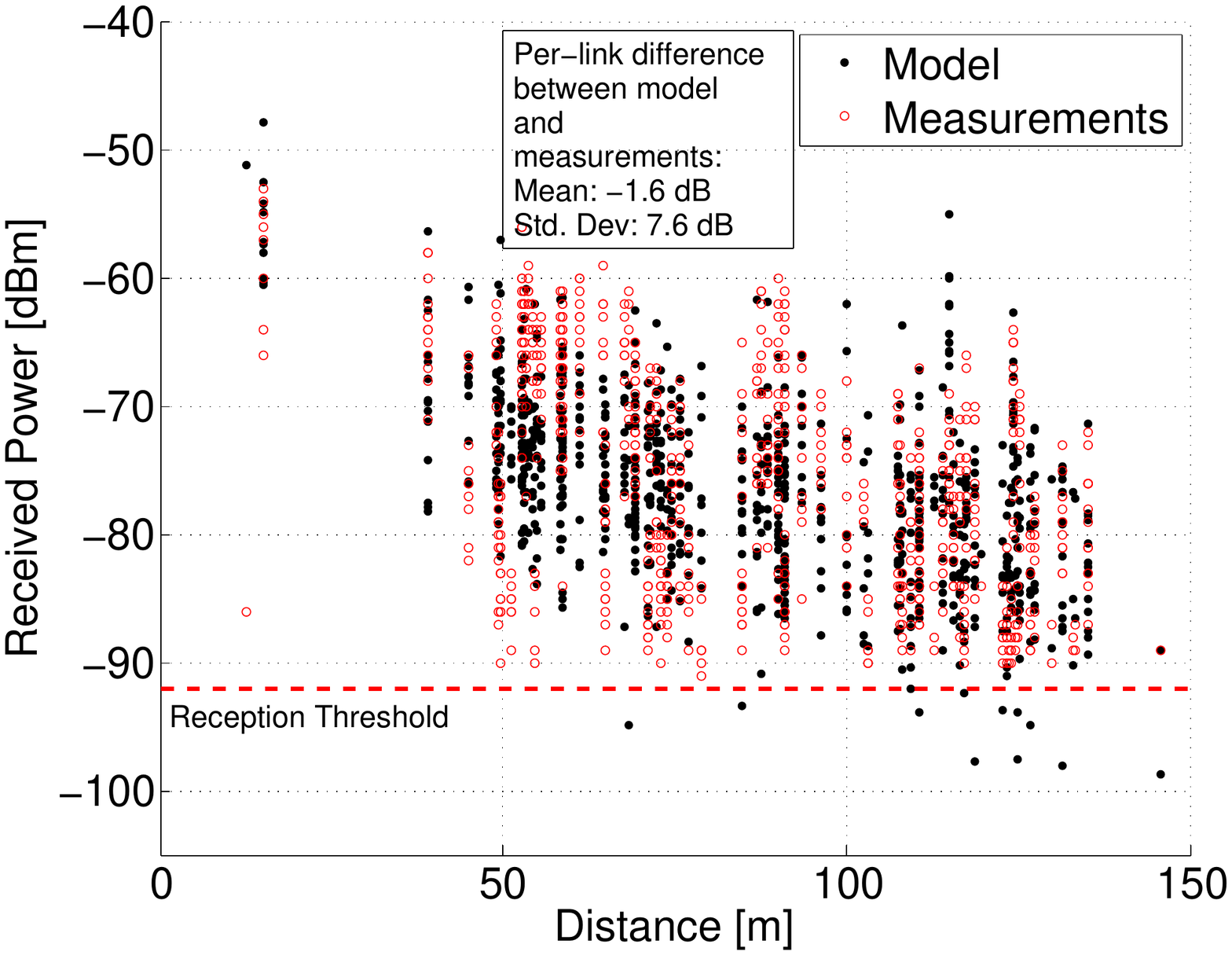}
     \caption[NLOSb data -- comparison of the received power generated by GEMV$^2$ and collected during the measurements.]{\small NLOSb data -- comparison of the received power generated by GEMV$^2$ and collected during the measurements in Porto Downtown location using passenger vehicles.
     \textbf{Measured averaged $\sigma$: 6.8~dB.} }
      \label{fig:NLOSbFits}
   \end{center}
\end{figure}

Figure~\ref{fig:OutletOverlayReflDiffr} shows the Porto Outlet location with the overlaid reflecting and diffracting rays as generated by GEMV$^2$. Once the vehicles are not in LOS, single-interaction reflections and diffractions become the predominant propagation mechanisms. Figure~\ref{fig:outletAnnotated} shows distinct transitions in the received power as the vehicles go from LOS to NLOSb conditions. GEMV$^2$ is able to capture the steep drop in the received power once the LOS is obstructed by building. At the same time, the log-distance path loss, because it is unable to capture the transition between LOS and NLOSb, underestimates the received power in LOS conditions and overestimates it in NLOSb conditions. This result highlights the importance of location-specific, link-level propagation modeling: the transition between different link types, which exhibit considerably different characteristics, %
 can only be performed by taking into account the objects in the specific location. On the other hand, models relying on the common parameters of an environment (such as the overall path-loss exponent in the case of log-distance path loss) are unable to model such transitions, which results in ``averaging'' of the received power across different link types. %
these results %
show that the different link types need to be identified and modeled separately.

Previous measurement studies concluded that, for communication in the 1~to~6~GHz frequency band, transmission through buildings does not play as important role as reflections and diffractions around buildings. %
For example, Anderson in~\cite{anderson98} performed measurements at 1.8~GHz and modeled the diffraction and reflection around an isolated building corner using uniform theory of diffraction (UTD). The author concluded that through-wall transmission is negligible compared to the corner diffraction and wall reflections. Durgin et.~al in~\cite{durgin98} performed measurements at 5.85~GHz and pointed out that ``transmission through the house was not as important as outdoor multipath scattering''. Similarly, as shown in Fig.~\ref{fig:Outlet}, our results showed a good match between the measurements and GEMV$^2$  including reflections and diffractions only (i.e., without modeling the transmission through buildings). For this reason, we do not consider the through-building transmission as an important effect and therefore do not include it in our model.

Figure~\ref{fig:NLOSbFits} shows NLOSb links for the measurements performed in downtown Porto. The difference between results generated by GEMV$^2$ and measurements is higher than in the case of LOS and NLOSv (mean difference is -1.6~dB, and the standard deviation is 7.6~dB). %
The increased difference is due to two reasons: 1) %
the variety of communication scenarios encompassed by NLOSb data is higher (e.g., slight obstruction by a building corner, deep obstruction by an entire building, obstruction by foliage, etc.), resulting in a standard deviation of the measured received power of 6.8~dB, which is considerably higher than LOS and NLOSv; 
and 2) along the measurement route, there was occasional foliage which was not recorded in the geographical database, thus it was not modeled. %

Furthermore, unlike the LOS and NLOSv data, where the packet delivery rate (PDR) was above 80\% for the observed distances, PDR for the collected NLOSb data was above 80\% for distances below 50~m, approximately 30\% between 50~m and 150~m, and below 15\% between 150~m and 500~m. %
Additionally, during measurements, only decodable data was recorded %
(i.e., only packets received above the reception threshold of -92~dBm were recorded). This made it impossible to compare the results generated by GEMV$^2$ with the measurement data below the reception threshold, thus increasing the difference between measurements and the model. %

\subsection{Small-scale signal variation}

\begin{figure}
\begin{center}
\includegraphics[trim=8cm 21.5cm 8cm 22.5cm,clip=true,width=0.41\textwidth]{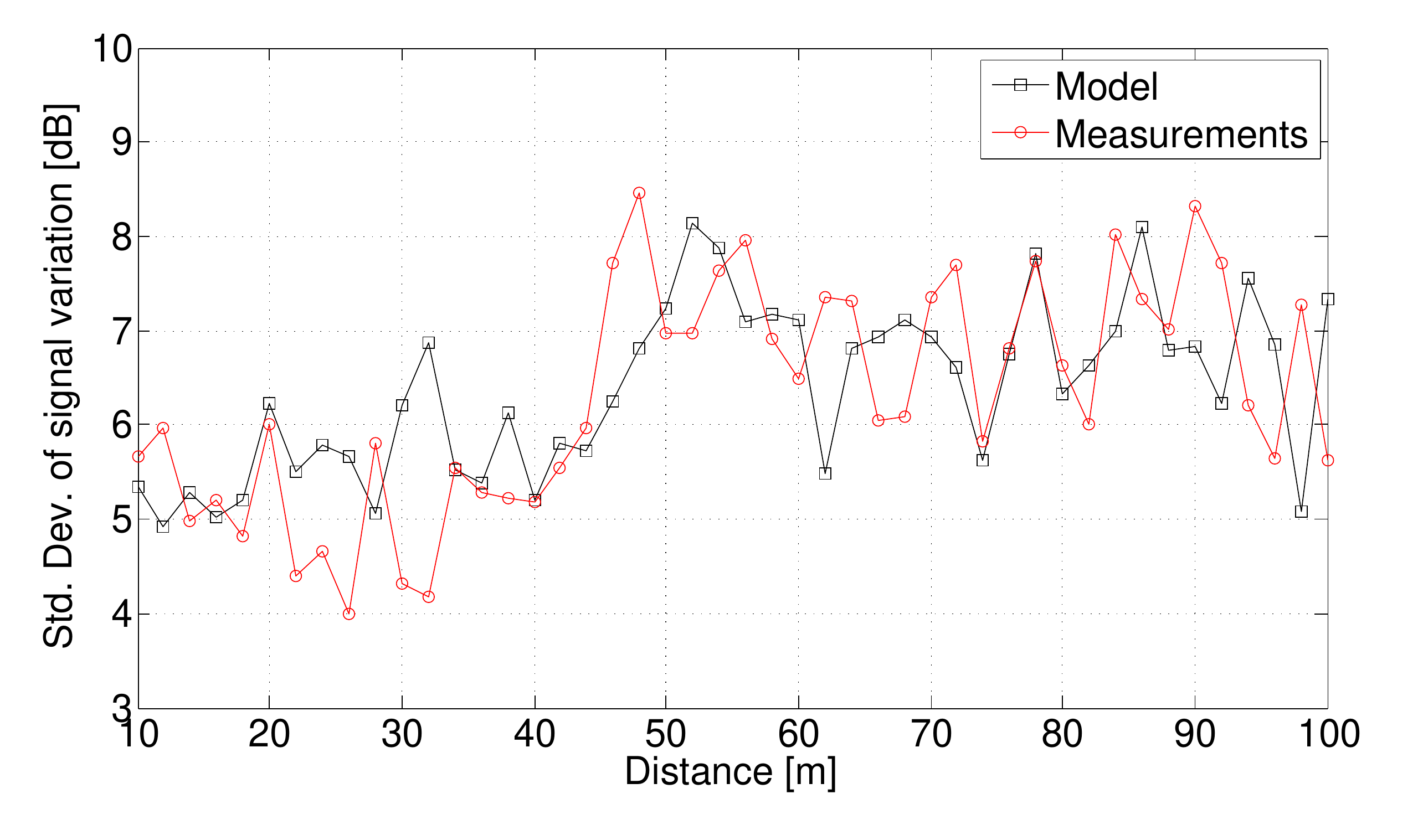}
\caption{Standard deviation of the small-scale signal variation generated by GEMV$^2$ and extracted from measurements in downtown Porto. All three link types (LOS, NLOSv, NLOSb) are combined and placed in two-meter distance bins. Only bins with more than 40 data points are included.}
\label{fig:fadingModelVsMeasurements}
\end{center}
\end{figure}

Figure~\ref{fig:fadingModelVsMeasurements} shows the standard deviation of the small-scale signal variation for GEMV$^2$ and measurements. 
For each two-meter bin, the variation is a composite result generated by the large-scale model (that also includes a part of small-scale effects through 
three diffracted rays in case of NLOSv links and reflections and diffractions in case of NLOSb links)
with the addition of the zero-mean, normally distributed variable with standard deviation $\sigma$ determined using eq.~\ref{eq:FadingCalc}. %
 GEMV$^2$ generates the overall signal variation comparable to that obtained through measurements, with %
 the variation across the distance bins for both GEMV$^2$ and the measurements of approximately 6.3~dB. %
 This result shows that GEMV$^2$ can capture the fast-varying signal changes in vehicular environment by considering the objects surrounding the communicating pair. %

\begin{figure}
\centering

\subfigure[ Snapshot showing a part of %
transmit-receive 
pairs in the city of Porto.]{\label{fig:GEPlotLarge}\includegraphics[width=0.45\textwidth]{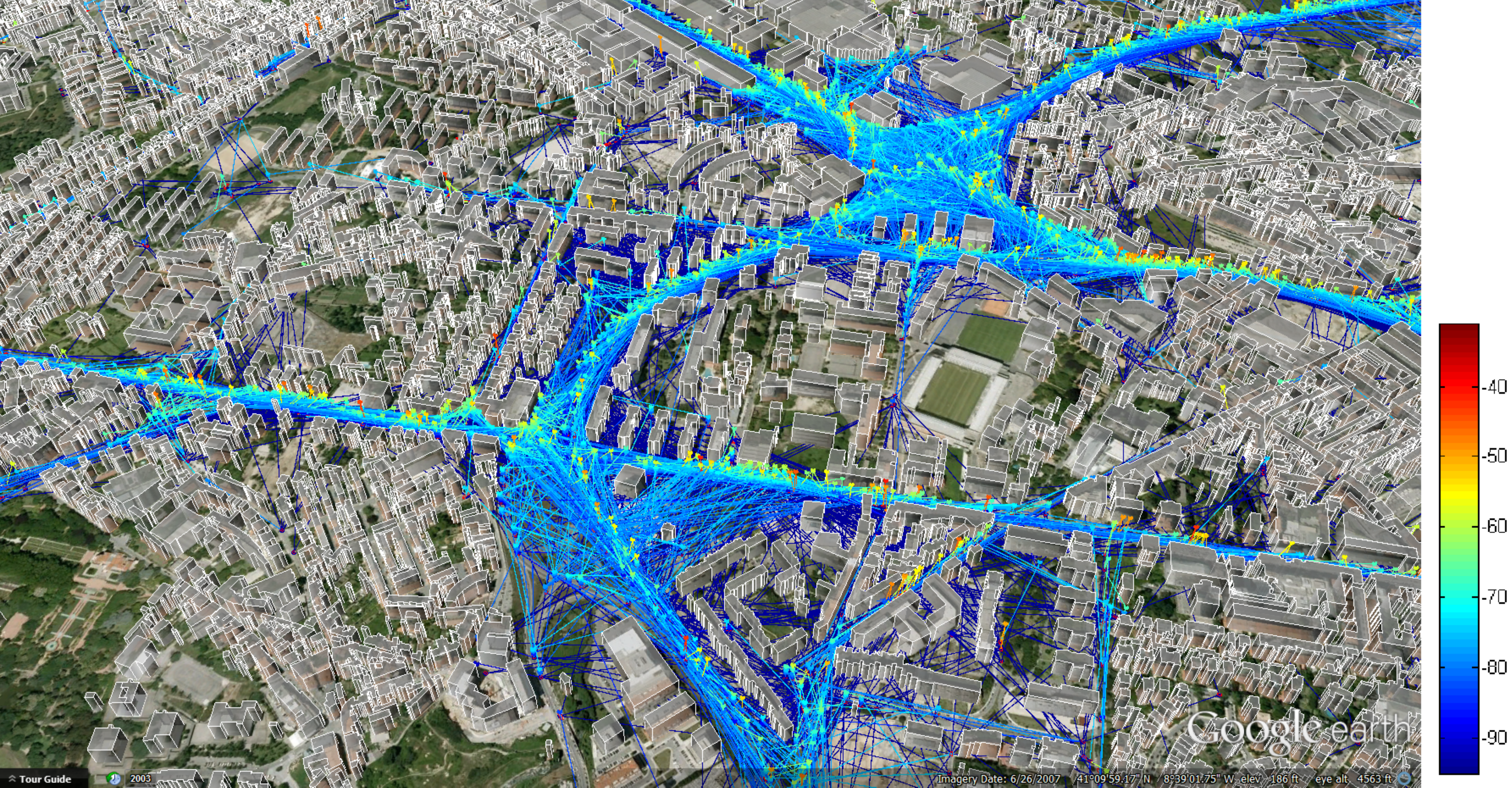}}
\subfigure[ Street-level view.]{\label{fig:GEPlotSmall}\includegraphics[width=0.45\textwidth]{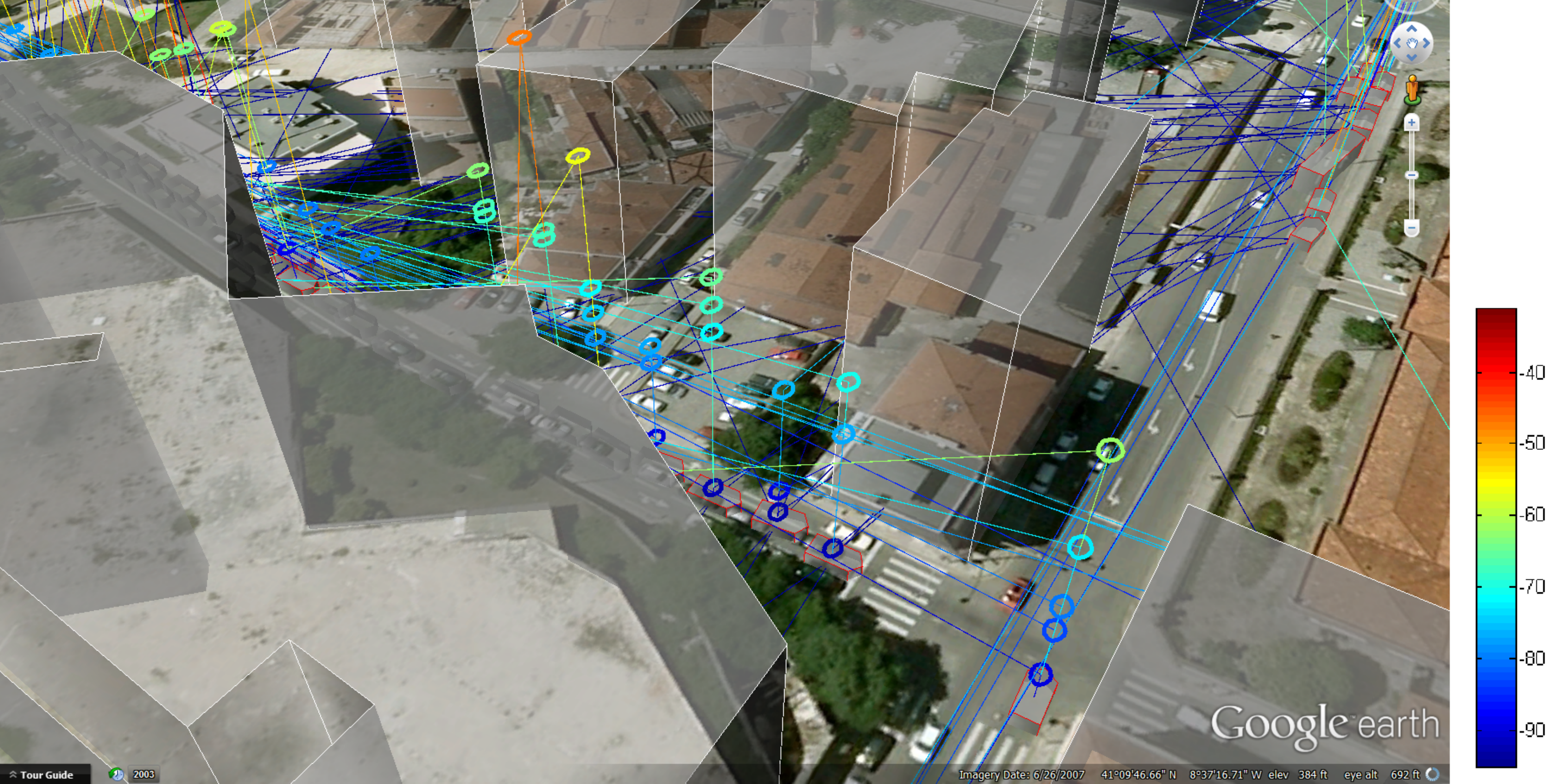}}
\caption{%
Google Earth visualization of the received power calculated by GEMV$^2$. Building outlines are colored white; vehicle outlines are colored red. Each communicating pair is connected with a line that has circles representing transmitter (dark blue circle at the ground level) and receiver (elevated circle). Warmer line colors and receiver circles positioned higher represent higher received power. Color bars are in dBm.}
\label{fig:GEPlots}
\end{figure}

%
%

%
%
%
%
%
%
%

%
%
%

\section{A Few Notes on MATLAB Implementation and Scalability of GEMV$^2$}\label{sec:Performance} 
We implemented GEMV$^2$ in MATLAB and made the source code  
freely available at \url{http://vehicle2x.net}. %
GEMV$^2$ was able to simulate propagation for the entire city of Porto (41~km$^2$ area containing 10566 vehicles and 17346 static objects) using commodity hardware and the communication ranges shown in Table~\ref{tab:rLOS}. Google Earth visualization of the received power in Porto as generated by GEMV$^2$ is shown in Fig.~\ref{fig:GEPlots}. %

\begin{figure}
\centering

\subfigure[ Processing time for different network sizes.]{\label{fig:totalSimTime}\includegraphics[trim=8cm 20cm 8cm 20cm,clip=true,width=0.23\textwidth]{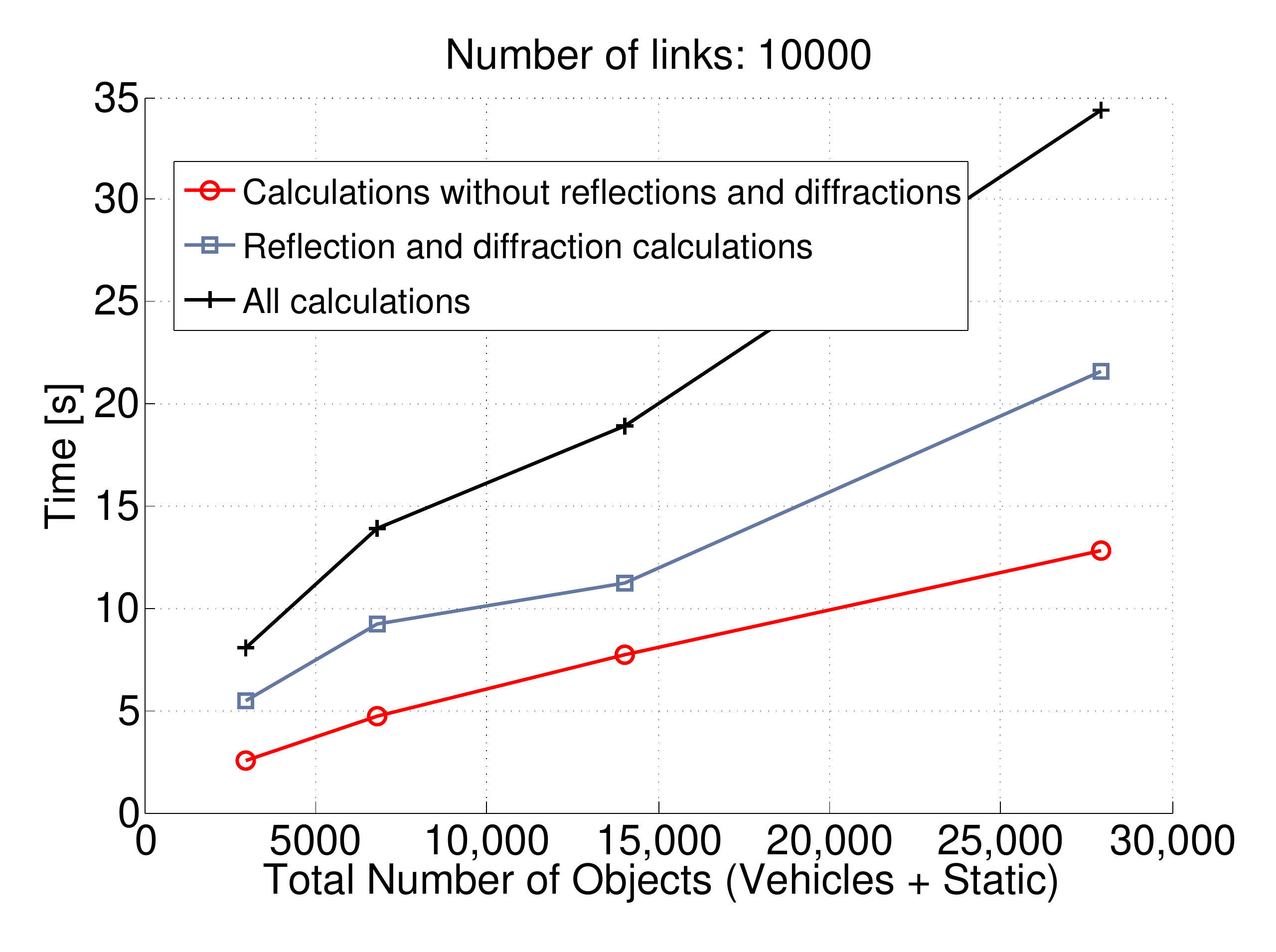}}
  \hspace{1mm}
\subfigure[ R-tree construction times. Note that static R-tree needs to be constructed once, whereas vehicle R-tree is constructed at each time step.]{\label{fig:RTreeTimes}\includegraphics[trim=8cm 20cm 8cm 20cm,clip=true,width=0.23\textwidth]{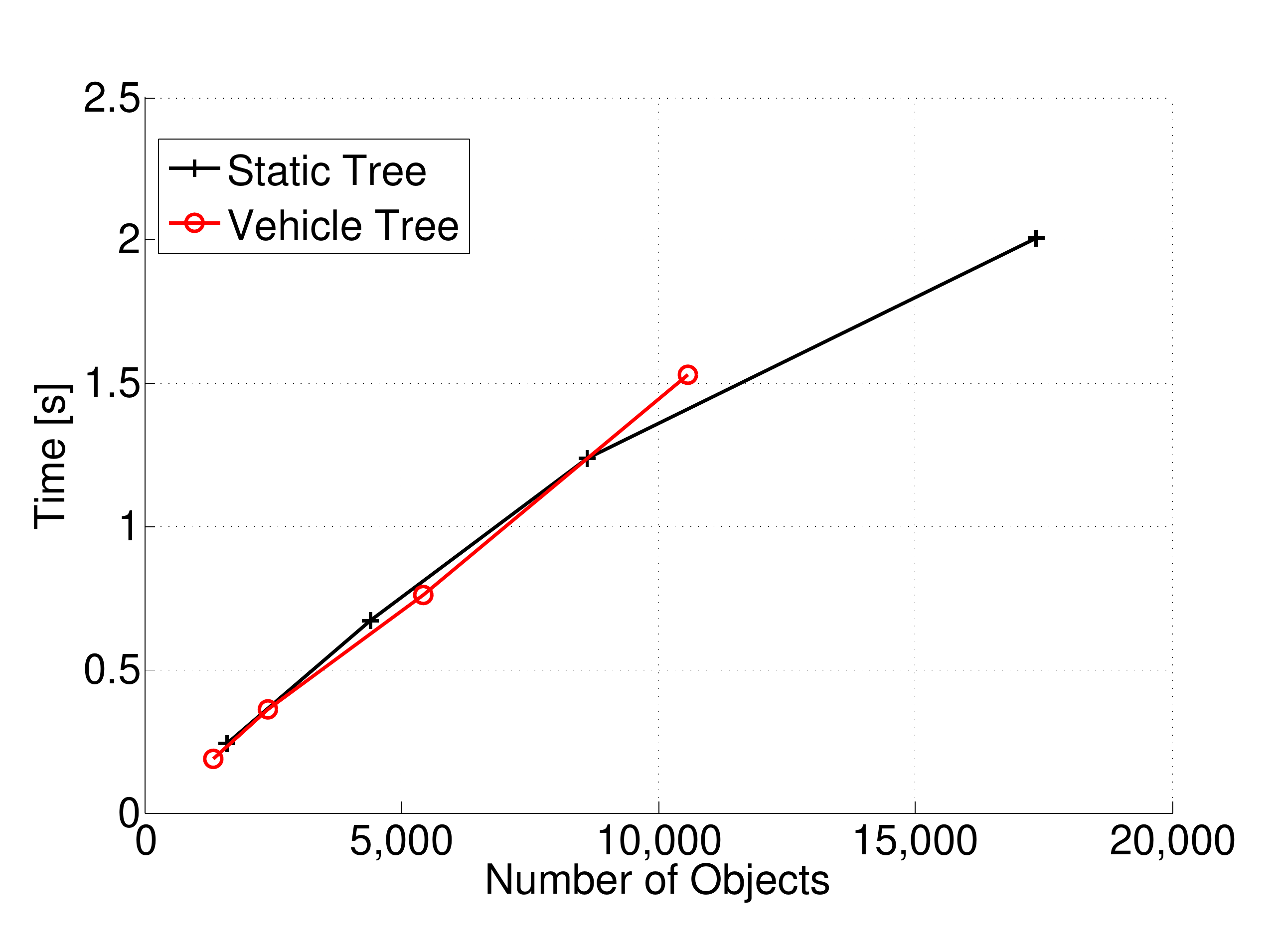}}
\subfigure[ Time to classify fixed number of links (10000) for different network sizes.]{\label{fig:10kLinkClassification}\includegraphics[trim=8cm 20cm 8cm 20cm,clip=true,width=0.23\textwidth]{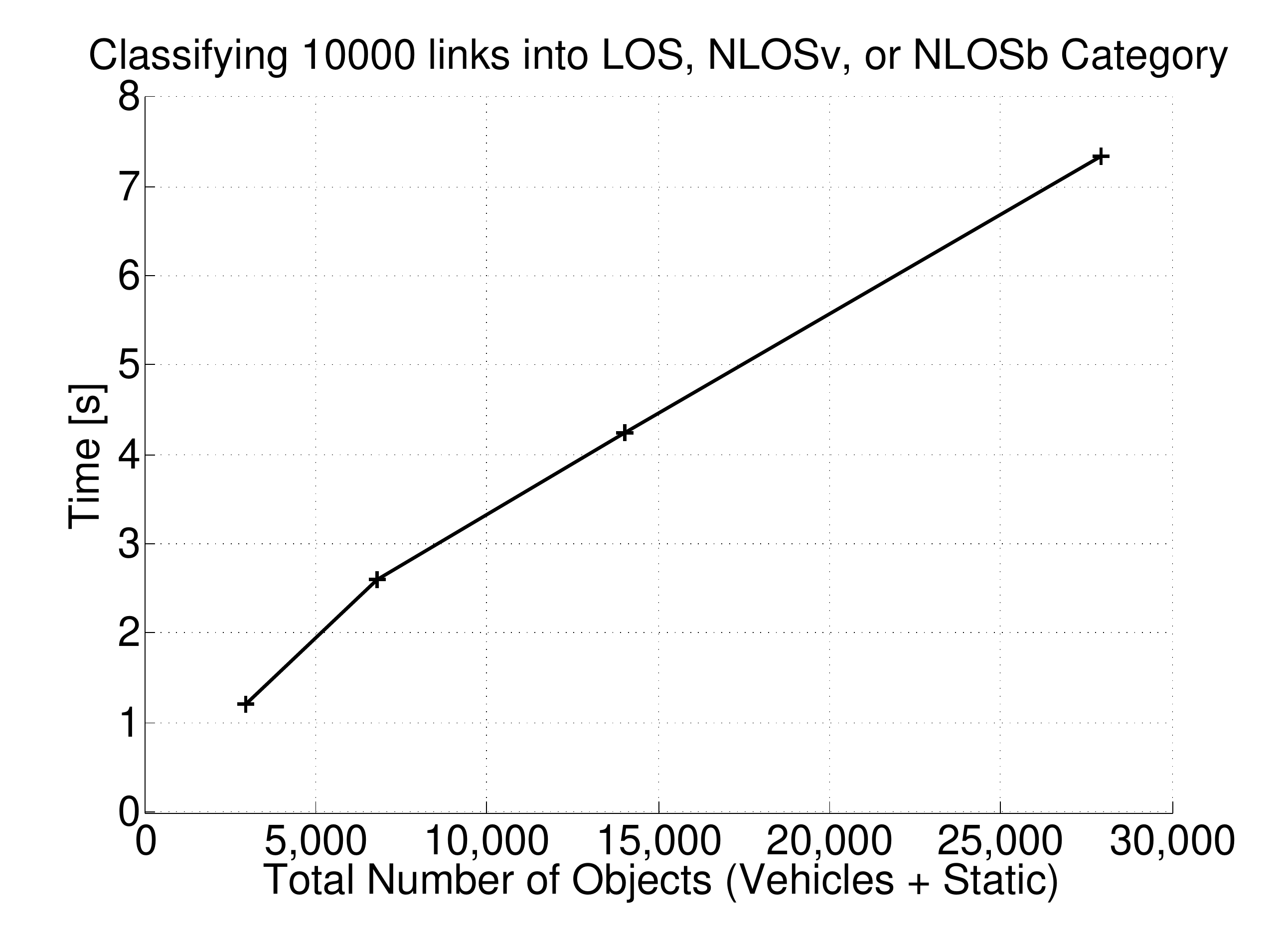}}
\hspace{1mm}
\subfigure[ Time to classify different number of links for a fixed network size.]{\label{fig:linkClassificationTime}\includegraphics[trim=8cm 20cm 8cm 20cm,clip=true,width=0.23\textwidth]{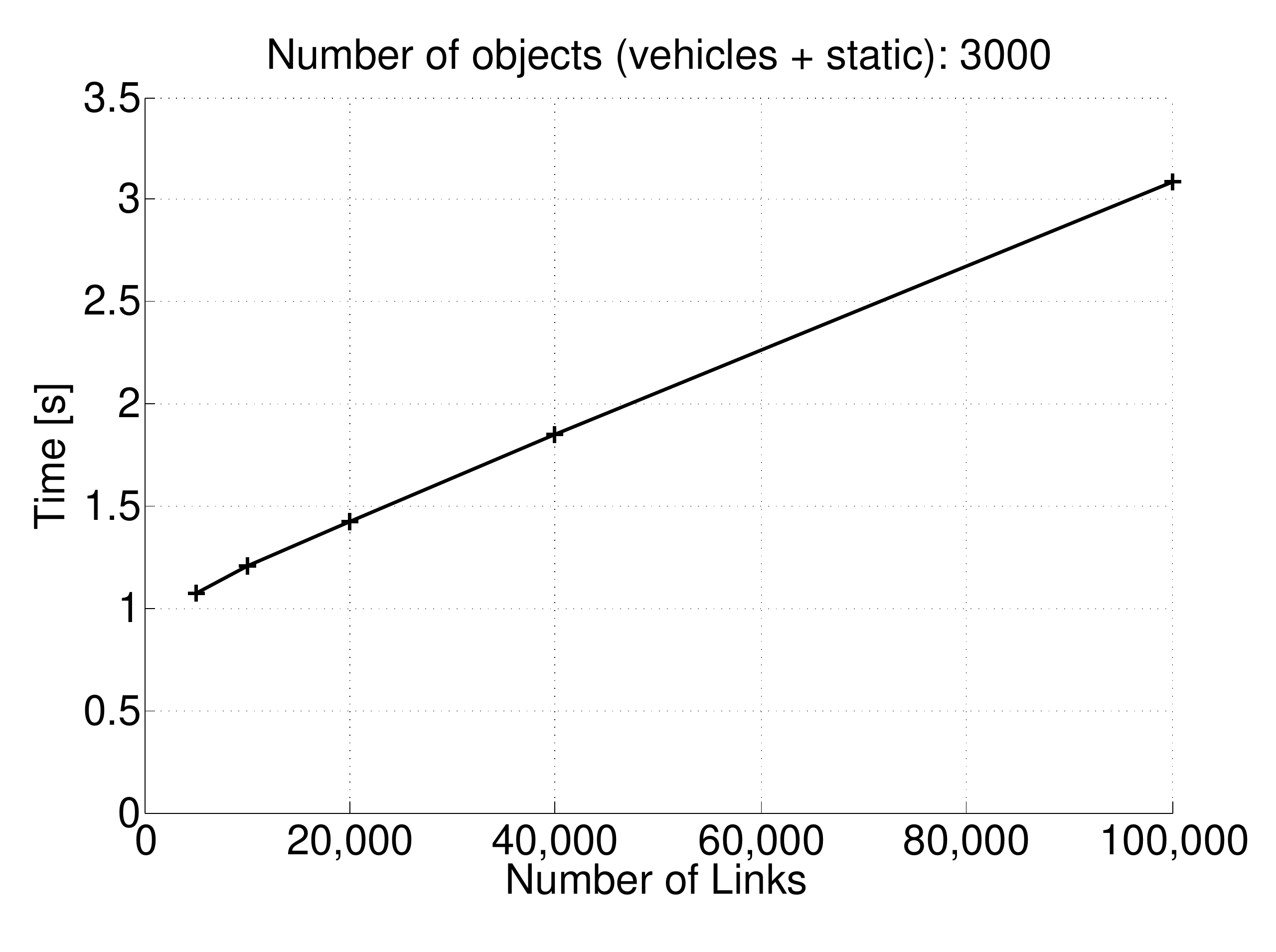}}

\caption[Performance of the Implemented Model]{Execution times of GEMV$^2$ implemented in MATLAB (version R2012a) under OS X v10.8 using the following hardware: 2011 Apple MacBook Air, 1.7GHz Core i5, 4~GB RAM. We used Porto Downtown dataset, which contains 10566 vehicles and 17346 static objects spread over 41 km$^2$. For simulations on smaller networks, %
we used half, quarter, and eighth of the entire area, which contained corresponding number of vehicles and static objects. All results were generated using a single CPU core (i.e., no parallel execution).  %
 }

\label{fig:modelPerformance}
\end{figure}

In terms of scalability, 
Fig.~\ref{fig:totalSimTime} shows the time it takes to process 10000 links (Tx-Rx communication pairs) in networks of varying size. By increasing the network size (i.e., the number of objects in the scene), the processing time increases linearly even for the largest network size with more than 28000 objects. 

Figure~\ref{fig:RTreeTimes} shows that the R-tree construction scales linearly with the number of objects that need to be stored in the tree. The results for constructing vehicle and static object R-trees are similar, since it takes marginally more time to fit the more complex static objects (outlines of buildings/foliage) in the minimum bounding rectangles. After that stage, the calculations per object are identical. Figure~\ref{fig:10kLinkClassification} shows the increase in link classification time when the network size (and therefore, the vehicle and static R-tree size) increases. Again, the increase is linear with the size of the network.

In order to calculate Signal to Interference plus Noise Ratio (SINR), network simulators need to take into account signal contributions from all currently transmitting neighboring vehicles (i.e., all vehicles concurrently transmitting within the collision domain). For this reason, we analyze the per-link processing time. Fig.~\ref{fig:linkClassificationTime} shows that, for a fixed network size, increasing the number of links results in a linear increase of processing time with a mild slope (e.g., to classify 5000 links, it takes 1.1 second, whereas for 100000 links it takes 3.1 second). 


Since calculating reflections and diffractions takes up two thirds of the overall processing time across all network sizes (see Fig.~\ref{fig:totalSimTime}), we investigate the performance of GEMV$^2$ when it relies on the log-distance path loss only while calculating the received power for NLOSb links (i.e., without calculating reflections and diffractions). We found out that, as the maximum transmission range for NLOSb links ($r_{NLOSb}$) increases, proportionally fewer links are affected by omitting reflections and diffractions. This is to be expected, since the larger transmission range results in a larger average distance between the communicating pairs, which in turn makes it less likely for NLOSb links to have single-interaction reflections and diffractions. Depending on the selected $r_{NLOSb}$, in the Porto Downtown environment, the proportion of links that are affected ranges between approximately 5\% (when $r_{NLOSb}=$~500~m) and 55\% (when $r_{NLOSb}=$~50~m). In other words, between 45\% and 95\% of NLOSb links relied on the log-distance path loss model in the first place, since the power contribution of single-interaction reflections and diffractions was low. 
Thus, a good approach might be to model reflections and diffractions for NLOSb links whose distance between the communicating pair is relatively small (e.g., up to 50~meters), thus accounting for majority of links that are likely to have single-interaction reflections and diffractions.

Furthermore, for the affected links, when compared against the measurement data, the log-distance path loss model performs well: in the Porto Downtown dataset, for 90\% of the links both the mean and standard error are within 2~dB of the results generated by the model implementing reflections and diffractions. While this might not be the case for all scenarios, it is a good indicator that the most important factor for V2V propagation modeling is the separation of different link types based on the obstruction type. In terms of the processing time, when not calculating reflections and diffractions, the processing time corresponds to the lowest (red) curve in Fig.~\ref{fig:totalSimTime}. 

Furthermore, it has to be noted that selecting the correct communication range is quite important in terms of the processing time. %
Specifically, increasing the range results in quadratic increase in the number of objects that need to be analyzed for a given communication pair. By design, in the extreme case, if the communication range is equal to the size of the simulated area, the number of neighboring objects (and therefore calculations) is quadratic with the number of communicating pairs. Therefore, the communication range for each of the LOS types needs to be carefully chosen so that it is minimized while accounting for potentially communicating pairs.

With regards to the scalability of GEMV$^2$, the trends shown in Fig.~\ref{fig:modelPerformance} are far more important than the actual processing times. The results show linear behavior even for large networks comprising tens of thousands of objects and communicating pairs. We also point out that most operations performed by the model (e.g., R-tree construction, classification of links through object querying, and intersection tests) can be parallelized. Since GEMV$^2$  relies on geometric manipulations of the objects that impact the propagation, analogies can be made to computer graphics problems, where parallel rendering techniques are utilized to perform occlusion/visibility and intersection testing. %
Parallelization techniques can be employed in both the object querying and intersection testing, as well as the R-tree construction. %
Since there is no dependency between different communication pairs (links), parallelizing the computations across different links is straightforward. Furthermore, recent advances in parallel R-tree construction, querying, and intersection testing (e.g., see Luo et.~al in~\cite{luo12}) indicate that significant speed increase can be obtained by using multicore graphics processing units.

\section{Related Work}\label{sec:relWorkComplete}

Several recent studies tackled efficient and realistic simulation of vehicle-to-vehicle channels in different VANET environments. Karedal et al.~\cite{karedal10} and Mangel et al.~\cite{mangel11_2} designed propagation models focused on street intersections, where buildings create NLOSb conditions. Both studies selected representative urban intersections and performed measurements which were then used to design the models and calibrate the path loss and fading parameters. Karedal et al.~\cite{karedal09} designed a %
V2V channel model based on measurements performed in highway and suburban environment at the 5.2 GHz frequency band. The model distributes the vehicles and static objects randomly and analyzes four distinct signal components: LOS, discrete components from vehicles, discrete components from static objects, and diffuse scattering. Based on the measurements, the authors propose a set of model parameters for highway and suburban environment. While it enables modeling of different propagation characteristics (path loss, multipath, Doppler spread, etc.), the proposed model assumes that the LOS component exists, therefore it does not specify how to determine the LOS conditions of the channel and the transitions between LOS, NLOSv, and NLOSb link types. %
Figure~\ref{fig:Outlet} shows that modeling the transitions between the LOS conditions is essential for obtaining realistic results, since the ensuing path loss is the most important component in determining the received power and, consequently, the decodability of the packet.

Sommer et al.~\cite{sommer11} performed measurements and used them for calibrating a computationally efficient path loss model aimed at distinguishing between the LOS and NLOSb conditions. In case of NLOSb, the model calculates the received power based on the length of transmission through buildings and the number of walls through which the transmitted ray travels, while %
diffracted and reflected rays are not accounted for. %
Conversely, empirical studies reported by Anderson in~\cite{anderson98} and Durgin et al.~\cite{durgin98} concluded that reflections and diffractions are the dominant propagation mechanisms for NLOSb links in the 1.9~GHz and 5.9~GHz frequency bands, whereas transmission through buildings was found not to contribute considerably. %

In terms of propagation modeling on a city-wide scale, studies reported by Giordano et al.~\cite{giordano10} and Cozzetti et al.~\cite{cozzetti12} focus on computationally efficient propagation modeling in grid-like urban environments, where streets are assumed to be straight and intersecting at a right angle. While such assumptions hold for certain urban areas, in others they might not (e.g., in the city of Porto -- Fig.~\ref{NLOSExplanation}).
With regards to improving the propagation modeling using location-specific information, Wang et al.~\cite{wang12} utilize aerial photography to determine the density of scatterers in the simulated area. By processing the aerial data to infer the scatterer density, the authors determine the fading level for a given location on the road.  

A number of studies were performed in various VANET environments to estimate the channel by performing measurements and fitting the measured data using well-known models (e.g., log-distance path loss~\cite{00parsons}). 
For example, Paschalidis et~al. in~\cite{paschalidis11} performed measurements in different environments (urban, suburban, rural, highway) and fitted the measurements data to the log-distance path loss model. The path loss exponent (PLE) varied considerably (between 1.83 and 3.59) for different locations and LOS conditions. The large range of PLE values goes to show that a single PLE value can not capture the characteristics of a channel, even for a single location/environment. Therefore, different link types (LOS, NLOSv, NLOSb) need to be distinguished and modeled separately.

When it comes to evaluating the impact of vehicular obstructions, apart from our previous studies described in~\cite{boban11,meireles10}, several experimental studies emphasized the importance of obstructing vehicles. 
Gallagher et al.~\cite{gallagher06} quantified the impact of vehicular obstructions on different parameters, such as packet reception, throughput, and communication range. Interestingly, Gonzalvez et~al. in~\cite{gozalvez12} performed measurements where the impact of vehicular traffic and tall vehicles (buses) also heavily influenced the vehicle-to-infrastructure (V2I) links, despite the roadside units being placed at elevated positions (between 3 and 10~meters) next to or above the roads. Tall vehicles decreased the effective communication range by 40\%, whereas the dense traffic reduced the range by more than 50\%. 
Furthermore, other studies suggested that obstructing vehicles could be an important factor in propagation modeling (e.g.,~\cite{tan08,matolak05}). 
 
The studies above were aimed at measuring the propagation channel characteristics and fitting the models to the already collected measurements. However, research aimed at incorporating the vehicles in the propagation model and therefore \emph{predicting} their effect has been scarce. Apart from our previous work reported in~\cite{boban11}, to the best of our knowledge, there have only been two studies aimed at explicitly introducing vehicular obstruction in propagation modeling. %
Abbas et al.~\cite{abbas12} performed V2V measurements and showed that a single vehicle can incur more than 10~dB attenuation, which is in line with the results reported in~\cite{boban11}. Based on the measurements, the authors designed a stochastic propagation model for highway environments that incorporates vehicular obstructions and determines the time duration of LOS, NLOSv, and NLOSb states using the measured probability distributions of each state. 
Wang et al.~\cite{wang09} perform isolated (``parking lot'') measurements and characterize the loss due to vehicles obstructing the LOS. Furthermore, they model the loss due to vehicles by employing a three-ray knife-edge model, where diffraction loss is calculated over the vehicles and on the vehicle sides. %
Their results show a good agreement between the isolated measurement results and the proposed method.

\section{Conclusions}\label{sec:conclusionsComplete}
We proposed GEMV$^2$, a computationally efficient, geometry-based V2V propagation model. %
Unlike the %
models currently used in VANET simulators, GEMV$^2$ utilizes the geographic descriptors to enable location-specific %
modeling of the V2V channel. Furthermore, the time-dependent component of the channel is accounted for: depending on the density of the vehicles in an area, the channel between two vehicles can change considerably as the surrounding vehicles move and create varying LOS conditions. %
Compared to the more complex geometry-based models (e.g., ray-tracing), the proposed model is beneficial in terms of: 1) computational complexity, since it performs only a subset of complex calculations required for full ray-tracing models; and 2) reduced requirements for geographical information -- the required information is limited to outlines and types of buildings and foliage, and locations and dimensions of vehicles, which are readily available through geographical databases and mobility traces. 

Furthermore, with limited (and often imperfect) geographical description of the simulated area, %
there is a point of diminishing returns in terms of simulation realism, where a marginal improvement in the realism requires a large computational effort. For this reason, 
we used VANET-specific information (e.g., the number of surrounding objects, the dimensions of objects %
and their propagation characteristics, etc.)
to limit the complexity of the model. 
To enable a more efficient modeling, GEMV$^2$ divides the links into three categories: 1) LOS; 2) NLOSv; and 3) NLOSb. 
The results regarding LOS and NLOSv conditions shown in Figs.~\ref{fig:LOSFits} and~\ref{fig:NLOSvFits} indicate that, in order to correctly model LOS and NLOSv links, it is sufficient to consider the main type of propagation mechanism for the respective link type (specifically, two-ray ground reflection model for LOS and vehicles-as-obstacles model for NLOSv~\cite{boban11}), with additional small-scale signal variation proportional to the number of vehicles and the area of buildings and foliage around the communicating pair. %
This allows for an efficient implementation in the simulation environments, as the required calculations are limited to link classification based on LOS conditions and determining the number and area of objects around the communication pair, both of which can be performed efficiently using spatial data structures. On the other hand, for NLOSb links, in certain scenarios, it is beneficial to consider reflections and diffractions as the main propagation mechanisms, since they enable a better estimation of the received power, particularly when vehicles are communicating ``around the corner'' (i.e., where vehicles are on two sides of a corner of a single building, as shown in Figs.~\ref{fig:OutletOverlayReflDiffr} and~\ref{fig:outletAnnotated}). 
However, reflections and diffractions incur a high computational cost: in case of our test data, they accounted for two thirds of the overall computation time. At the expense of losing some accuracy, considerable speed improvement can be achieved if a simpler model (e.g., log-distance path loss) is used to calculate the power for NLOSb links. 

The main goal of GEMV$^2$ is efficient propagation modeling for VANET simulations involving large number of vehicles over large geographic areas. However, since it inherently distinguishes both the link types (LOS, NLOSv, NLOSb) as well as different environments (urban, suburban, highway), GEMV$^2$ can also be used as a basis for fine-grained channel modeling. For example, by using impulse response measurements of LOS, NLOSv, and NLOSb channels in different environments such as those described by Bernard{\'o} et al.~\cite{bernardo13} and Alexander et al.~\cite{alexander11}, GEMV$^2$ can be used to correctly assign the appropriate channel statistics (e.g., delay and Doppler spreads) to each link.   

Finally, we implemented GEMV$^2$ in MATLAB and showed that it can simulate networks with hundreds of thousands of communicating vehicle pairs in different environments (highway, suburban, urban). %
\section*{Acknowledgements}
We are grateful to Rui Meireles and Carlos Pereira for their invaluable help with the measurements and data processing.
\bibliographystyle{IEEEtran}
\bibliography{references}
\clearpage 

\end{document}